\documentclass[structabstract,oldversion]{aa}

\usepackage{graphicx}
\usepackage{txfonts}
\usepackage{natbib}
\usepackage{subfigure}
\usepackage{color}
\usepackage{xspace}
\usepackage{epsfig}
\usepackage{epsf}
\usepackage{comment}
\usepackage{ulem}

\usepackage{hhline}
\usepackage{multirow}
\usepackage{textcomp} 
\usepackage{amssymb}  

\definecolor{gris}{gray}{0.4}

\newcommand{\feh}{$\rm{[Fe/H]}$\xspace}
\newcommand{\GEAR}{\texttt{GEAR}\xspace}

\begin{document}

\title{Pushing back the limits: detailed properties of dwarf galaxies in a $\Lambda$CDM universe}

\author{Yves Revaz\inst{1} \and Pascale Jablonka\inst{1,2}}

\institute{Institute of Physics, Laboratory of Astrophysics, \'Ecole Polytechnique F\'ed\'erale de Lausanne (EPFL), 1290 Sauverny, Switzerland 
\and GEPI, CNRS UMR 8111, Observatoire de Paris, PSL University, F-92125, Meudon, Cedex, France}

\date{Received -- -- 20--/ Accepted -- -- 20--}

\abstract {We present the results of a set of high resolution chemo-dynamical
  simulations of dwarf galaxies in a $\Lambda$CDM cosmology. Out of an original
  $(3.4\,\rm{Mpc}/h)^3$ cosmological box, a sample of 27 systems are re-simulated from
  z=70 to z=0 using a zoom-in technique.  Gas and stellar properties are confronted to the observations in
  the greatest details: in addition to the galaxy global properties, we
  investigated the model galaxy velocity dispersion profiles, half-light radii,
  star formation histories, stellar metallicity distributions, and [Mg/Fe] abundance
  ratios. The formation and sustainability of the metallicity gradients and
  kinematically distinct stellar populations are also tackled. We show how the
  properties of six Local Group dwarf galaxies, NGC 6622, Andromeda II,
  Sculptor, Sextans, Ursa Minor and Draco are reproduced, and how they pertain
  to three main galaxy build-up modes.  Our results indicate that the
  interaction with a massive central galaxy could be needed for a handful of
  Local Group dwarf spheroidal galaxies only, the vast majority of the systems
  and their variety of star formation histories arising naturally from a
  $\Lambda$CDM framework.  We find that models fitting well the local Group
  dwarf galaxies are embedded in dark haloes of mass between $5\times 10^8$ to a
  few $10^9\,\rm{M_\odot}$, without any missing satellite problem. We
  confirm the failure of the abundance matching approach at the mass scale of
  dwarf galaxies. Some of the observed faint however gas-rich galaxies with
  residual star formation, such as Leo T and Leo P, remain challenging.
  They point out the need of a better understanding of the UV-background
  heating.}


   \titlerunning{Detailed properties of dwarf galaxies in a $\Lambda$CDM universe}
   \authorrunning{Revaz, Jablonka}
   \maketitle



\section{Introduction}
\label{introduction}


Dwarf galaxies are the least luminous galactic systems with the
largest dark-to-stellar mass ratio in the Universe. They are very
sensitive to any perturbation, such as stellar feedback (radiative and
thermal), environmental processes (gravitational interactions,
ram pressure stripping), and heating by the cosmic UV-background. As
such they constitute unique laboratories with which to infer the physics leading
to the formation and evolution of galaxies.  Last but not least, in
the hierarchical $\Lambda$CDM paradigm, dwarf galaxies are the most
numerous as well as first galaxies to be formed, playing a key role in the
re-ionization of our Universe
\citep{choudhury2008,salvadori2014,wise2014,robertson2015,bouwens2015,atek2015}.


The first generation of dark matter only (DMO) $\Lambda$CDM cosmological
simulations conclude serious tensions between their results and the
observations 
(see \citet{bullock2017} for a recent review)
: over-prediction of small mass systems around the Milky Way, the
missing satellite problem \citep{klypin1999,moore1999} and the
too big to fail problem \citep{boylankolchin2011,boylankolchin2012}.
These DMO simulations also predicted the dark haloes to follow a universal cuspy
profile \citep{navarro1996,navarro1997}.  while the observations seemed to
favour cored ones \citep{moore1994}.  More recently
\citet{pawlowski2012,pawlowski2013,ibata2013} have advocated a correlation in
phase-space between both the Milky Way and the Andromeda satellites as if they
were all located within a rotationally-supported disc. If definitively confirmed,
such a configuration would be at odds with the $\Lambda$CDM predictions in which
satellites are expected to be isotropically distributed.

On the one hand, these tensions have been a strong motivation for exploring
alternative cosmologies that could directly impact the formation and evolution
of dwarf galaxies \citep[see for
  example][]{spergel2000,lovell2012,rocha2013,vogelsberger2014}.  On the other
hand, they have led to studies revealing the strong impact of the baryonic
physics on the faintest galaxies \citep{zolotov2012,brooks2014}.
While the treatment of baryons in $\Lambda$CDM simulations remains challenging
\citep{revaz2016b}, significant improvements have recently been achieved to address the above issues.

High resolution hydro-dynamical simulations of the Local Group such as
APOSTLE \citep{sawala2016b} or Latte \citep{wetzel2016} demonstrated
that the ram pressure and tidal stripping of the haloes orbiting their
host galaxy, together with the reduction of their mass due to the
UV-background heating and stellar feedback, strongly reduces the
amplitude of the faint tail of the luminosity function. This
potentially solves the missing satellite and too big
to fail problems.  While its reality is still debated
\citep{pineda2017}, the core-cusp problem may be solved by a
very bursty star formation history induced by a strong feedback,
transforming the central cusp into a large density core
\citep{pontzen2014}. This theoretical prediction has been reproduced
in some cosmological hydro-dynamical simulations
\citep{onorbe2015,chan2015,fitts2017}.


We are now at a stage where simulations are mature enough to allow detailed
investigations of the properties of the systems arising from a $\Lambda$CDM
universe at the level offered by present day observations.  This last decade has
seen a wealth of studies focused on the reproduction of the mean or integrated
physical quantities of dwarfs such as their central velocity dispersion, mean
metallicity, total luminosity, gas mass, and half light radius
\citep{valcke2008,revaz2009,sawala2010,schroyen2011,revaz2012,cloet-osselaer2012,sawala2012,cloet-osselaer2014,sawala2016b,wetzel2016,fitts2017,maccio2017}.
Following the semi-analytical cosmological models  \citep[e.g.][]{salvadori2008}, more and more hydrodynamical simulations include some stellar abundance 
patterns in their calculations \citep[e.g.][]{escala2018, hirai2017}.
To definitely validate any theoretical framework, we not only need to
confront the model predictions to the mean galaxy properties via their scaling
relations but also to go one step further and consider their structure, that is,
the galaxy stellar velocity dispersion profiles, stellar metallicity
distributions, stellar abundance ratios as well as more subtle properties such as
their metallicity gradients or kinematically distinct stellar populations.





To do so, we can take advantage of a plethora of accurate
observations in the Local Group dwarf galaxies.  In addition to the standard
morphological, gas and luminosity measurements \citep{mcconnachie2012}, we have
access to line-of-sight velocities revealing the galaxy stellar dynamics
\citep[ex.][]{walker2009,battaglia2008,fabrizio2011,fabrizio2016}.  Deep colour
magnitude diagrams allow to infer the galaxy star formation histories
\citep{dolphin2002,deboer2012,deboer2012b,weisz2014,deboer2014,weisz2014b,santana2016,kordopatis2016,bettinelli2018}.
High resolution spectroscopic observations of individual stars allow us to
determine accurate stellar abundances providing strong constraints on the galaxy
chemical evolution
\citep{shetrone2001,shetrone2003,fulbright2004,sadakane2004, koch2008,aoki2009,cohen2009,frebel2010b,norris2010,cohen2010,tafelmeyer2010,letarte2010,kirby2010,venn2012,lemasle2012,starkenburg2013,jablonka2015,tsujimoto2015}.

Stellar population gradients, in other words a variation of the population properties
with galacto-centric distances, provide constraints on the interplay between
dynamics and chemical evolution
\citep{harbeck2001,tolstoy2004,koch2006,battaglia2006,faria2007,gullieuszik2009,kirby2011,battaglia2011,vargas2014,ho2015,lardo2016,spencer2017,suda2017,okamoto2017}.


In the present study, we follow the formation and evolution of dwarf galaxies in
a $\Lambda$CDM cosmological context, confronting the above observational
constraints and the model predictions.  This work leverages our previous and
preparatory work on simulations of galaxies either in isolation
\citep{revaz2009,revaz2012,revaz2016}, or interacting with a Milky-Way-like
host, through tidal or ram pressure stripping \citep{nichols2014,nichols2015}.
While the present simulations are fully cosmological, they do not include a
massive galaxy.  Nevertheless the environmental impact on galaxy evolution is
taken into account.  Indeed, the model dwarfs experience accretion and merger
events. They are also affected by the strong ionizing UV-background radiation,
which has been identified as being capable of evaporating most of the gas in  
dwarf galaxies 
\citep{efstathiou1992,quinn1996,bullock2000,noh2014}
and forms a natural
constraint to the period of reionization
\citep{susa2004,ricotti2005,okamoto2010,shen2014}.

This paper is organized as follows: In Section~\ref{methods} we briefly describe
our code, \GEAR, its recent improvements and the settings of our cosmological
zoom-in simulations. The results are presented in Section~\ref{results}. They
include the multi-phase structure of the gas (Section~\ref{multiphase}), the
global properties of the dwarf galaxies (Section~\ref{obsglob}), and their star
formation histories (Section~\ref{sfrh}), as well as the line of sight stellar velocity dispersion
profiles, the stellar metallicity distribution function and the [Mg/Fe] abundance ratios
(Section~\ref{detailed_stellar_properties}).  The formation of the age and metallicity gradients are discussed in
Section~\ref{metallicity_gradient} and the origin of the kinematically distinct stellar
populations in Section~\ref{stellar_populations}. We summarize our main results
in Section~\ref{conclusions_discussions}.





\section{Methods}\label{methods}



%
%

All simulations presented in this work have been run with the
chemo-dynamical Tree/SPH code \GEAR, developed by \citet{revaz2012}
and \citet{revaz2016}. We summarize below its main features, which are presented in detail in
the above references.  \GEAR is fully parallel based on
\texttt{Gadget-2} \citep{springel2005}. It includes gas cooling, star
formation, chemical evolution, and Type Ia and II supernova yields
\citep{kobayashi2000,tsujimoto1995} and thermal blastwave-like
feedback \citep{stinson2006}, for which 10\% of the SNe explosion
energy, taken as $10^{51}$ erg, is released in the interstellar medium
($\epsilon=0.1)$. \GEAR follows the pressure-entropy SPH formulation
of \citet{hopkins2013} and operates with individual and adaptive time
steps as described in \citet{durier2012}. In the following, the
initial mass function (IMF) of each model stellar particle is sampled
with the random discrete scheme (RIMFS) of \citet{revaz2016}, i.e., a
stochastic approach that reproduces well the discretization of the
IMF. We employ the Smooth Metallicity Scheme that permits chemical
mixing between stellar particles
\citep{okamoto2005,tornatore2007,wiersma2009} and matches the observed
low level of chemical scatter at very low metallicity in galactic
haloes. The V-band luminosities are derived following the stellar
population synthesis model of \citet{vazdekis96} computed for the
revised \citet{kroupa01} IMF used in our model.

The simulations of galaxies in isolation of \GEAR have successfully
reproduced the main observable properties of dSphs \citep{revaz2012},
including their stellar metallicity distribution function and chemical
abundance trends. We recently conducted an extensive study of the
early pollution of the galaxy interstellar medium (ISM) and of the
resulting dispersion in chemical abundance ratios, [$\alpha$/Fe], at
low metallicity \citep{revaz2016}.  \GEAR has also been tested in the
context of the Assembling Galaxies Of Resolved Anatomy
(AGORA) project \citep{kim2016} where it demonstrated its ability to
simulate Milky Way-like galaxies.
The next section describes some of the recent improvements implemented
in \GEAR.


\subsection{Cooling, UV-background heating and self-shielding}
%

The radiative cooling of the gas has been updated with the
\texttt{Grackle} cooling library \citep{smith2017} adopted by the
AGORA project \citep{kim2014,kim2016}.  We used the equilibrium
cooling mode of the library (metals are assumed to be in ionization
equilibrium) in which cooling rates have been pre-computed by the
\texttt{Cloudy} photo-ionization code \citep{ferland2013}.  Cooling
rates are tabulated as a function of the gas density, temperature and
metallicity. They contain
the contribution of the primordial cooling as well as metal line
cooling determined for solar abundances. The latter contribution is
scaled to the metallicity of the gas.  The heating resulting from the
presence of a redshift-dependent uniform UV-background radiation
\citep{haardt2012} is also taken into account.

At high densities, the gas may no longer be considered as optically thin
and hydrogen becomes self-shielded against the ionizing UV-background
radiation.  
The critical density at which hydrogen becomes
self-shielded is model dependent. It is sensitive to the
method used (ray tracing, moment-based), the photo-ionization rates as well as to the 
exact spectrum of the ionizing photons considered and the adopted hydrogen cross section.
At some level, it may also depend on the  fitting formula used.
Values ranging from
$n_{\rm{H}}=10^{-3}$ to $n_{\rm{H}}=6\times 10^{-2}\,\rm{atom/cm^3}$ have been derived
through different types of radiative transfer models \citep[see for example][]{tajiri1998,aubert2010,yajima2011,rahmati2013}. 
In the present work, we simulated the shielding multiplying the UV-heating by
an Heaviside function centred on the density threshold
$n_{\rm{H}}=0.007$, following \citet{aubert2010}.

\subsection{Star formation}
\label{sfr}


The high resolution of our simulations leads us to deal with gas at
densities higher than $1\,\rm{atom/cm^3}$ which reaches temperatures
lower than $100\,\rm{K}$.  These low temperatures result from the
combined effect of the hydrogen self-shielding against the ionizing
UV-background radiation \citep{tajiri1998,aubert2010,yajima2011} and a
very short cooling time.  In such conditions, the Jeans length of the
gas becomes very small, possibly smaller than the model spatial resolution
and is no longer resolved. Consequently, the noise
inherent to the model discretization may generate perturbations,
leading to a spurious fragmentation of the gas
\citep{truelove1997,bate1997,owen1997}.  A classical solution
to avoid this spurious fragmentation is to include an additional
non-thermal pressure term in the equation of state of the gas, which
ensures that the Jeans length is comparable to the resolution of the
system \citep{robertson2008,schaye2008}.  This additional pressure can
be interpreted as the non-thermal pressure due to the ISM turbulence, which
is unresolved.
Here, we have used the following pressure floor, which is a modified version of the formulation proposed by \citet{hopkins2011}:
\begin{equation}
\label{pressure}
P_{\rm{Jeans},i} = \frac{\rho_i}{\gamma} \left(  \frac{4}{\pi} G h_i^2 \rho_i N_{\rm{Jeans}}^{2/3} - \sigma_i^2  \right),
\end{equation}

in which $G$ is the universal gravitational constant and $\gamma$, the
adiabatic index of the gas fixed to $5/3$.  $h_i$, $\rho_i$ and
$\sigma_i$ are respectively the SPH smoothing length, density, and
velocity dispersion of the gas particle $i$.  This equation is
obtained by requesting the SPH mass resolution of a particle $i$,
$\Delta m_i = 4/3\,\pi\,\rho_i h_i^3$, to be smaller by a factor
$N_{\rm{Jeans}}$ than the Jeans mass.  The difference between our formulation and that of
\citet{hopkins2011} is in the inclusion of an estimation of the local
turbulence $\sigma_i$ summed over the neighbouring particles in the
Jeans mass.
\begin{equation}
\sigma_i = \frac{1}{\rho_i} \sum_{j=1}^{N_{\rm{ngb}}} m_j\, W(r_{ij}, h_i)\, \left|\vec{v}_i-\vec{v}_j \right|^2,
\end{equation}
where $\vec{v}_i$ is the velocity of the particle $i$ and $W(r_{ij}, h_i)$ the SPH kernel.

The new negative term in Eq. \ref{pressure} slightly reduces the level
of the pressure floor.  One drawback of this approach though is that in
high-resolution cosmological simulations designed to study small scale
structures, this additional non-thermal term can dominate the gas
pressure.  In that case, the small clumps of gas are artificially kept
at hydro-static equilibrium as the direct consequence of the
additional pressure term which counterbalances self-gravity. Despite a
strong cooling, the collapse of these clumps and  star formation are
therefore hampered.  Even more problematic is the fact that those
clumps can eventually merge with larger systems, bringing fresh gas
(that should have form stars in realistic conditions), strongly
biasing the star formation history by producing short star bursts.

The solution adopted here is to set a star formation density threshold
based on the Jeans polytrope \citep{ricotti2016}, written as a
function of the density and temperature, given that the SPH smoothing
length $h_i$ directly correlates with the density $\rho_i$:
\begin{equation}
\rho_{\rm{SFR},i} = \frac{\pi}{4} G^{-1} N_{\rm{Jeans}}^{-2/3} h_i^{-2} \left( \gamma \frac{ k_{\rm{B}}}{\mu m_{\rm H}} T  + \sigma_i^2  \right).
\label{eq:rho_sfr}
\end{equation}

This critical density corresponds to the density above which, at given
temperature $T$, the Jeans pressure dominates over the thermal one,
or, in other words, the system is strongly dominated by unresolved
physics.  Above this density threshold, stars may form with a
probability parameterized by the star formation efficiency parameter
$c_\star$ which set from our past experience in modelling dwarf
galaxies and equal to $0.01$ in this study.

\subsection{Simulations}\label{simulations}


All simulations have been run in a cosmological $\Lambda$CDM context from z=70 to z=0.
The cosmological parameters are taken from the full-mission Planck observations \citep{planck2016}, namely :
$\Omega_{\Lambda} = 0.685$,
$\Omega_{\rm{m}} = 0.315$,
$\Omega_{\rm{b}} = 0.0486$,
$H_0 = 67.3\,\rm{[km/s/Mpc]}$,
$n_s = 0.9603$ and
$\sigma_8 = 0.829$.

\subsubsection{Setting up the resolution}

\citet{revaz2016} have demonstrated that in order to properly follow
the physics involved in the formation of dwarf galaxies, in particular
their chemical evolution, the stellar mass should be larger than about
$1'000\,\rm{M_\odot}$.  Therefore our fiducial simulations have a
stellar particle initial mass of $1'024\,\textrm{M}_\odot/h$   . The gas
particles can form up to four stellar particles and an initial mass of
$4'096\,\textrm{M}_\odot/h$. This results in a good sampling of the star
formation rate. The mass of the dark matter particles is
$22'462\,\textrm{M}_\odot/h$, corresponding to a baryonic fraction of
$0.154$ \citep{planck2016}.  Choosing a resolution of $2\times 512^3$
($(2^9)^3$, known also as level 9) particles, equivalent to a total of
$134'217'728$ dark matter particles, this fixed the size of our
cosmological box to $3.4\,\textrm{Mpc}/h$.

\subsubsection{The dark matter only and zoom-in simulations}

We first ran a dark matter only simulation (DMO) with the above
resolution.  Using a simple friend-of-friend halo finder, we extracted
all haloes which at $z=0$ have virial masses between $10^{10}$ and $4
\times 10^8\,\rm{M_\odot}$$/h$. We then traced back with time the positions of the particles of the
extracted haloes and computed their corresponding convex hull
\citep{onorbe2014}. This hull defines the refined region where baryons
are subsequently included.  Outside these areas, the resolution was
gradually degraded down to level 6.  All initial conditions have been created
using the \texttt{MUSIC}\footnote{https://www-n.oca.eu/ohahn/MUSIC/index.html}
code \citep{hahn2011}.

From these 198 haloes extracted at $z=0$, we only kept the 62 ones
whose initial mass distribution was compact enough to allow a zoom-in
re-simulation.  In the  \texttt{MUSIC} code, the  refined region cannot be larger than half the box size.
Each simulation has been simulated 
with the same resolution and the same physics.
Starting from $z_{\rm{init}}=70$, they have all been run
down to $z=0$. $z_{\rm{init}}$ has been
determined by ensuring that the rms variance of the initial density
field is between $0.1$ and $0.2$ \citep{knebe2009}.  The initial gas
temperature is $80\,\rm{K}$. The softening length of the particles are fixed in
comoving coordinates up to $z=2.8$ and kept constant in physical
coordinates afterwards.  At $z=0$, gas particles and dark matter
particles have softening lengths of $10$ and
$50\,\rm{pc}/h$  respectively, which is sufficient to resolve the inner structure of
dSph galaxies.


\section{Results}\label{results}


Among the 62 simulations, 27 have a final V-band luminosity larger than $10^5\,\rm{L_\odot}$, corresponding to Local Group classical spheroidal dwarfs (dSph) or irregular dwarfs (dIrr). The remaining 35 fall in the domain of the ultra-faint dwarf galaxies (UFD) and will be analysed in a forthcoming paper.

\subsection{Catalogue of simulated galaxies}
\label{final_galaxies}

\begin{figure*}[!h]
        \centering
        \leavevmode   
        \includegraphics[width=0.85\hsize]{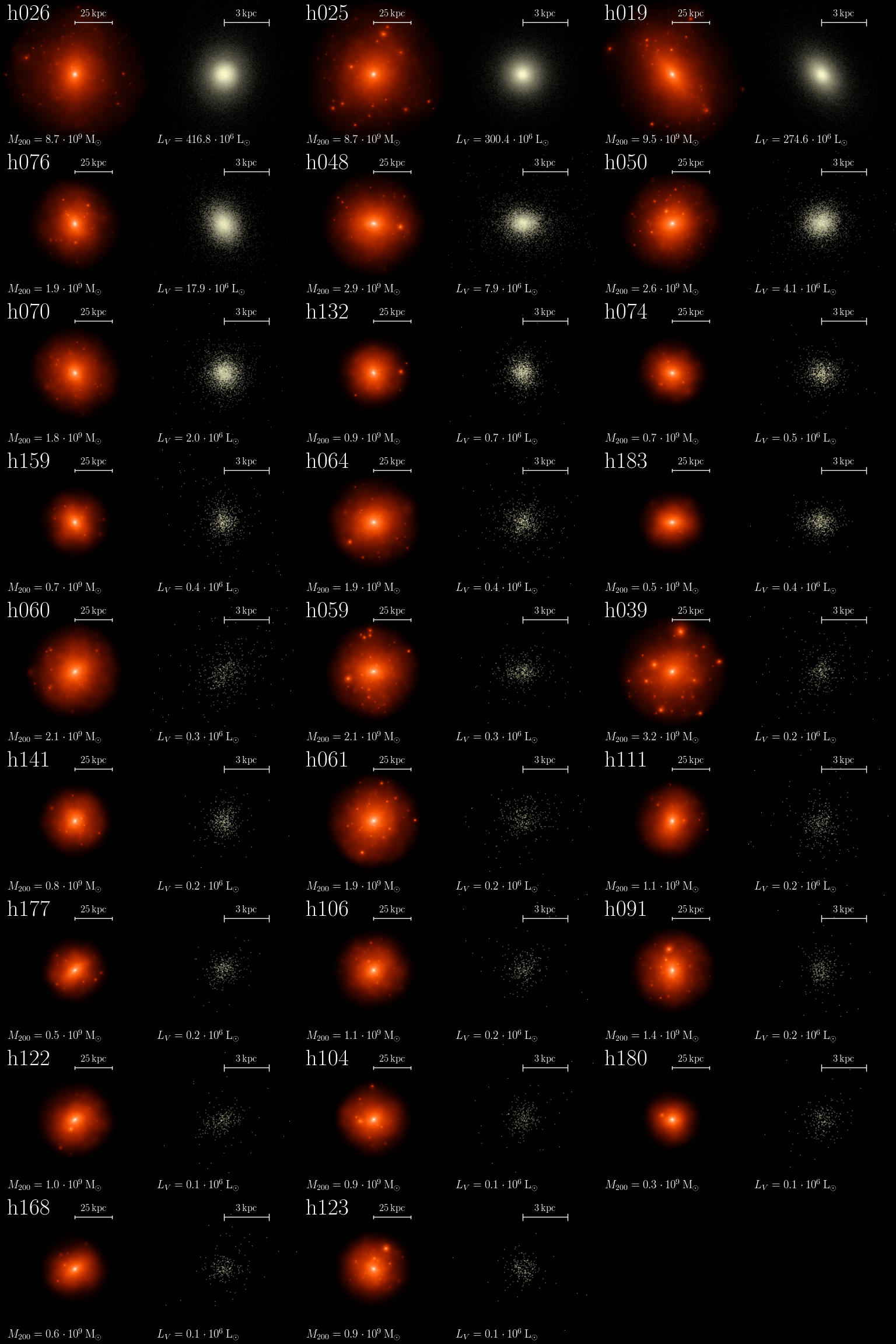}
        
        \caption{Catalogue of the 27 zoom-in haloes at $z=0$ with a final luminosity higher than $10^5\,\rm{L_\odot}$ corresponding to Local Group classical dwarfs. 
                The dark halo surface density of each model truncated to its respective $R_{200}$ radius is shown with orangeish colours. 
                The size of the box is $100\,\rm{kpc}$.
                A zoom on the corresponding stellar component, more precisely, the projection of the luminosity of each stellar particle
                is represented in yellowish colours. Here, the size of the box is $10\,\rm{kpc}$.
                The models are sorted from left to right and top to bottom according to their total luminosity.
                }
        \label{fig:maps}
\end{figure*}

The catalogue of our 27 dwarf galaxies, ranked by their final V-band
luminosity, is given in Fig.~\ref{fig:maps}. For each galaxy, it
displays the dark matter surface density and the stellar
distribution. The identification of each model that is used
further in this paper is given in the upper left quadrant. This
catalogue is supplemented with the full list of the extracted galaxy
physical properties in Table~\ref{tab:catalogue}.
\begin{table*}
  \begin{center}
  \begin{tabular}{c c c c c c c c c c c c c}
    Model ID & $L_{\rm{V}}$           & $M_{\star}$            & $M_{\rm{200}}$         &  $M_{\rm{gas}}$         & $M_{\rm{cold\,gas}}^{r<r_t,T<10^4\,\rm{K}}$ & $M_{\rm{hot\,halo\,gas}}^{T\ge 10^4\,\rm{K}}$    & $R_{\rm{200}}$ & $r_{1/2}$   & $\sigma_{\rm{LOS}}$ & \feh \\
       & $[10^6\,\rm{L_\odot}]$ & $[10^6\,\rm{M_\odot}]$ & $[10^9\,\rm{M_\odot}]$ &  $[10^6\,\rm{M_\odot}]$ &  $[10^6\,\rm{M_\odot}]$ & $[10^6\,\rm{M_\odot}]$ & $[\rm{kpc}]$   & $[\rm{pc}]$ & $[\rm{km/s}]$ & dex \\ 
        \hline
\texttt{h026}  & $451.12$ & $537.84$ & $8.71$ & $460.04$ & $195.37$ & $262.64$ & $49.2$ & $  1$ & $32.3$ & $-0.75$  \\
\texttt{h025}  & $309.37$ & $520.16$ & $8.67$ & $246.10$ & $83.15$ & $162.50$ & $49.1$ & $  1$ & $30.6$ & $-0.61$  \\
\texttt{h019}  & $291.39$ & $433.92$ & $9.53$ & $280.17$ & $99.06$ & $180.72$ & $50.7$ & $  0$ & $30.5$ & $-0.64$  \\
\texttt{h021}  & $233.79$ & $334.81$ & $6.83$ & $183.01$ & $142.43$ & $235.68$ & $45.4$ & $  1$ & $28.1$ & $-0.59$  \\
\texttt{h076}  & $18.56$ & $38.33$ & $1.88$ & $22.96$ & $14.09$ & $8.70$ & $29.5$ & $  0$ & $13.3$ & $-1.04$  \\
\texttt{h048}  & $8.02$ & $13.71$ & $2.92$ & $26.53$ & $9.70$ & $16.75$ & $34.2$ & $  0$ & $11.2$ & $-1.10$  \\
\texttt{h050}  & $4.16$ & $9.56$ & $2.62$ & $15.11$ & $3.48$ & $11.62$ & $33.0$ & $  0$ & $10.4$ & $-1.37$  \\
\texttt{h070}  & $2.02$ & $5.83$ & $1.83$ & $0.01$ & $0.00$ & $0.01$ & $29.2$ & $  0$ & $10.8$ & $-1.61$  \\
\texttt{h132}  & $0.75$ & $2.10$ & $0.90$ & $0.52$ & $0.00$ & $0.52$ & $23.1$ & $  0$ & $9.3$ & $-1.86$  \\
\texttt{h074}  & $0.50$ & $1.35$ & $0.69$ & $0.39$ & $0.00$ & $0.39$ & $21.2$ & $  0$ & $9.1$ & $-2.06$  \\
\texttt{h159}  & $0.42$ & $1.07$ & $0.67$ & $0.01$ & $0.00$ & $0.01$ & $21.0$ & $  0$ & $8.9$ & $-2.24$  \\
\texttt{h064}  & $0.41$ & $1.07$ & $1.88$ & $5.16$ & $0.00$ & $4.92$ & $29.5$ & $  0$ & $9.7$ & $-2.00$  \\
\texttt{h183}  & $0.37$ & $1.00$ & $0.54$ & $0.02$ & $0.00$ & $0.02$ & $19.5$ & $  0$ & $8.6$ & $-2.09$  \\
\texttt{h060}  & $0.27$ & $0.69$ & $2.06$ & $3.70$ & $0.00$ & $3.69$ & $30.4$ & $  0$ & $10.4$ & $-2.45$  \\
\texttt{h039}  & $0.27$ & $0.64$ & $3.24$ & $17.18$ & $0.00$ & $17.18$ & $35.4$ & $  1$ & $11.7$ & $-2.55$  \\
\texttt{h059}  & $0.26$ & $0.67$ & $2.07$ & $4.25$ & $0.00$ & $4.24$ & $30.5$ & $  0$ & $9.3$ & $-2.25$  \\
\texttt{h141}  & $0.22$ & $0.58$ & $0.76$ & $0.47$ & $0.00$ & $0.47$ & $21.8$ & $  0$ & $8.3$ & $-2.17$  \\
\texttt{h061}  & $0.22$ & $0.53$ & $1.93$ & $3.45$ & $0.00$ & $3.44$ & $29.8$ & $  0$ & $9.1$ & $-2.26$  \\
\texttt{h111}  & $0.20$ & $0.49$ & $1.09$ & $0.11$ & $0.00$ & $0.07$ & $24.6$ & $  0$ & $10.4$ & $-2.57$  \\
\texttt{h177}  & $0.19$ & $0.49$ & $0.54$ & $0.01$ & $0.00$ & $0.01$ & $19.4$ & $  0$ & $7.7$ & $-2.32$  \\
\texttt{h091}  & $0.17$ & $0.40$ & $1.36$ & $0.84$ & $0.00$ & $0.83$ & $26.5$ & $  0$ & $10.1$ & $-2.38$  \\
\texttt{h106}  & $0.15$ & $0.37$ & $1.09$ & $0.08$ & $0.00$ & $0.08$ & $24.6$ & $  0$ & $9.9$ & $-2.62$  \\
\texttt{h122}  & $0.14$ & $0.36$ & $0.97$ & $0.01$ & $0.00$ & $0.01$ & $23.7$ & $  0$ & $9.1$ & $-2.41$  \\
\texttt{h104}  & $0.14$ & $0.34$ & $0.92$ & $0.28$ & $0.00$ & $0.27$ & $23.3$ & $  0$ & $9.0$ & $-2.36$  \\
\texttt{h123}  & $0.13$ & $0.34$ & $0.91$ & $0.05$ & $0.00$ & $0.05$ & $23.2$ & $  0$ & $7.6$ & $-2.35$  \\
\texttt{h180}  & $0.12$ & $0.30$ & $0.29$ & $0.01$ & $0.00$ & $0.01$ & $15.9$ & $  0$ & $6.9$ & $-2.45$  \\
\texttt{h168}  & $0.12$ & $0.29$ & $0.56$ & $0.02$ & $0.00$ & $0.02$ & $19.7$ & $  0$ & $8.3$ & $-2.66$  \\
  \end{tabular}         
  \caption{Properties of the 27 zoom-in dwarf galaxy models.} 
  \label{tab:catalogue}
  \end{center}
\end{table*}
%

\subsection{Generic evolution and properties of the multi-phase gas}
\label{multiphase}

\begin{figure*}
        \centering
        \leavevmode   
        \includegraphics[width=\hsize]{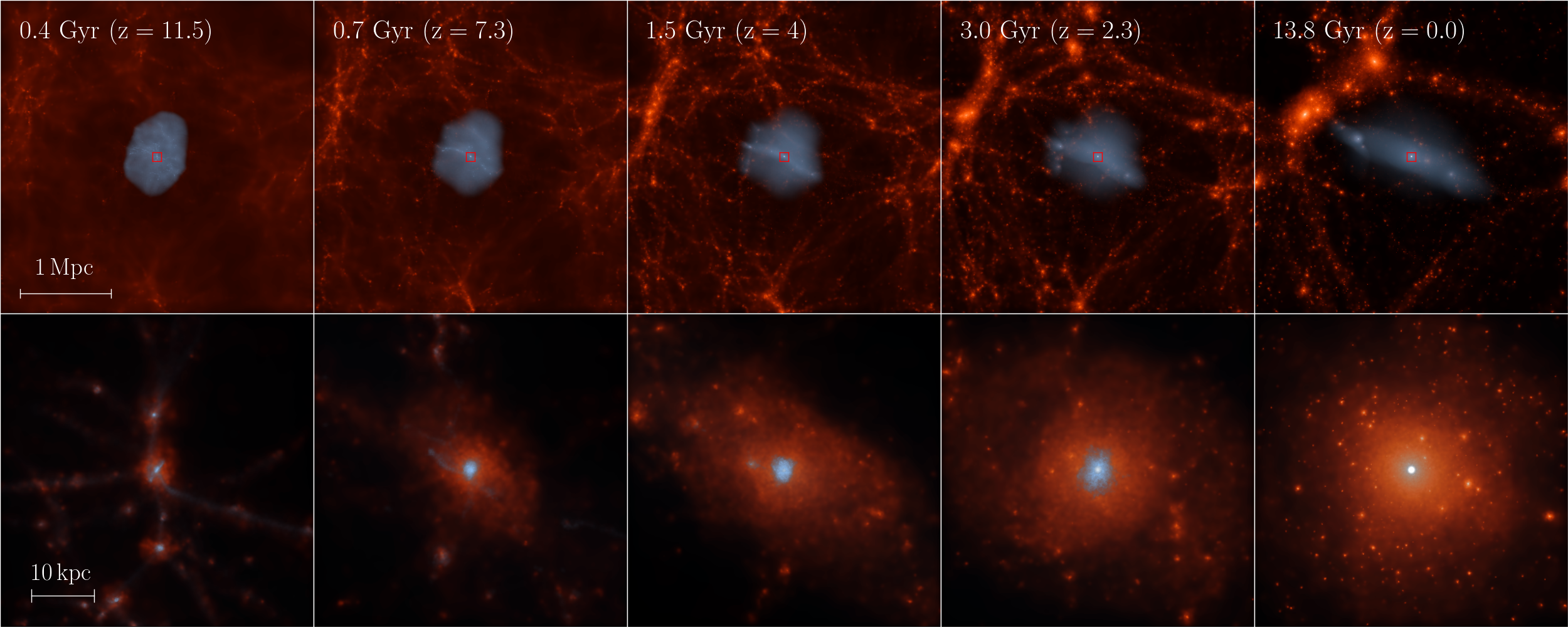}
        \caption{Evolution with time of the halo \texttt{h025}. The top panels display the evolution of the full  $(3.4\,\rm{Mpc}/h)^3$  cosmological box. The red central squares identify the areas displayed in the bottom panels. In all quadrants, stars are shown in white, the gas is displayed in blue, and the dark matter in orange. The gas is only present in the regions which have been re-simulated at the highest resolution.}
        \label{fig:evol}
\end{figure*}

To illustrate the formation and evolution of dwarf galaxies in our
zoom-in simulations, Fig.~\ref{fig:evol} displays a few snapshots of
the model \texttt{h025} from $z=11$ down to $z=0$. The top panels show
the evolution in the full cosmological box, with the refined
regions identified by the gas distribution in blue. The bottom panels
are magnifications of the dwarf itself.
%



\begin{figure}
        \centering
        \leavevmode   
        \includegraphics[width=\hsize]{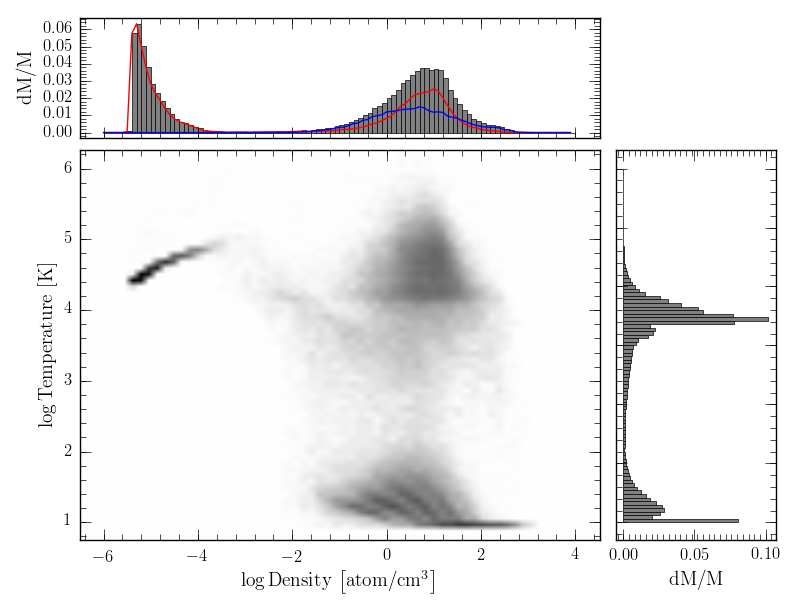}
        \caption{Distribution of the gas in $\log \rho$ versus $\log T$ for the model  \texttt{h025} at a redshift z=0. 
                The intensity of the grey colour scales with the logarithmic of the 2-D mass-weighted histogram of the gas.  
                The upper and right panels display the gas fraction as a function of its density and temperature, respectively.
                The red and the blue curves in the upper panel correspond to the warm/hot and cold gas distribution.}
        \label{fig:logRhologT}
\end{figure}

Figure~\ref{fig:logRhologT} shows the distribution of the gas of model
\texttt{h025}, in a $\log \rho$ vs $\log T$ diagram at a redshift of
zero.  Three gas phases are considered: 

(i) the cold dense gas ($\rho \gtrsim 10^{-1}\,\rm{atom/cm^3}$, $T<1000\,\rm{K}$), 
(ii) the low density warm gas ($\rho \lesssim 10^{-3}\,\rm{atom/cm^3}$) forming a
halo that can extend up to the virial radius of the dwarf, and (iii)
the dense warm or hot gas ($\rho \gtrsim 10^{-1}\,\rm{atom/cm^3}$,
$T>1000\,\rm{K}$).  The cold phase of the gas arises from compression
during the first collapse of the dark halo. Because of the hydrogen
self-shielding, for density higher than $\sim 10^{-2}\,\rm{atom/cm^3}$, the cooling dominates and the gas
temperature drops to $10\,\rm{K}$ which corresponds to the floor
temperature adopted in the simulations.  At this high density, the cold gas
can form stars which in turn, via the feedback of the supernovae
explosions, heats the gas up to $10^5-10^6\,\rm{K}$ forming the
warm/hot phase.  This over-pressurized gas expands and cools back adiabatically to
very low temperature. A cycle between the cold and warm/hot phase is
naturally maintained.  For densities higher than about
$10^{-1}\,\rm{atom/cm^3}$ the gas is then consequently multi-phase.

\begin{figure}
        \centering
        \leavevmode   
        \includegraphics[width=\hsize]{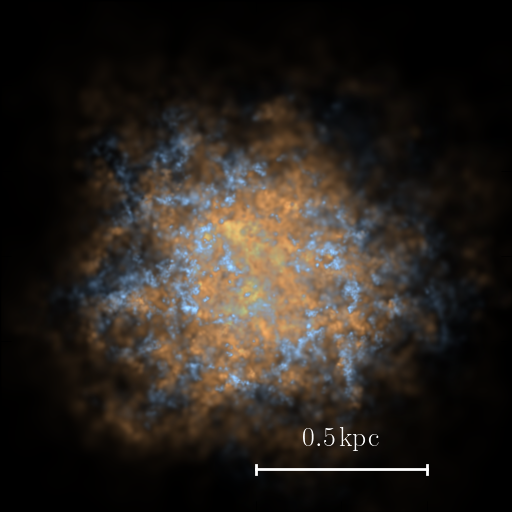}
        \caption{Surface density of the dense cold (in blue) and dense warm/hot (in orange) gas extracted from  the model \texttt{h025} at z=0.}
        \label{fig:gas}
\end{figure}

Figure~\ref{fig:gas} displays the surface density of the dense cold  and dense warm/hot gas in a box of size of $1.5\,\rm{kpc}$.
The cold gas appears very clumpy compared to the warm/hot one which is smoother.
This underlines the fact that the gas is far from equilibrium as being continuously shaken by supernovae explosions that 
drives the formation of bubbles, inter-penetrating each-other.
The gas distribution strongly homogeneous and stars are not restricted to form only in the inner few parsecs.
A quick look a the line of sight of velocity dispersion of the gas show values of about $14\,\rm{km/s}$ for the cold phase, up to $17\,\rm{km/s}$ for the
warm/hot phase. Less-massive systems display cold gas velocity dispersion lower than $10\,\rm{km/s}$.

\subsection{Scaling relations}
\label{obsglob}

Our simulations are run in a cosmological context, however, they do
not evolve under the influence of a massive host such as the Milky Way
or Andromeda.  Nevertheless, while the most detailed observational
constraints can only be gathered in the Local Group, we see in
the following that most of their observed features result from
physical processes independently of any interaction with a central
galaxy.

Our comparison sample is essentially based on the list of galaxies listed in the
compilation of Local Group galaxies by \citet{mcconnachie2012}. We restrict the
sample to the satellites of the Milky Way and Andromeda galaxies brighter than
$10^5\,\rm{L_{\odot}}$.  We did not include the Sagittarius dSph as it is
strongly stripped, making it difficult to get a census of its global properties.





\subsubsection{Luminosity - velocity dispersion}
\label{luminosity_velocity}

\begin{figure}
        \centering
        \leavevmode   
        \includegraphics[width=9cm]{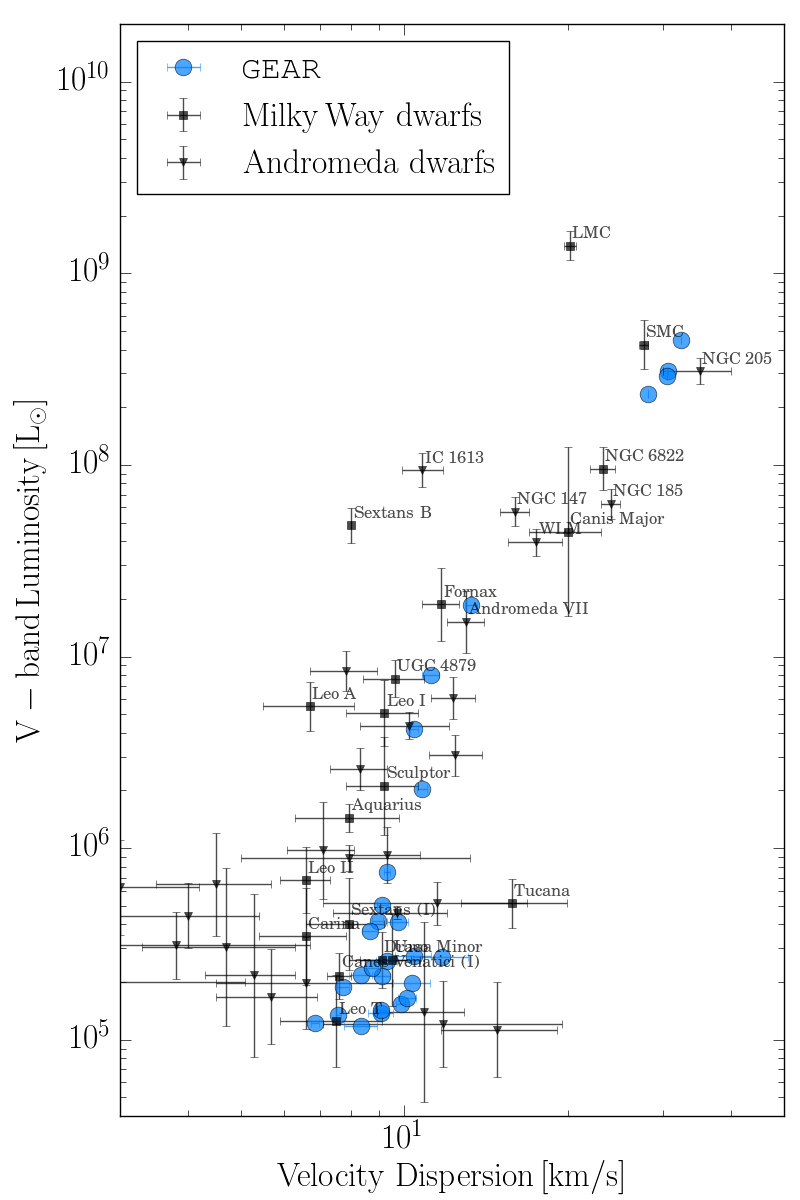}
        \caption{Galaxy V-band luminosity versus its stellar line-of-sight velocity dispersion.  The blue circles correspond to the predictions of the simulations, while the back squares and triangles stand for the Milky Way's and M31 satellites respectively (see Section~\ref{obsglob} for the references).  For clarity, we only label the M31 satellites brighter than $10^7\,\rm{L_{\odot}}$.}
        \label{fig:LvvsSigma}
\end{figure}

Figure~\ref{fig:LvvsSigma} displays the V-band luminosity of the simulated
dwarf galaxies with respect to their line of sight stellar velocity dispersion,
$\sigma_{\rm{LOS}}$, calculated as the velocity dispersion along a given line of sight of
our stellar particles inside a $1\,\rm{kpc}$ cylindrical radius. For each
galaxy, $\sigma$ is computed seven times, choosing different lines of sight. The
quoted velocity dispersion is taken as the mean of the seven realizations and the
error corresponds to their standard deviation.  The predictions of our models
are compared with a compilation of observations of both Milky Way and Andromeda
galaxies as defined above.


Our model galaxies nicely reproduce the observed L$_{\rm{V}}$-$\sigma$ relation
over four order of magnitudes, from nearly $10^9\,\rm{L_{\odot}}$ down to
$10^5\,\rm{L_{\odot}}$, corresponding to velocity dispersions of $35\,\rm{km/s}$
down to $7\,\rm{km/s}$. Thereby they provide insights on the origin of the
diversity of formation histories, hence very different luminosities, of the
systems with $\sigma_{\rm{LOS}}\lesssim 10\,\rm{km/s}$. Indeed, this results from
the interplay between the build-up of the galaxy dark halo, the stellar feedback
and the intensity of the UV-background heating during the re-ionization epoch.
The primordial structures in the densest regions of the cosmic web merge on
shorter timescales than those in more diffuse ones.  Before the re-ionization
epoch, the gas density of the former haloes is already high enough to be
self-shielded against the coming UV-background.  Moreover their potential well
are deep enough to retain the fraction of gas which will be heated.  These two
factors prevent the star formation to be rapidly quenched (see
Section~\ref{sfrh}) and the galaxies can reach luminosities up to
$10^7\,\rm{L_{\odot}}$.  In contrast, the formation timescale of galaxies in
more diffuse regions is longer. As a consequence, the final halo is not yet
assembled at the onset of the re-ionization epoch. It leaves the constitutive
sub-haloes with shallow potential well. The gas is not self-shielded anymore and
can evaporate, hence the star formation quickly stops,
building very low luminosity systems down to $10^5\,\rm{L_{\odot}}$ or even
lower in the regime of UFDs.  In the most extreme cases, the stellar feedback by
itself is sufficient to considerably reduce the star formation rate even before
the re-ionization epoch.

These mechanisms lead to a non-monotonic relation between dark mater haloes and
their stellar content.  Consequently, the assumption of abundance matching
between dark halo and stellar masses
\citep{moster2010,guo2010,sawala2011,behroozi2013,moster2013} breaks down below
$ \sim 10\,\rm{km/s}$.  In the simulations of \citet{sawala2015} the dispersion
in luminosity at given halo mass originated from the interaction between the
the Milky Way and its satellites.  Here, we show that the assembly of the dwarf
systems itself is sufficient to generate this scatter.

While we do not produce systems with velocity dispersions below $5\,\rm{km/s}$ such as
the six Andromeda satellites, we note the large uncertainties on their velocity
measurement and the absence of similar systems around the Milky Way.
{
It is worth noting that there is a dearth of models in the luminosity range 
$2\times 10^7$ to $2\times 10^8\,\rm{L_{\odot}}$. We can only attribute this dearth to a lack of
statistics.
}

\subsubsection{Metallicity - luminosity}

\begin{figure}
        \centering
        \leavevmode   
        \includegraphics[width=9cm]{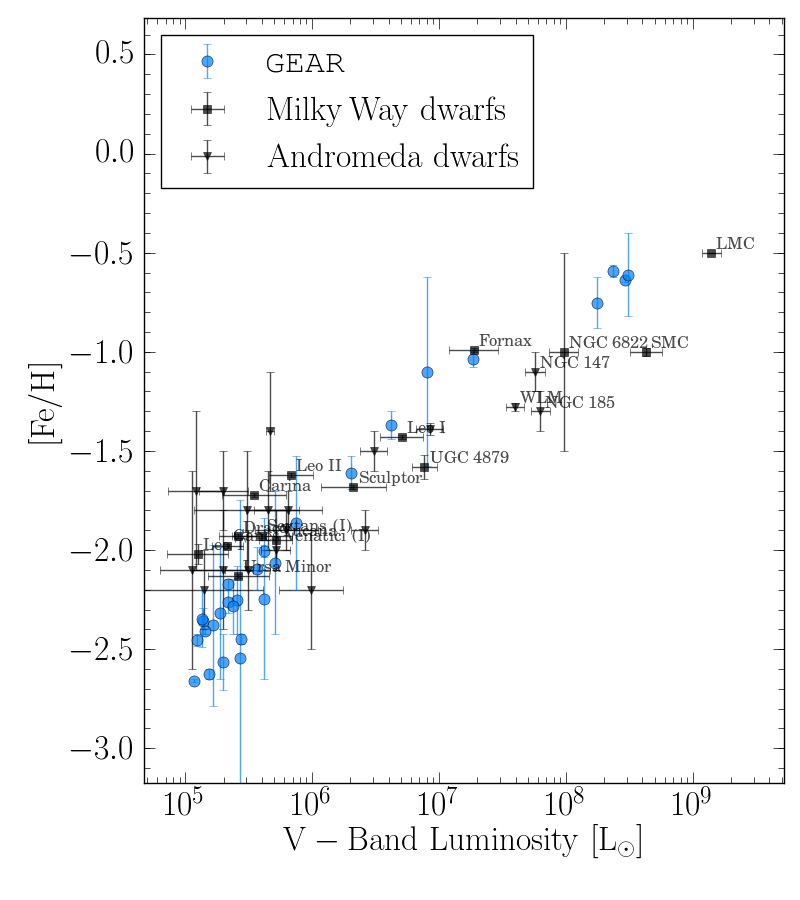}
        \caption{Galaxy mean metallicity (\feh) as a function of
          its final V-band luminosity ($L_{\rm{V}}$). The blue circles
          corresponds to the model predictions, for which \feh is
          computed as the median of the galaxy stellar metallicity
          distribution function.  The error bars correspond to the
          difference between the mode and the median values of this
          distribution.  The Milky Way and M31 satellites, with
          reliable \feh determination (see text for details), are identified with black
          squares and triangles, respectively.  For clarity, we only label the M31 satellites brighter than $10^7\,\rm{L_{\odot}}$.}
        \label{fig:FeHvsLv}
\end{figure}

Figure~\ref{fig:FeHvsLv} displays the median of the stellar metallicity distribution
(\feh) of each dwarf model galaxy, as a function of its V-band luminosity
($L_{\rm{V}}$).  The error bars correspond to the difference between the mode
and the median of the \feh distribution.  Our models are compared with our Local
Group sample of galaxies, restricting to galaxies that benefit from medium
resolution spectroscopy with metallicity derived either from spectral synthesis
or Calcium triplet (CaT) calibration.

Our models again match the observations over four dex in luminosities.  Though,
below $10^6\,\rm{L_{\odot}}$ there is a tail of faint dwarfs which have a median
metallicity below the observed range.  Given the fact that those systems have
correct luminosities and velocity dispersions, this discrepancy most likely
arises from the fact that 
our IMF and yields are kept constant all along the simulations. A more detailed
model should consider POP III stars with a different IMF \citep{bromm2013} and yields \citep{iwamoto2005,heger2010},
as assumed in some other studies \citep{verbeke2015,salvadori2015,jeon2017}.
This could play a role, in particular, for the very first generation of
stars and more strongly impact the smallest systems. Another point is that we
have here some of the lightest haloes which could well be disrupted by the
proximity of a massive galaxy through tidal stripping.  A similar trend has been
found in \citet{onorbe2015,wetzel2016} and \citet{maccio2017} at even higher
luminosities.

\subsubsection{Size - luminosity}\label{half_light_radius}

\begin{figure}
        \centering
        \leavevmode   
        \includegraphics[width=9cm]{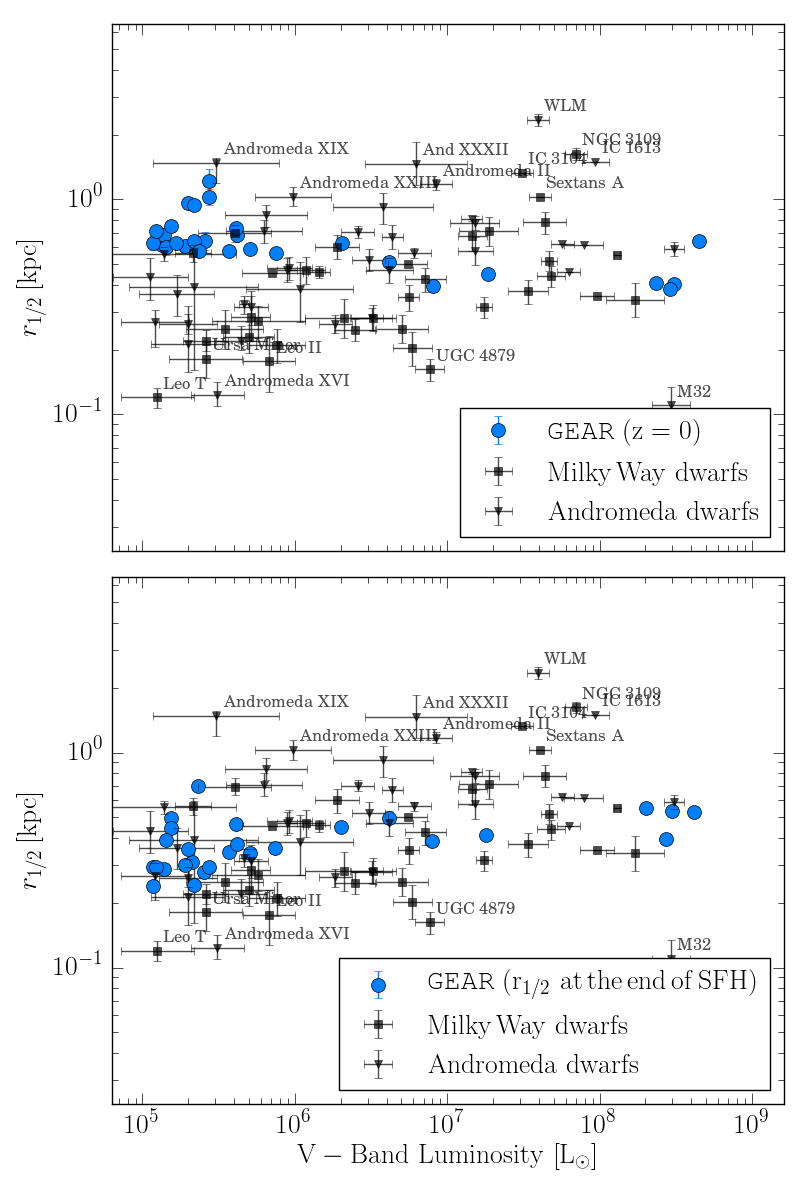}
        \caption{Galaxy half-light radius as a function of its
          final V-band luminosity ($L_{\rm{V}}$).  The blue hexagons
          show the model predictions at z=0 (top panel) or soon after
          the end of the star formation activity (bottom panel).
          The Milky Way and M31 satellites are identified with black squares and triangles, respectively. 
              For clarity, we only label galaxies with extreme half light radius, 
              larger than $1\,\rm{kpc}$ or smaller than $200\,\rm{pc}$.
          }
        \label{fig:r12vsLv}
\end{figure}

Figure~\ref{fig:r12vsLv} presents the relation between the half light radius
($r_{1/2}$) and the V-band luminosity for our model galaxies.
For each of our dwarf, we first removed any contamination from
sub-haloes containing stars, falling within its $r_{200}$.  This was performed
with a halo finder following the \texttt{SUBFIND} algorithm 
\citep{springel2001}.  Then, we fitted a modified King model to the
projected radial stellar profile:
\begin{equation}
I(R) = \frac{I_0}{\left( 1 + (R/R_c)^2 \right)^a} +  b,
\end{equation}
where $I_0$ is the central surface brightness of the standard King model, $R_c$
is the core radius, $a$ sets the outer slope, and $b$ represents some background
noise.  We measured the total luminosity within a radius which is defined as the
one at which the background noise dominates. Concretely, this is set when the fitted profile
deviates from more than $0.5\%$, in logarithmic scale, from the profile without
noise ($b=0$).  The half light radius is then the projected radius containing
half the total luminosity.

The upper panel of Fig.~\ref{fig:r12vsLv} shows the result of this
calculation at $z=0$.  Above $10^6\,\rm{L_\odot}$, our models fall
within the range of the observed values.  Moving below
$10^6\,\rm{L_\odot}$, the mean $r_{1/2}$ of the models is kept nearly
constant, thereby staying at the level of the upper envelope of the
galaxy observed sizes. The lower panel of Fig.~\ref{fig:r12vsLv} shows
the same $r_{1/2}$ - luminosity relation when the size of the
galaxies are calculated soon after the quenching of their star
formation when this happens, that is essentially for the faintest systems. 


The agreement between the models and the observations is clearly
improved, for the following reason: both dark matter and stellar
components are sampled with live particles in the simulations. These
particles interact and exchange linear and angular momentum.  As stars
originates from the gas, they inherit their velocity dispersion, which
is slightly smaller than that of the dark halo.  The energy exchange
between the dark matter and the stars is subsequently possible leading
to dynamical heating of the stars. This results in the secular
flattening of the central stellar component and therefore increases
$r_{1/2}$.  This effect is a direct consequence of the resolution of
our simulations and more generally of any similar modelling.  Would
the dark matter be simulated as a perfectly smooth component, either
as a fixed potential or with an extremely high resolution, this
heating would disappear.  We could also have set the gravitational
smoothing length to very large values, however this would have
affected all the dynamics of the systems.  It is worth noting that
such heating mechanism is akin to the one described by
\citet{jin2005}, where a dark halo is composed of massive black holes.
We stress that we have discarded all other possible source of heating,
either spurious numerical ones, namely (i) dynamical heating due to
too small a gravitational softening length, (ii) inaccuracy of the
gravity force in the treecode, (iii) non-conservation of
momentum at the time of the supernova feedback, as well as physical
ones such as (i) a continuous accretion of small dark haloes, (ii) an
adiabatic expansion due to the quick gas removal resulting from the
UV-heating, (iii) a continuous mass loss of the stellar population.

It is still true that our models do not reach the same compactness as some
of the observed dwarfs, such as Leo T or Andromeda XVI with a half
light radius as small as $100\,\rm{pc}$.  This discrepancy is also
faced by other groups as for example \citet{fitts2017},
\citet{maccio2017}, or \citet{jeon2017} who explore the domain of
ultra-faint dwarf galaxies.  Hence, it seems that this is a genuine
problem of $\Lambda$CDM simulations which deserves further
investigation.

\subsubsection{Gas mass - luminosity}
\label{gas}

\begin{figure}
        \centering
        \leavevmode   
        \includegraphics[width=9cm]{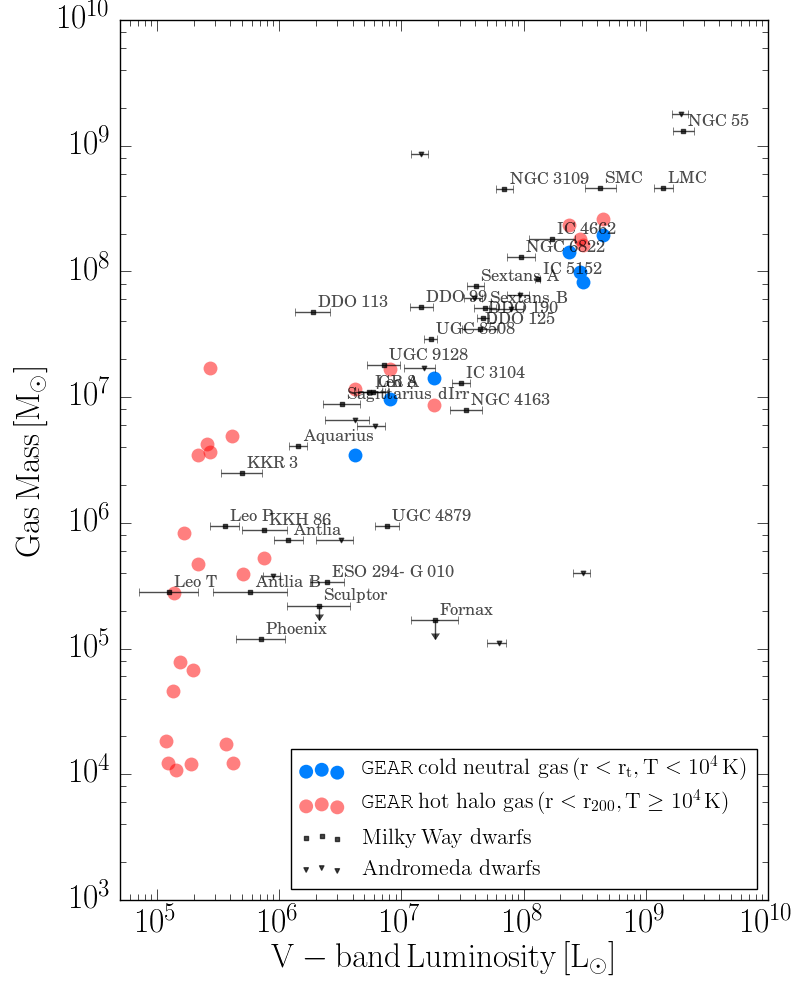}
        \caption{Galaxy gas mass content versus its V-band luminosity at z=0. For all galaxies, if present, 
                        the cold neutral gas phase ($T < 10'000\,\rm{K}$) found inside the tidal radius $r_{\rm{t}}$ is shown with blue circles.
                        The red circles corresponds to the hot halo gas ($T > 10'000\,\rm{K}$) which extend up to the virial radius $r_{200}$.
                        For comparison, the mass of neutral hydrogen (HI)  observed in some of the Milky Way and  M31 satellites is shown in black.  
                        For Sculptor and Sextans we used the upper limits detection values as given by \citet{mcconnachie2012}.}
        \label{fig:MgvsLv}
\end{figure}

Figure \ref{fig:MgvsLv} displays the relation between the mass of the different gas phases and
the final V-band luminosity of the model galaxies.  The models are
compared to the HI gas masses of the Local Group galaxies provided by
\citet{mcconnachie2012}. Although denser and colder components can be
present, HI remains the dominant gas component, by a factor of approximately five,
of the dwarf systems \citep{cormier2014}.

Extracting the different gas phases, H2, HI, and HII, requires targeted procedures,
which are not yet implemented in \GEAR  \citep[see for example][]{derijcke2013,crain2017}.
For the sake of our present purpose, we extracted for each model galaxy
cold and hot halo gas component.
The cold gas component ($T<10'000\,\rm{K}$) is define here as the neutral gas component
and shares the same extension as the stars.
The hot halo gas ($T \ge 10'000\,\rm{K}$) extends much further and we have derived its mass
up to the virial radius, as reported in Table~\ref{tab:catalogue}.
While all models show the presence of hot gas,
we stress that this gas will be easily removed by ram pressure stripping while dwarfs orbit around
their host galaxy.

Model galaxies above $L_{\rm{V}}>10^7\,\rm{L_\odot}$ are able to keep their cold gas and can still form stars up to z=0
(see Section~\ref{sfrh}).
We only have three models in the luminosity range $L_{\rm{V}}=10^6$ to $L_{\rm{V}}>10^7\,\rm{L_\odot}$ (\texttt{h048},\texttt{h050},\texttt{h070}) 
and two of them still have cold gas in agreement with the observations. 
Between  $L_{\rm{V}}=10^5$ and $L_{\rm{V}}>10^6\,\rm{L_\odot}$ our models are lacking the gas which is expected from the observations.
Larger dark matter haloes would have deeper potential wells and would therefore retain their gas more easily. However, correct modelling of the
luminosities and metallicities of these galaxies would then require the use of a stronger
supernovae feedback. This would result in velocity dispersions too high to be compatible with the observations.
Given the fact that our feedback is already low, the most likely source of this dearth of cold gas is the intensity of the UV-background
at the origin of the gas heating.

%
%


\subsection{Star formation histories}\label{sfrh}

 The galaxy star formation rates have a $\sim 100\,\rm{Myr}$
 periodicity which results from the global pulsation of the systems as
 a direct consequence of the supernovae feedback.  All galaxies
 display an additional short timescale variation. For the faintest
 systems, the periods of star formation interruption occur over
 a timescale of about $10\,\rm{Myr}$. This timescale is reduced to
 $0.5\,\rm{Myr}$ for the most luminous systems, as a consequence of
 the strong inhomogeneity of the ISM (see
 Section~\ref{multiphase}). Indeed, despite a strong supernova
 feedback boosted by an adiabatic phase \citep{stinson2006}, some
 regions of the dwarfs include dense and cold clumps of gas that can turn into stars at
 nearly all times.

\begin{figure*}
        \centering
        \leavevmode   
        \subfigure[$L_{\rm{V}}>10^8\,\rm{L_{\odot}}$ : sustained]{\resizebox{0.33\hsize}{!}{\includegraphics[angle=0]{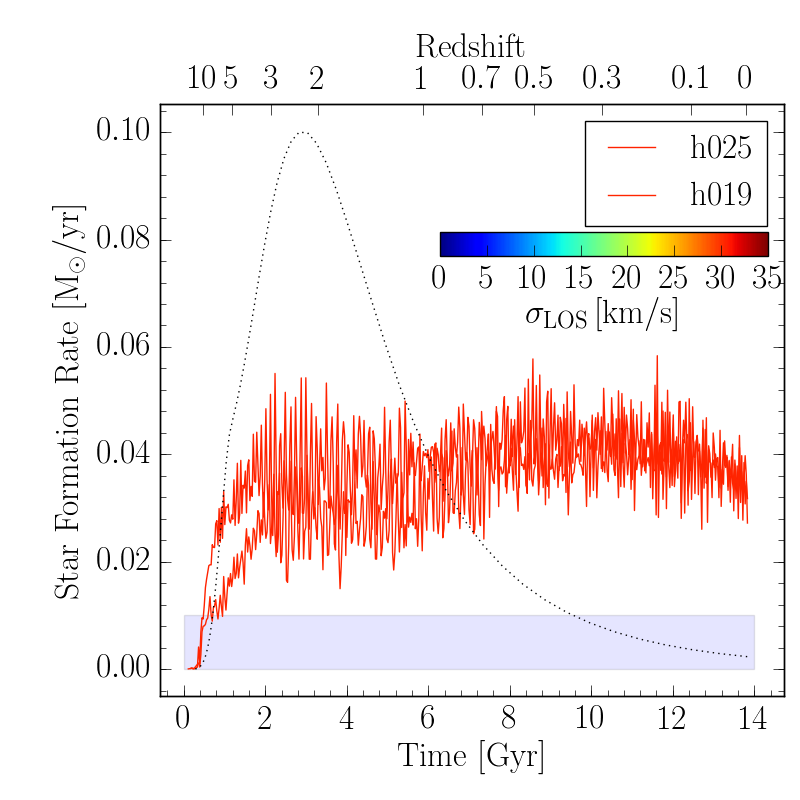}}}   
        \subfigure[$10^6\,\rm{L_{\odot}}<L_{\rm{V}}<10^8\,\rm{L_{\odot}}$ : extended]{\resizebox{0.33\hsize}{!}{\includegraphics[angle=0]{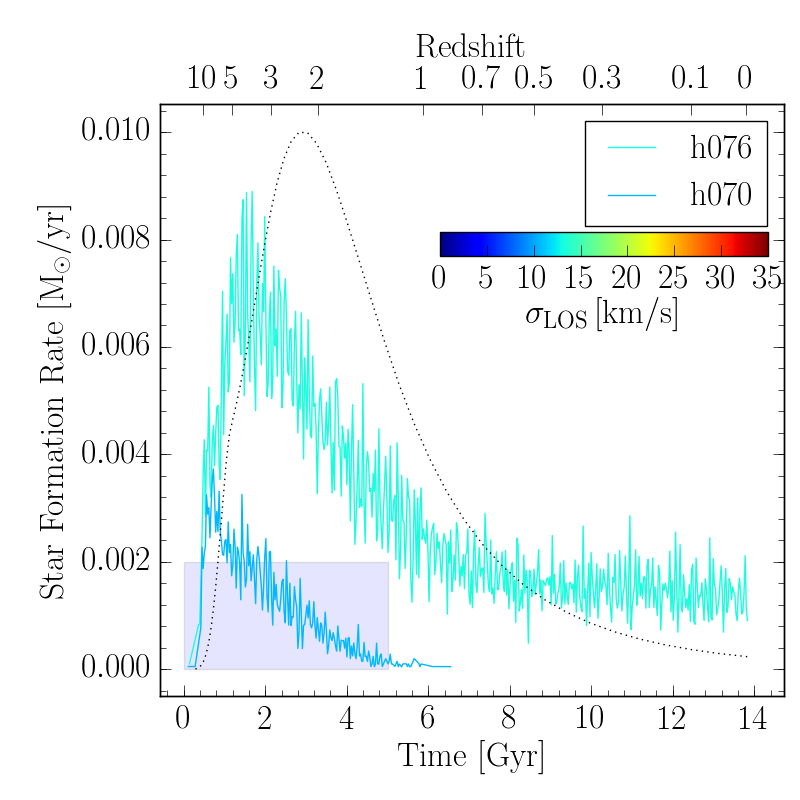}}}
        \subfigure[$L_{\rm{V}}<10^6\,\rm{L_{\odot}}$ : quenched]{\resizebox{0.33\hsize}{!}{\includegraphics[angle=0]{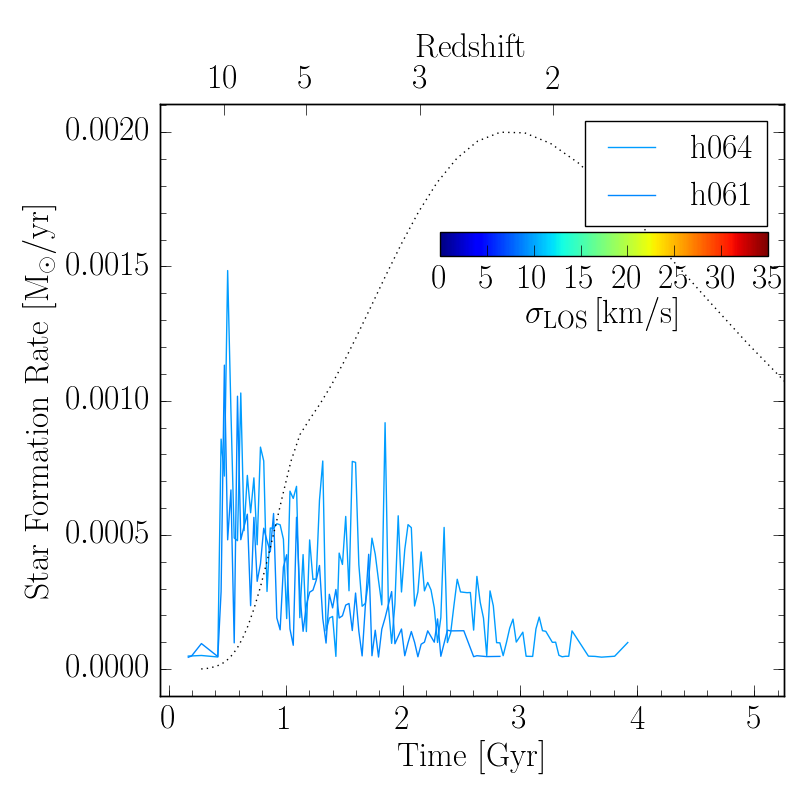}}}
        \caption{The three categories of star formation histories observed: sustained, extended and quenched. The colours code the model galaxy velocity dispersion. The dotted black curves show the redshift dependent hydrogen photo-heating rate due to the UV-background photons.
The curves are normalized so that the peak of the photo-heating rate coincides with the top of the y-axis. 
To illustrate the change in intensity and length of the star formation histories between the
three categories, the shaded rectangles in panel (a) and (b) have dimensions corresponding to the maximal x and y-axis extensions of the panels (b) and (c), respectively.}
        \label{fig:SFR}
\end{figure*}

As expected, the final luminosity ($L_{\rm{V}}$) of our model dwarfs
strongly correlates with the shape of their formation histories.  We
divide our models into three categories dependent on their $L_{\rm{V}}$
range. In the following we will refer to them as sustained, extended
and quenched.  A few representative cases of each of these three
categories are shown in Fig.~\ref{fig:SFR}.  The strength of the
UV-background heating is indicated by the dotted black curve. It
represents the hydrogen photo-heating rate due to the UV-background
photons following the model of \citet{haardt2012}.

\begin{itemize}
        
        \item{(a) $L_{\rm{V}} > 10^8\,\rm{L_{\odot}}$, sustained:} 
        The star formation rate of those massive and luminous dwarfs increases over $1$ to $2\,\rm{Gyrs}$. This period
        is followed by a rather constant SFR plateau. 
        These systems are massive enough to resist
        the UV-background heating and, once formed, to form stars
        continuously . This sustained star formation activity that
        lasts up to $z=0$ results from the self-regulation between
        stellar feedback and gas cooling \citep{revaz2009,revaz2012}.
        
        \item{(b) $10^6\,\rm{L_{\odot}}<L_{\rm{V}}<10^8\,\rm{L_{\odot}}$, extended:} In this luminosity range, the star formation
        is clearly affected by the UV-background. After a rapid increase, the star formation activity is damped
        owing to the increase of the strength of the UV-heating. However, at the exception of the \texttt{h070} halo which is definitively quenched after $6.5\,\rm{Gyr}$, 
        the potential well of those dwarfs is sufficiently deep to avoid a complete quenching. The star formation activity extends to $z=0$,
        however, at a much lower rate than the original one.
        
        \item{(c) $L_{\rm{V}}<10^6\,\rm{L_{\odot}}$, quenched:} The
          potential well of those galaxies is so shallow that the gas
          heated by the UV photons escape the systems. Star formation
          is generally rapidly quenched after $2$ or $3\,\rm{Gyr}$.
          Only halo \texttt{h064} shows signs of activity up to
          $4\,\rm{Gyr}$.  Those galaxies may be considered as true
            fossils of the re-ionization in the nomenclature of
          \citet{ricotti2005}.  They are all faint objects with only
          old stellar populations.
        
\end{itemize}

%

\subsection{Detailed stellar properties of local group dSphs}
\label{detailed_stellar_properties}

In this section, we go one step further by investigating whether the
models emerging from our zoom-in $\Lambda$CDM simulations not only
reproduce the galaxy integrated properties over several dex in
luminosities, but also match their detailed stellar properties, such
as velocity dispersion profiles, stellar metallicity distribution, and
abundance ratios.

For each class of model star formation history (sustained, extended,
quenched), we have looked for well documented Local Group dwarf
galaxies.  One of our stringent criterion was the availability of a
significant number of stars observed at high resolution ($R>15'000$).
This is indeed essential to derive accurate chemical abundances. In
the following, we will particularly focus on magnesium and iron.


For the quenched galaxies, we have selected Sextans, Ursa Minor and
Draco\footnote{Leo I has only two stars measured at high resolution 
\citep{suda2017}, which are insufficient to constrain a model.}. 
Andromeda II and Sculptor, for which a $6$ to
$8\,\rm{Gyr}$-long star formation history has been derived
\citep{weisz2014,skillman2017,deboer2012} are our comparison galaxies
for the extended star formation cases.  Only few observed systems fall
in the range of luminosity and metallicity of the sustained class of
our models.  NGC\,6822 seems to be the best match in the global
relation diagrams. Unfortunately these massive systems are also more
distant and do not possess high resolution spectroscopic follow-up.
We have not considered the LMC and SMC because of their mutual
interaction which evidently affected their evolution.

\begin{figure*}
        \centering
        \leavevmode  
        \subfigure[NGC\,6822]{\resizebox{0.49\hsize}{!}{\includegraphics[angle=0]{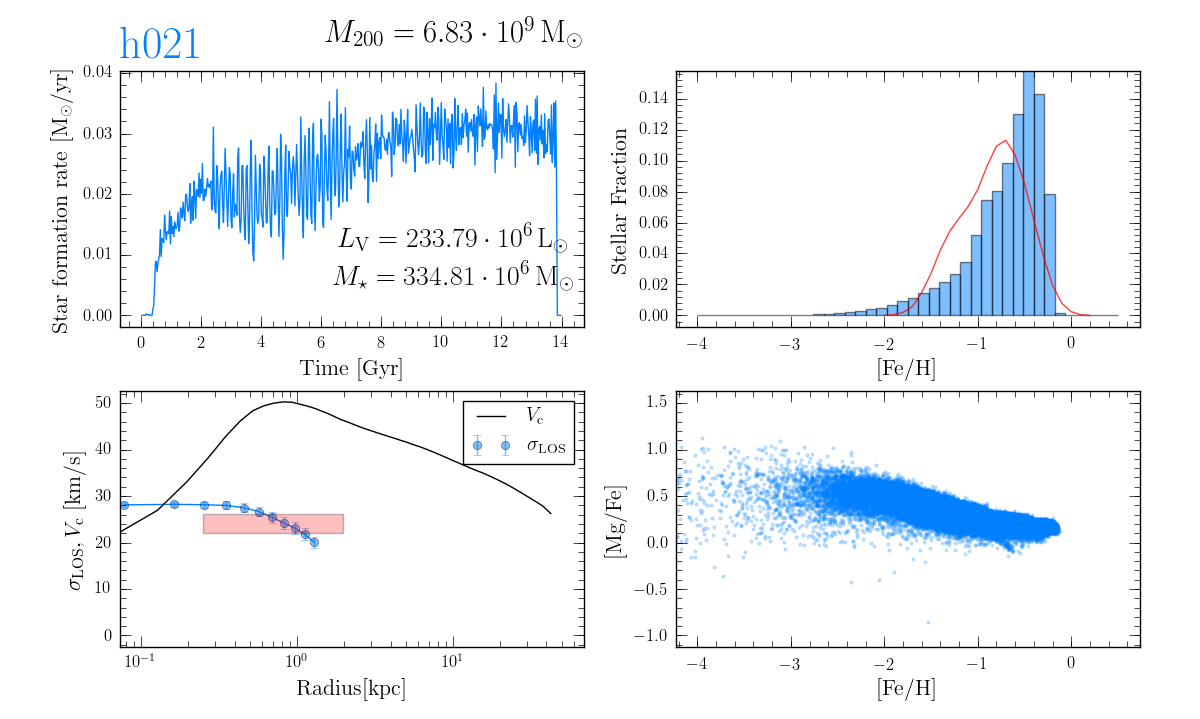}}}         
        \subfigure[Andromeda II]{\resizebox{0.49\hsize}{!}{\includegraphics[angle=0]{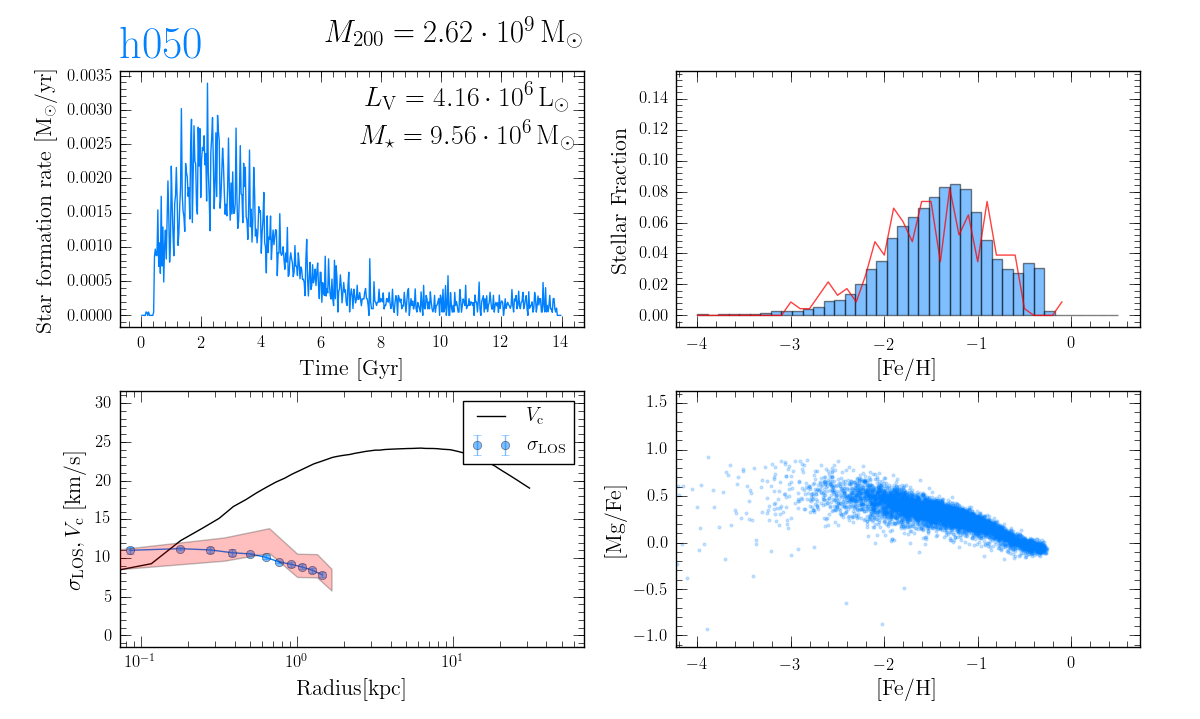}}}          
        \subfigure[Sculptor]{\resizebox{0.49\hsize}{!}{\includegraphics[angle=0]{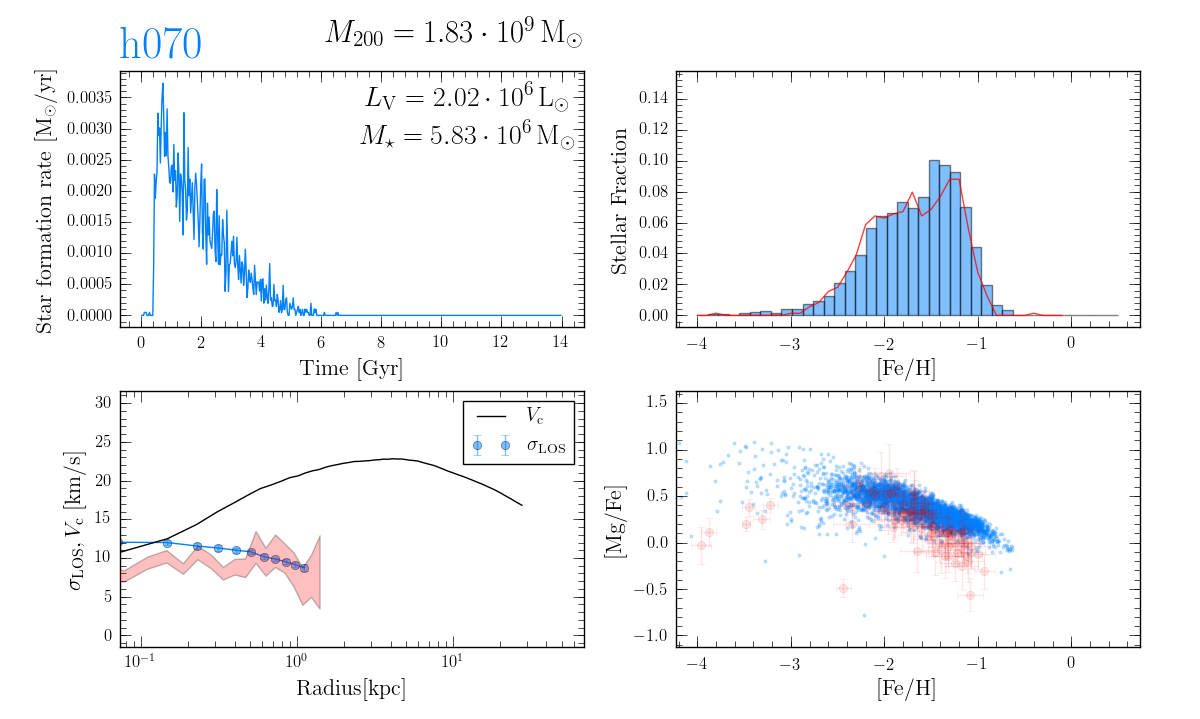}}}   
        \subfigure[Sextans]{\resizebox{0.49\hsize}{!}{\includegraphics[angle=0]{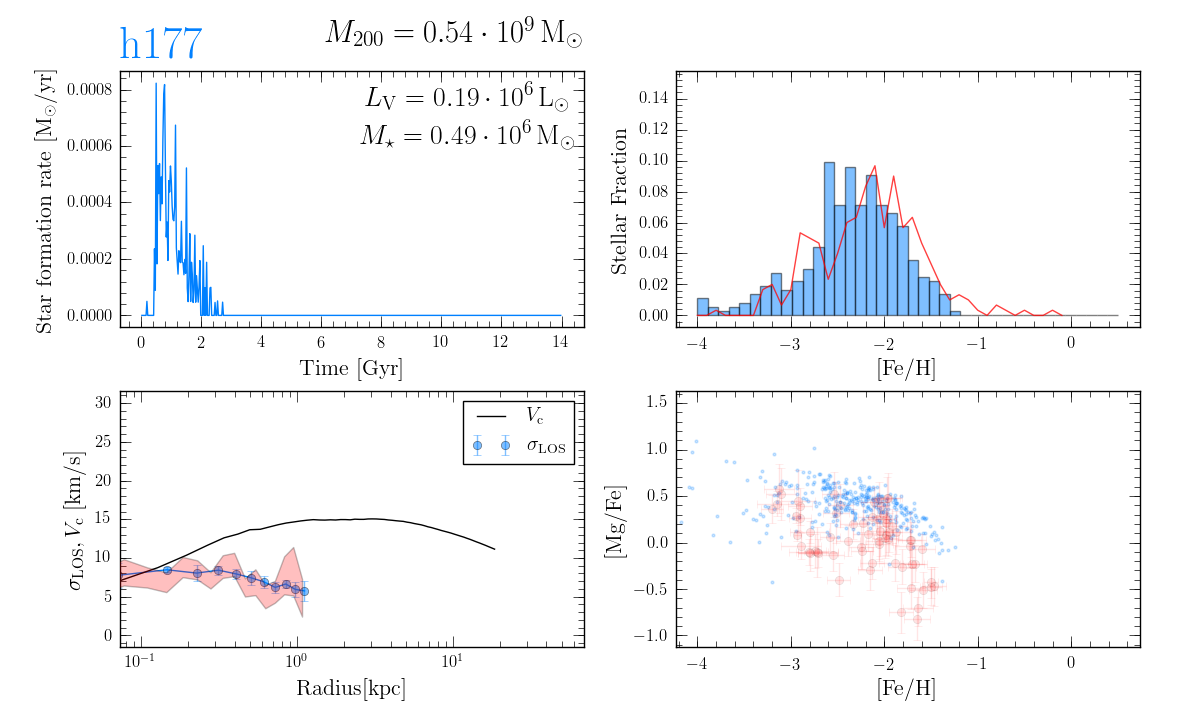}}}   
    \subfigure[Ursa Minor]{\resizebox{0.49\hsize}{!}{\includegraphics[angle=0]{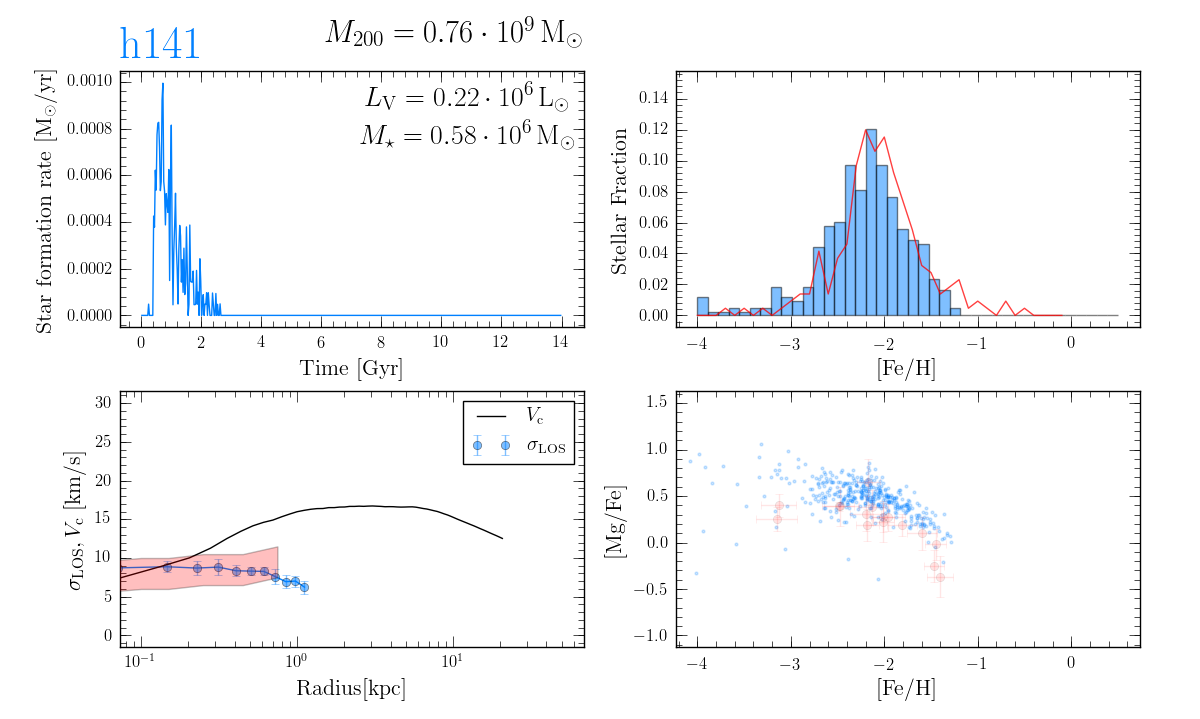}}}   
        \subfigure[Draco]{\resizebox{0.49\hsize}{!}{\includegraphics[angle=0]{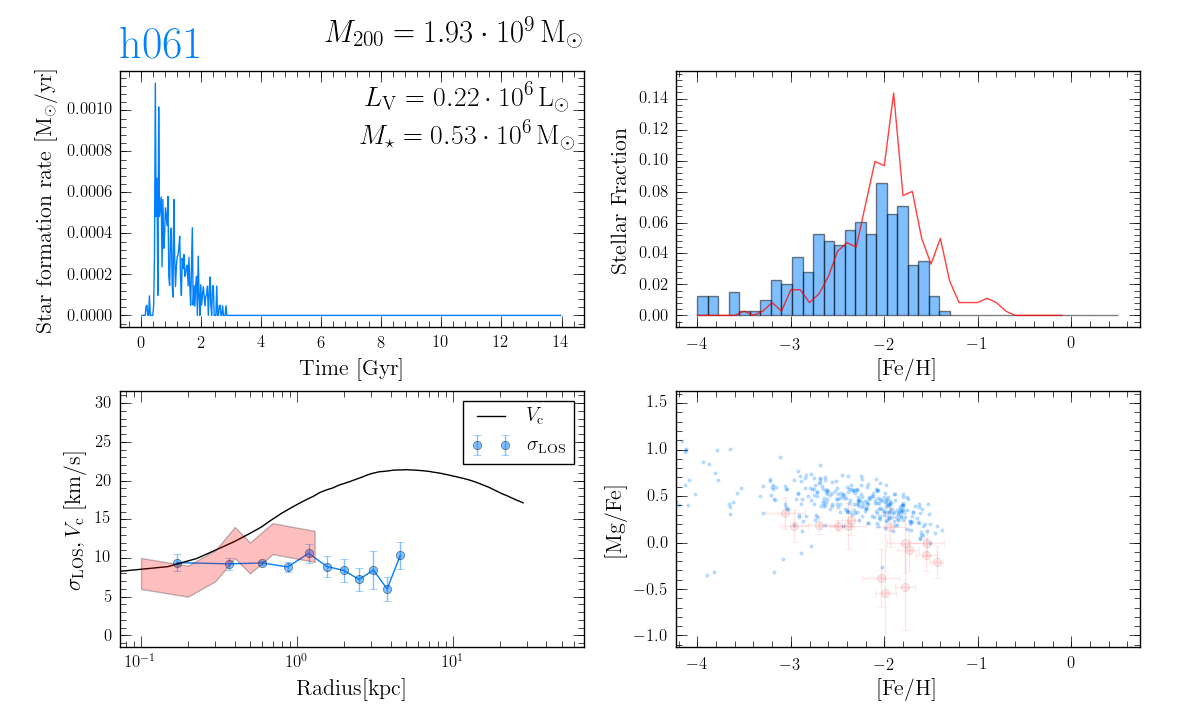}}}         
        \caption{
Comparison of the stellar properties of the six Local Group dwarf galaxies: NGC\,6822, Andromeda II, Sculptor, Sextans, Ursa Minor and Draco.
The model id,   the model galaxy total mass ($M_{200}$), total luminosity ($L_{\rm{V}}$), and total stellar mass ($M_\star$), which are calculated
inside the virial radius $R_{200}$ are indicated on the upper left hand side of each series of four panels.
For each galaxy, the upper left panel displays the model star formation history. The lower left panel shows the line of sight velocity dispersion of the model galaxy in blue, as compared to the observations (red shaded area). The black line corresponds to the total circular velocity.
The upper right panel displays the model galaxy stellar metallicity distribution. The lower right panel gives the galaxy [Mg/Fe] vs [Fe/H] distribution. In both quadrants, the model is seen in blue and the observations in red.}
        \label{fig:lgdSphs}
\end{figure*}

Failures can be as informative as successes. In this regard, we point
out that two well documented galaxies do not match any of our model
star formation histories, the Carina and Fornax dSphs.  Carina is
known for its distinct episodes of star formation with the majority of
its stellar mass formed at intermediate ages
\citep{hurley-keller1998,weisz2014,deboer2014,kordopatis2016,santana2016}. Moreover,
The Carina Project has recently demonstrated the complex structure of
its velocity dispersion, potentially arising from strong gravitational
disturbances \citep{fabrizio2011,fabrizio2016}. The population of
Fornax is dominated by intermediate age stars and stars as young as
$200\,\rm{Myr}$ are also present
\citep{coleman2008,saviane2000,deboer2012b}. The very peculiar star
formation histories of these Carina and Fornax could be the
consequence of a strong interaction with a massive host galaxy as
proposed by \citet{pasetto2011} for Carina, and indeed we have none in
our cosmological volume.
The comparison between the six selected dSphs and the models matching
the galaxy total V-band luminosity, stellar line of sight velocity
dispersion, stellar metallicity distribution, and when available,
their [Mg/Fe] ratio (where Mg is used as a proxy for
$\alpha$-elements), is provided in Fig.~\ref{fig:lgdSphs}.

For each dSph, the lower left panel compares the line of sight
velocity dispersion profile of the model (blue curve) with the
observations in red.  In addition, we display in black the model
total circular velocity.  For NGC\,6822 the line of sight velocity
dispersion have been determined by \citet{tolstoy2001},
\citet{kirby2013} and \citet{swan2016} with values ranging between
$23.2\,\rm{km/s}$ and $24.5\,\rm{km/s}$. We adopted a value of
$24\,\rm{km/s}$ with a reasonable error bar of $\pm 2\,\rm{km/s}$.
For Andromeda II, we took the x-axis velocity dispersion profile of
\citet{ho2012}.  For Sculptor and Sextans, we used the publicly
available radial velocity measurement of \citet{walker2009} and only
kept stars with a membership probability larger than $90\%$.  For
those two galaxies, we could, just as for the models, directly derive
the luminosity weighted line of sight velocity dispersion in circular
logarithmic radial bins.  Because the size of the bins increases with
galacto-centric distances, the noise is reduced at larger radii.  For
Draco and Ursa Minor, the profile is obtained by by-eye fitting the
plots of \citet{walker2009b} assuming a constant error of
$2\,\rm{km/s}$.

The upper right panel of Fig.~\ref{fig:lgdSphs} provides the
comparison between the model and the observed stellar metallicity distribution
functions.  For NGC\,6822, the metallicity distribution comes from
\citet{swan2016}. For Andromeda II we inferred it ourselves from the
sample of \citet{vargas2014}.  Finally, for Sextans, Sculptor, Ursa
Minor and Draco, we retrieved the \feh values from the SAGA database
\citep{suda2008}, thanks to a series of original works on
Ursa Minor
\citep{shetrone2001,sadakane2004,cohen2010,kirby2010},
Draco \citep{shetrone2001,cohen2009,kirby2010, tsujimoto2015},
Sculptor \citep{battaglia2008,kirby2010, lardo2016}, 
 Sextans \citep{kirby2009,kirby2010, battaglia2011}.

The lower right panel compares the model and observed [Mg/Fe] vs
\feh trends.  The abundance of magnesium and iron have been retrieved
from the works quoted above for Ursa Minor and Draco. For Sculptor we
compiled the works from
\citet{shetrone2003, tafelmeyer2010, tolstoy2009, jablonka2015}. For Sextans,
we benefited from \citet{shetrone2001, aoki2009, tafelmeyer2010,
  theler2018}.

Admittedly, this is the first time that models arising from a
$\Lambda$CDM cosmological framework and their physical ingredients are
checked in such a detail and they do indeed match the observations
well. The six observed dwarfs, chosen as test-beds, which sample a
large variety of star formation and merger histories, chemical
evolutions, and velocity dispersions, are remarkably well reproduced.
Some matches could have most likely been slightly improved had we
benefited from a larger sample of simulated galaxies, in other words, a larger
cosmological volume. An obvious example is provided by NGC\,6822 for
which the model luminosity and peak of the stellar metallicity distribution
are both slightly too high. As we mentioned earlier we faced a lack of
models in this luminosity and metallicity range.
It is worth mentioning that our Sculptor model predicts a maximal
velocity of about $23\,\rm{km/s}$ in perfect agreement with the recent
predictions of \citet{strigari2017}.  

Figure~\ref{fig:lgdSphs} also points out how challenging it is to observationally derive
reliable total halo masses for dwarf galaxies. Indeed,
while model \texttt{h141} and \texttt{h061} share similar luminosities, star formation
histories, mean metallicities and velocity dispersions, their virial halo masses differ by more than a
factor of 2.5. 
Likewise, the models  \texttt{h061} and \texttt{h070}  have similar virial masses, however
very different stellar properties.
To shed light on these diverging properties, we computed the time evolution of the virial and stellar mass of those three models.

During the first Gyr, \texttt{h141} has a slightly higher virial mass allowing to form
about twice as much stars as \texttt{h061}. However later on, \texttt{h061} experiences two
important accretion events. The first event double the dwarf dark mass, without
gaining neither gas nor stars, as only small dark haloes devoted of baryons are swallowed.
Later on, the dwarf double its mass once more by merging with a smaller dwarf which also adds stars and gas. 
Those two events occurs after the decrease of the star formation due to UV-background
heating. Hence,  despite a different assembly history, the star formation history of models \texttt{h061} and
\texttt{h141} are similar.
Compared to model \texttt{h061}, model \texttt{h070} is assembled much quicker which helps it to resist the UV-background heating 
and extend its star formation history .
This different assembly histories echoes the large scatter of the luminosity - velocity
dispersion relation found in Section~\ref{luminosity_velocity}.

\subsection{Metallicity and age gradients}
\label{metallicity_gradient}

\begin{figure*}
        \centering
        \leavevmode  
        \resizebox{0.245\hsize}{!} {\includegraphics[angle=0]{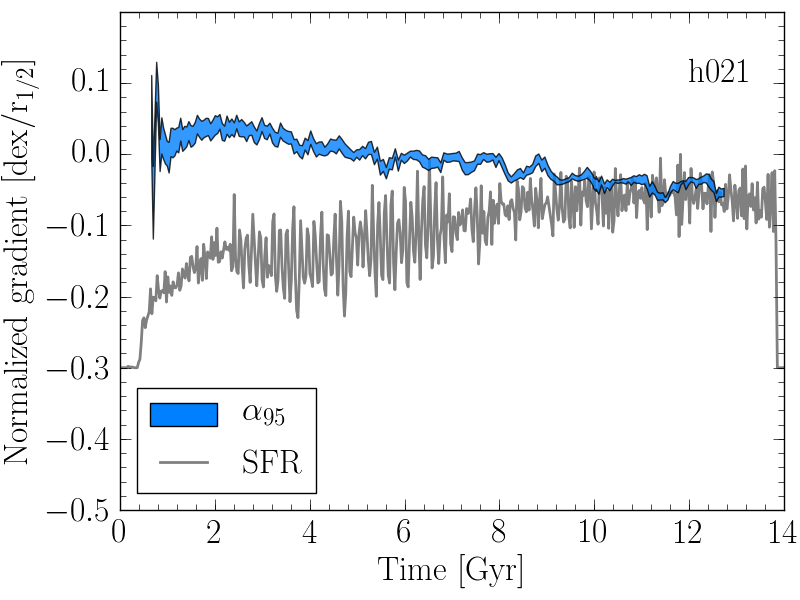}}
        \resizebox{0.245\hsize}{!} {\includegraphics[angle=0]{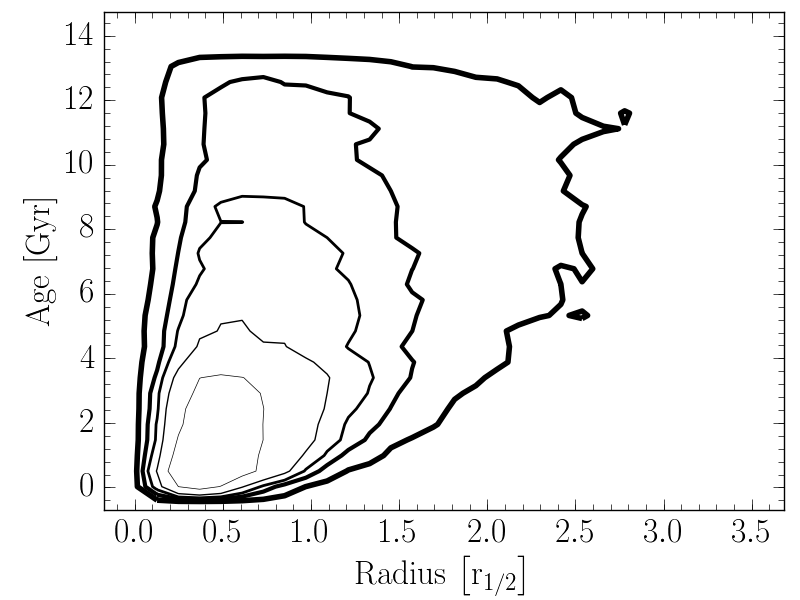}}
        \resizebox{0.245\hsize}{!} {\includegraphics[angle=0]{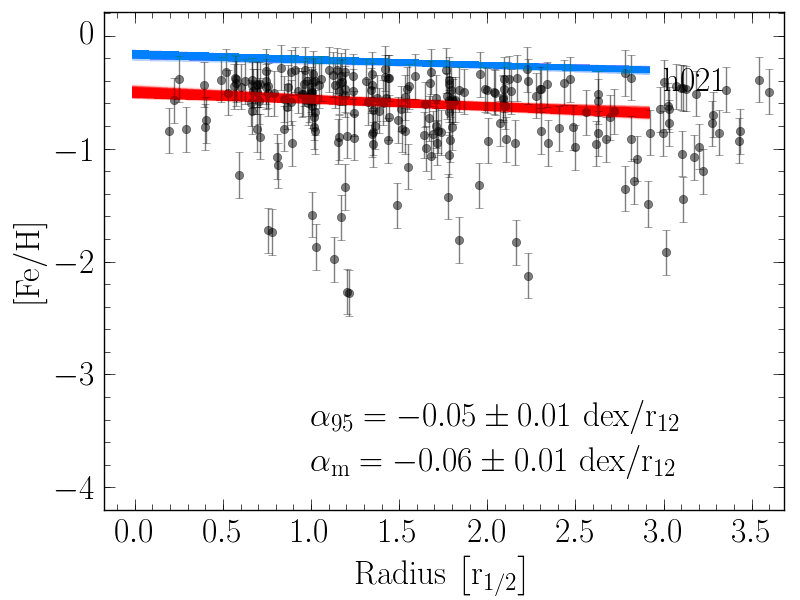}}
        \resizebox{0.2459\hsize}{!}{\includegraphics[angle=0]{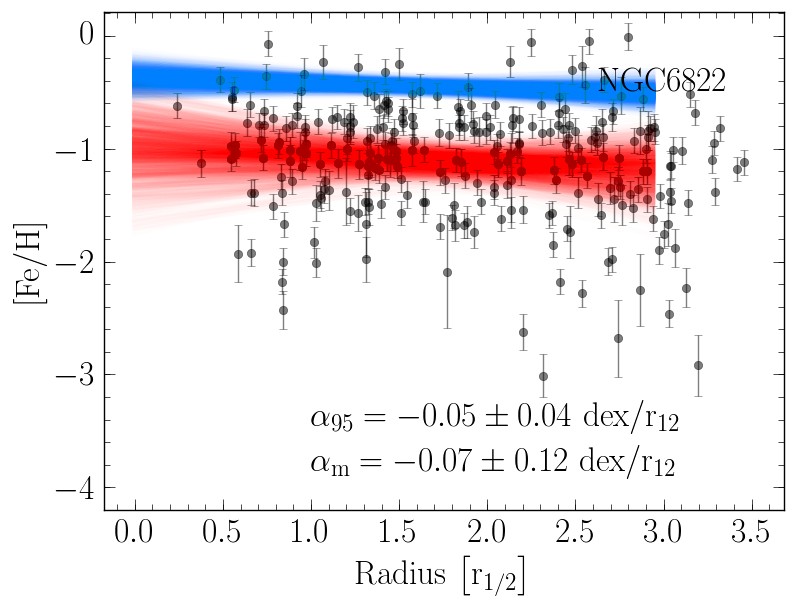}}
        \resizebox{0.245\hsize}{!} {\includegraphics[angle=0]{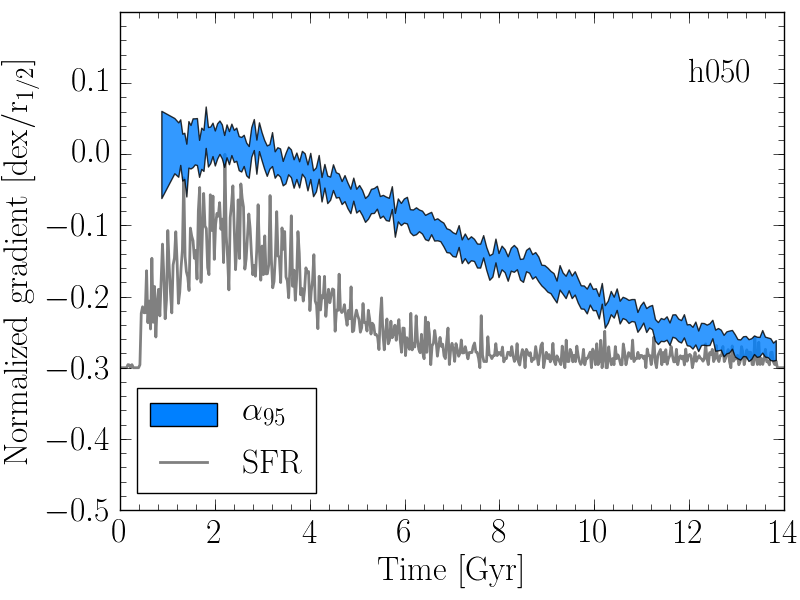}}
        \resizebox{0.245\hsize}{!} {\includegraphics[angle=0]{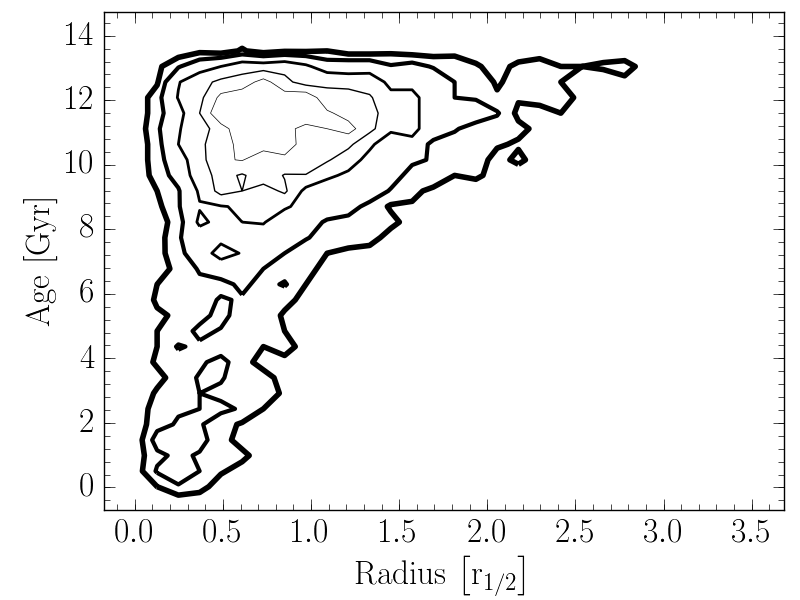}}
        \resizebox{0.245\hsize}{!} {\includegraphics[angle=0]{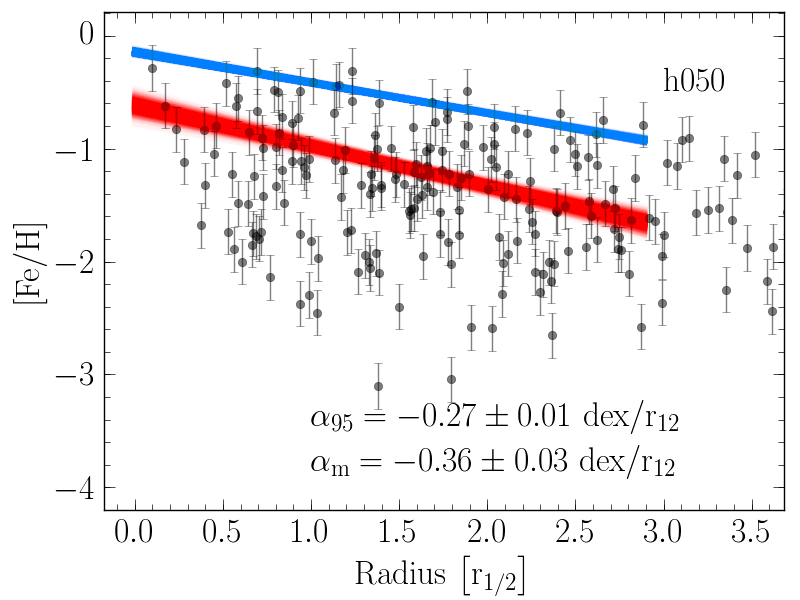}}
        \resizebox{0.2459\hsize}{!}{\includegraphics[angle=0]{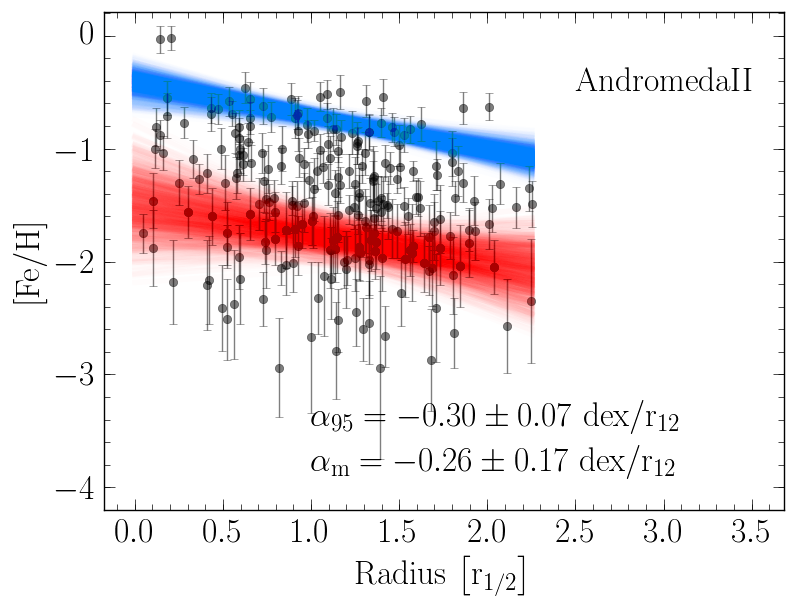}}
        \resizebox{0.245\hsize}{!} {\includegraphics[angle=0]{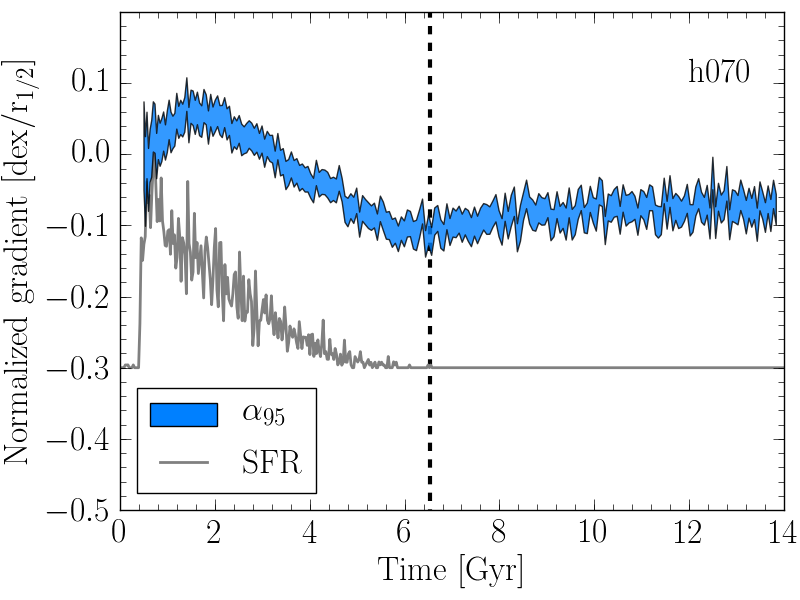}}
        \resizebox{0.245\hsize}{!} {\includegraphics[angle=0]{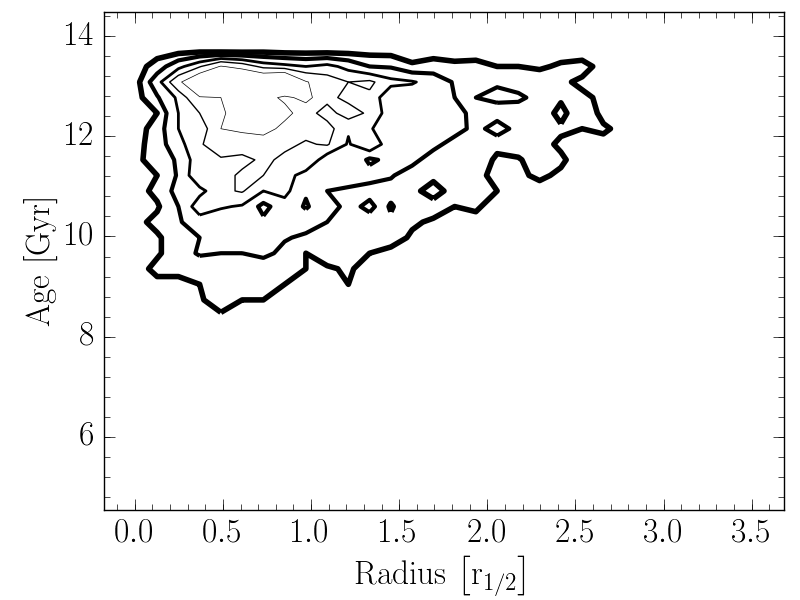}}
        \resizebox{0.245\hsize}{!} {\includegraphics[angle=0]{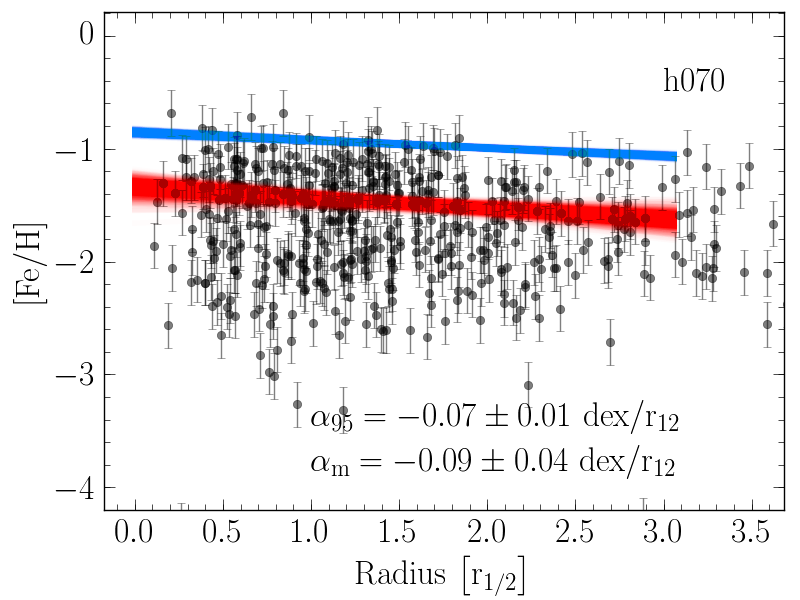}}
        \resizebox{0.245\hsize}{!} {\includegraphics[angle=0]{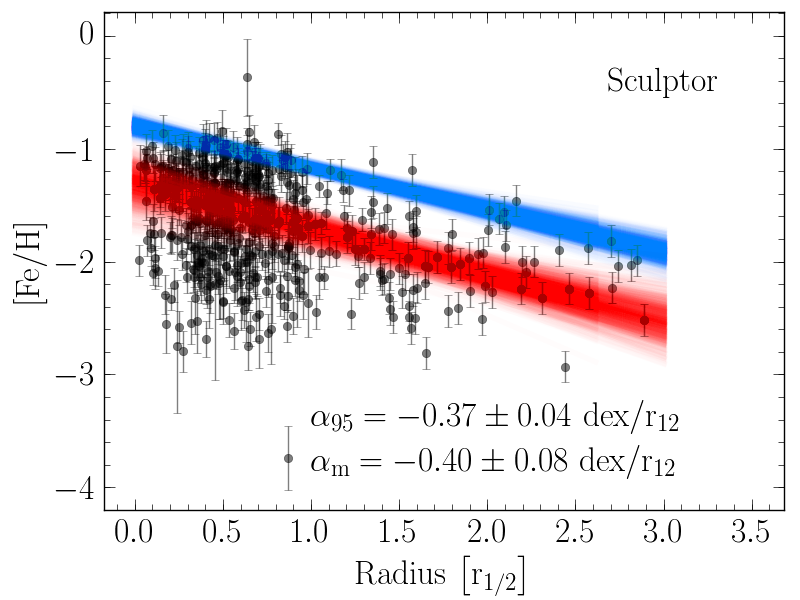}}
        \resizebox{0.245\hsize}{!} {\includegraphics[angle=0]{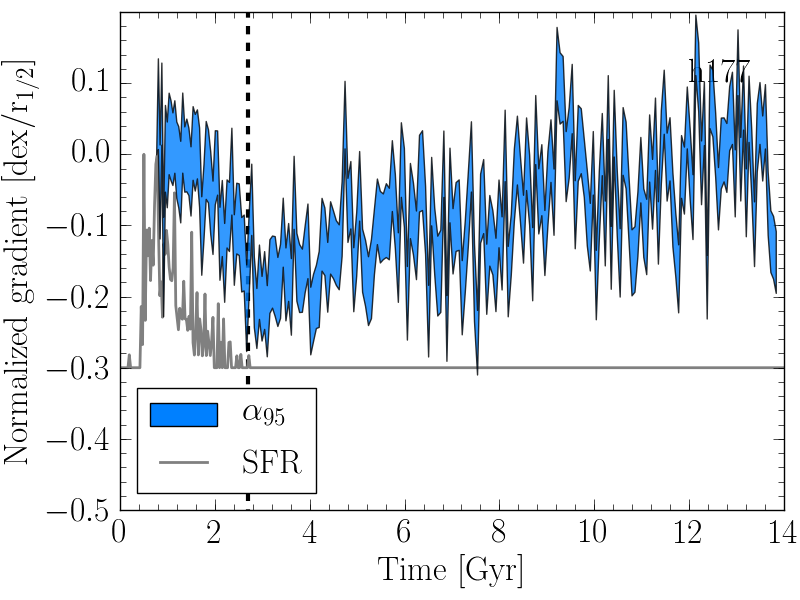}}
        \resizebox{0.245\hsize}{!} {\includegraphics[angle=0]{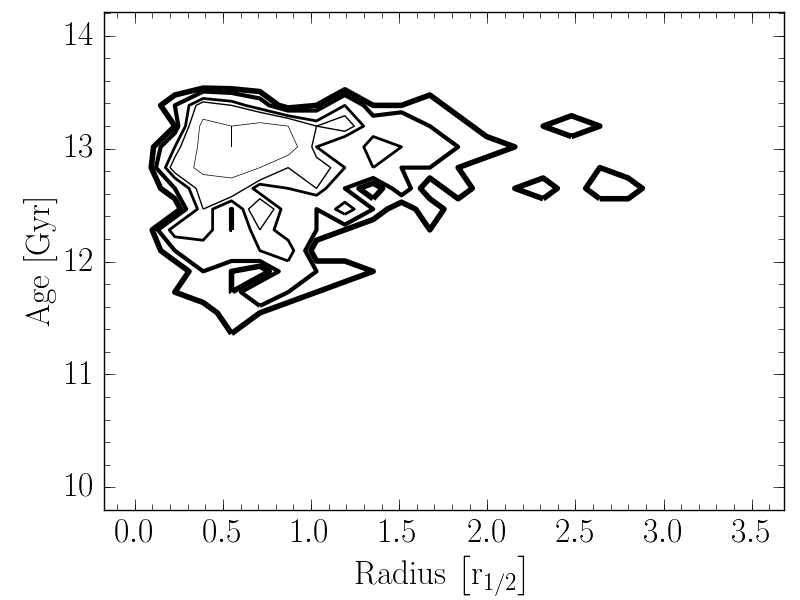}}
        \resizebox{0.245\hsize}{!} {\includegraphics[angle=0]{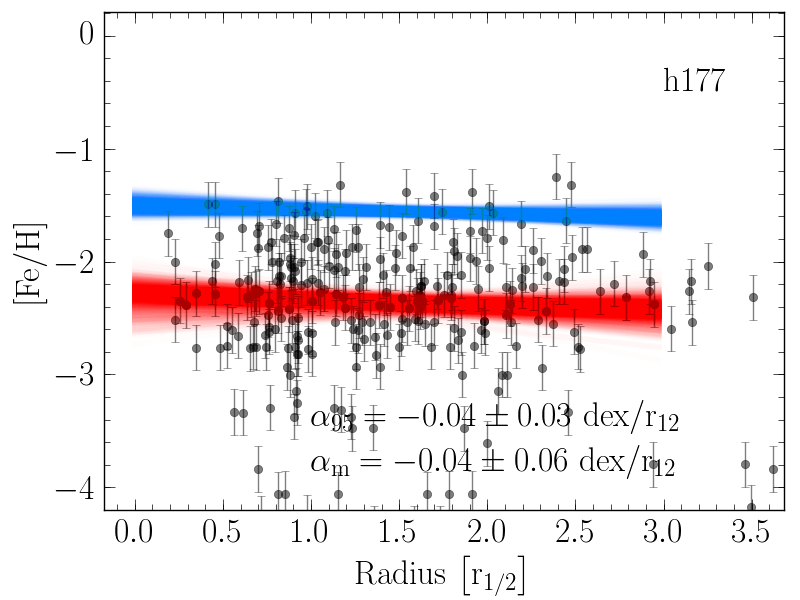}}
        \resizebox{0.245\hsize}{!} {\includegraphics[angle=0]{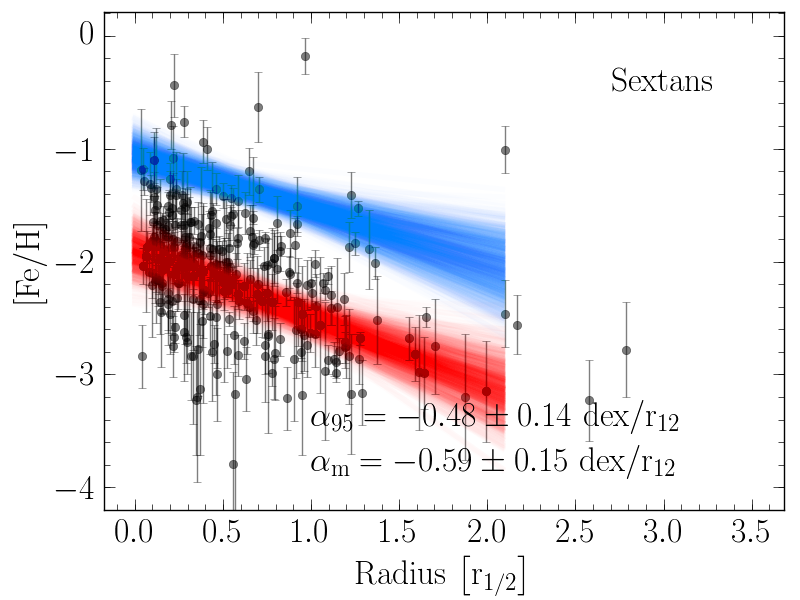}}
        \resizebox{0.245\hsize}{!} {\includegraphics[angle=0]{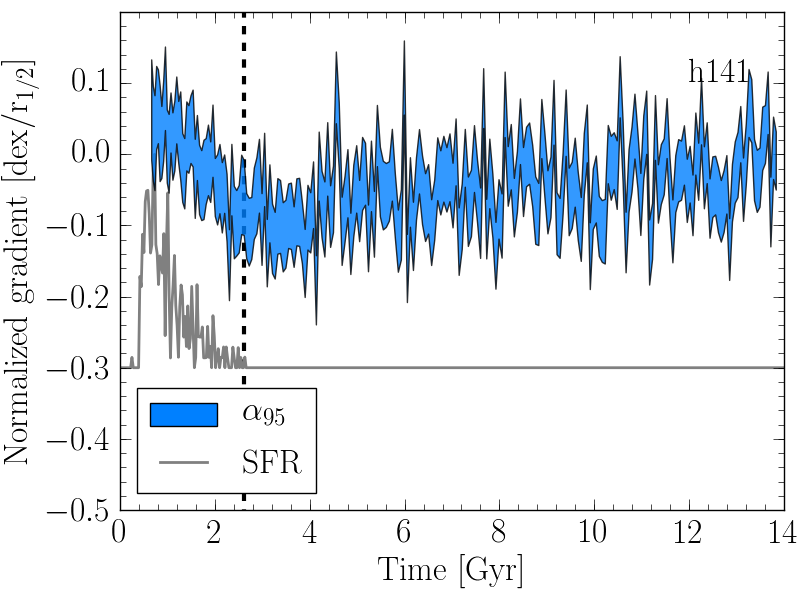}}
        \resizebox{0.245\hsize}{!} {\includegraphics[angle=0]{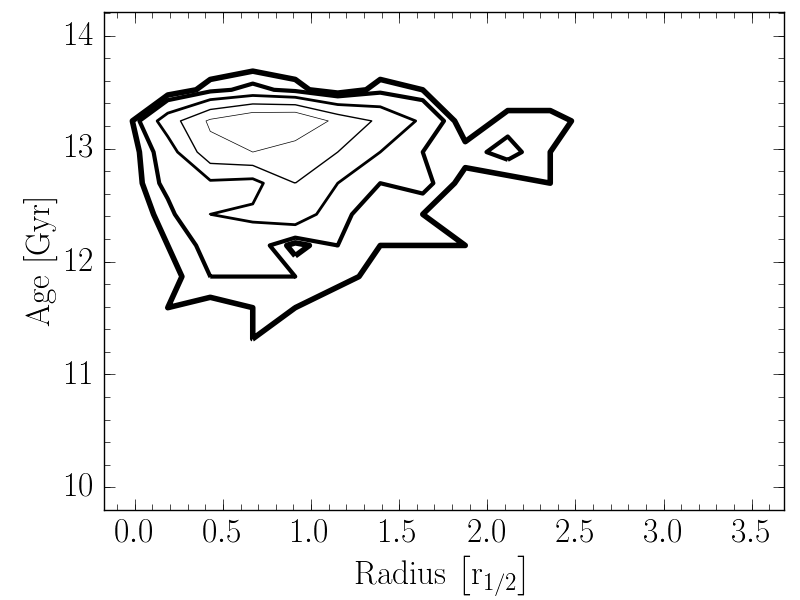}}
        \resizebox{0.245\hsize}{!} {\includegraphics[angle=0]{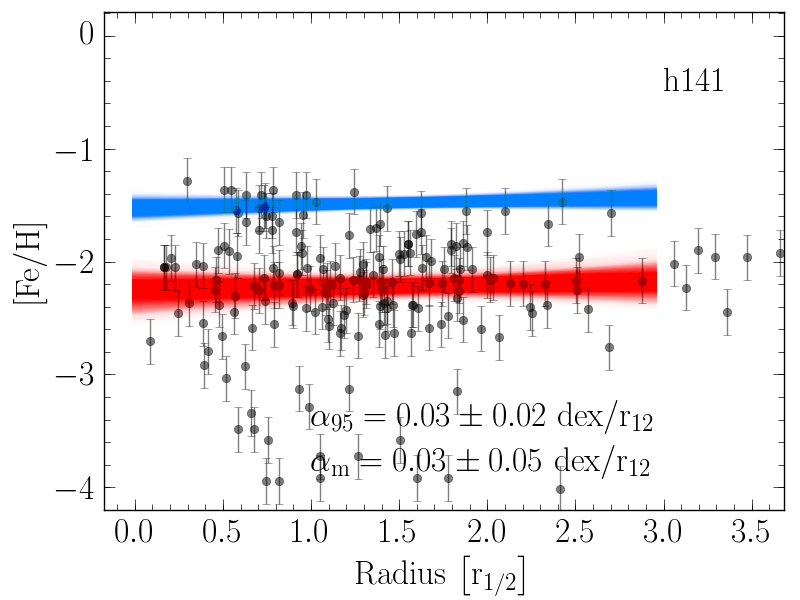}}
        \resizebox{0.245\hsize}{!} {\includegraphics[angle=0]{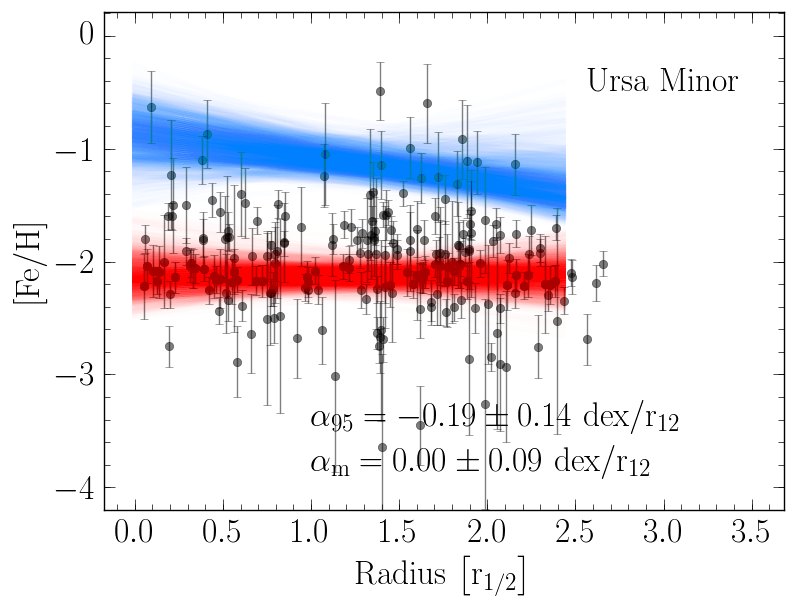}}
        \resizebox{0.245\hsize}{!} {\includegraphics[angle=0]{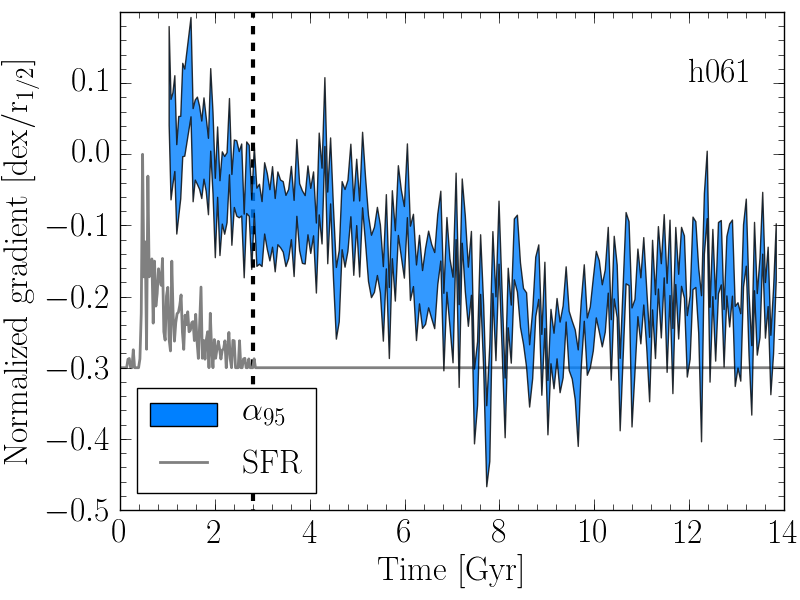}}
        \resizebox{0.245\hsize}{!} {\includegraphics[angle=0]{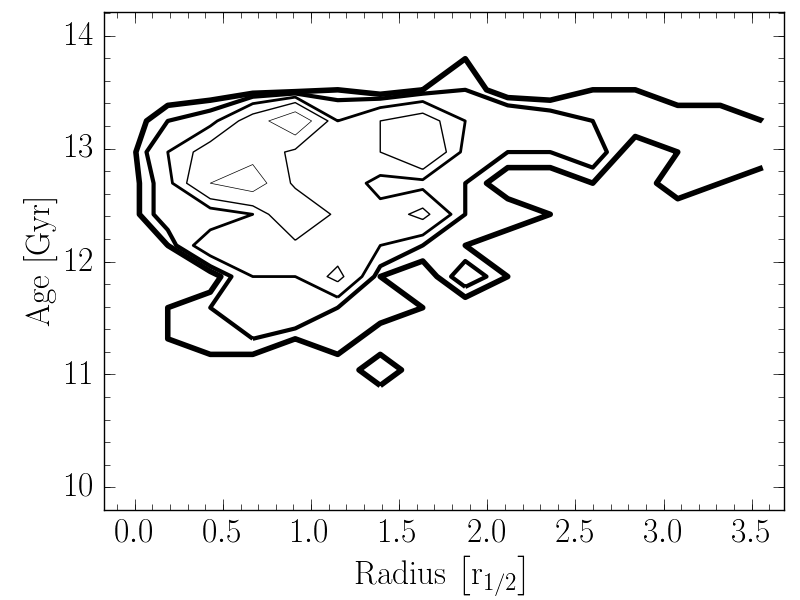}}
        \resizebox{0.245\hsize}{!} {\includegraphics[angle=0]{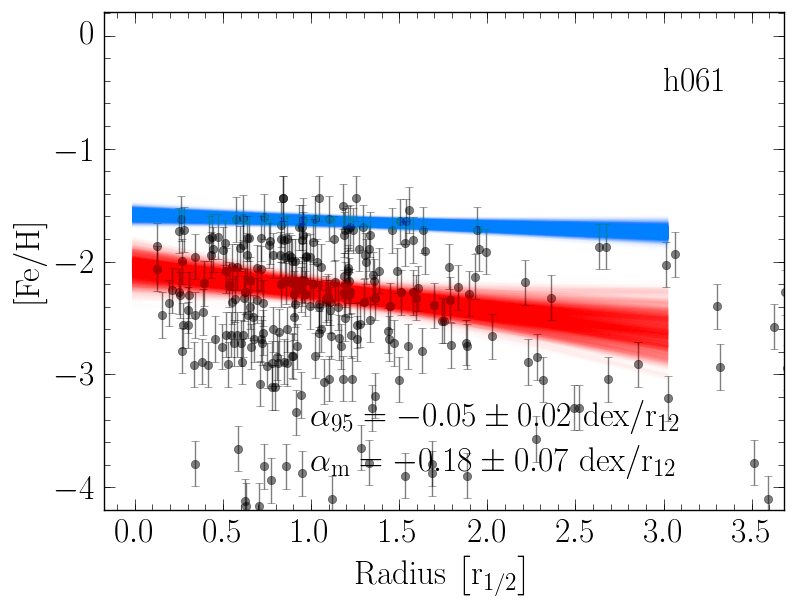}}
        \resizebox{0.245\hsize}{!} {\includegraphics[angle=0]{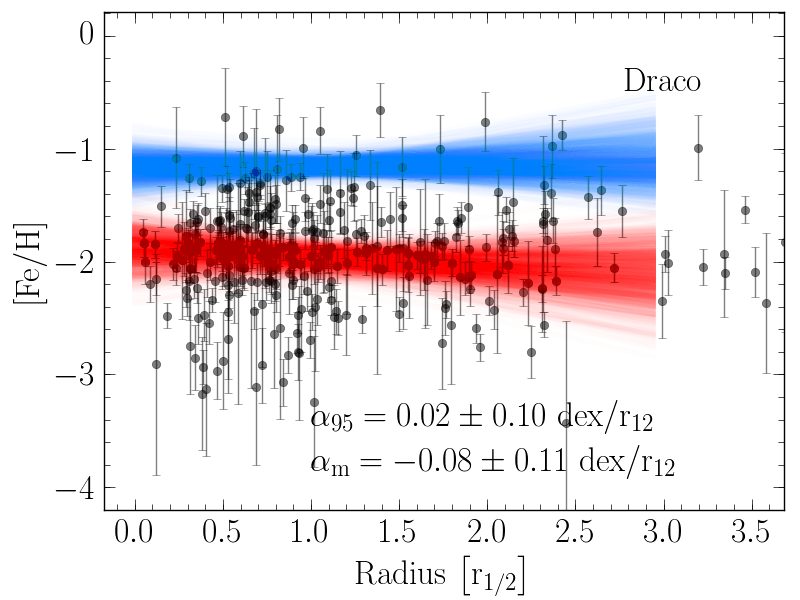}}
        \caption{Metallicity and age gradients of the simulated
          dwarf galaxies as compared to the observations. The first
          column shows the time evolution of the normalized \feh
          gradients determined by the 95th percentile ($\alpha_{95}$,
          in blue) of the model galaxy stellar metallicity distribution. The
          width of the curve traces the error on the slopes (see text
          for details). The grey curve corresponds to the normalized
          star formation rate and, when displayed, the vertical dashed
          line indicates the end of the star formation period.  The
          second column shows the age distribution of the stars in as
          function of the galaxy projected radius. The third and
          fourth columns compare the metallicity gradients of
          each model at z=0 , to a corresponding observed galaxy. The red and
          blue lines correspond to the fit of the mode (red) and 95th
          percentile (blue) of the observed and model stellar metallicity distributions. Their width reflect the uncertainty
on the slope. The values of $\alpha_{\rm{m}}$ and $\alpha_{95}$ together with their
          error bars are provided. For a fair comparison, the
          number of stars in the model and in the observed are equal.}
        \label{fig:FevsR}
\end{figure*} 

Stellar metallicity and gradients are quite common in the Local Group dSphs. For example they have been reported in
Sculptor, Fornax, Sextans, Carina, Leo I, Leo II, NGC 185, and Andromeda I-VI  \citep{harbeck2001,tolstoy2004,battaglia2006,battaglia2011,kirby2011, deboer2012,deboer2012b, vargas2014,ho2015, lardo2016, okamoto2017,spencer2017}

However, only a few physical interpretations have been put forward to explain these
structured  age and metallicity radial changes. Suggestions have been made that they
could form either secularly, from the centrally concentrated star formation
\citep{schroyen2013} or more recently  by a late gas accretion which could
reignite star formation in the galaxy central regions
\citep{benitez-llambay2016}. The metal-poor stars could have been scattered over
large galacto-centric distances either by successive merger events at high
redshift \citep{benitez-llambay2016} or by recurrent strong stellar feedback,
pushing away the dark matter and driving stars on less bound orbits
\citep{pontzen2014}.

Figure~\ref{fig:FevsR} illustrates the formation of these gradients in
our models in comparison with the observations.  We first describe the
procedure we applied to both observed and model galaxies.  Stars or
stellar particles were first considered in a \feh vs projected and
normalized galacto-centric radius space. The normalization was done to
the half-light radius of the model dwarf, as often the case for the
observations. As to our simulations, this also avoids to introduce a
bias owing to the dynamical heating described in
Section~\ref{half_light_radius}.

While we use the terminology gradient, a more accurate
description should probably be that the metal-poor stars are evenly
distributed from the central galaxy regions to the outer ones, while
the most metal-rich populations are more spatially concentrated
\citep[see also][]{battaglia2011}. Therefore characterizing the radial
variation of the upper envelope of the stellar metallicity distribution is
often more appropriate than a classical radial linear fit.
Consequently, we computed the metallicity distribution function in 20
radial bins and, for each of them, determined both their mode and 95th
percentiles. We further performed a linear fit of those values versus
the projected radii.  The slopes of these fits are $\alpha_{\rm{m}}$
and $\alpha_{95}$, respectively. They are taken as a proxy for the
normalized metallicity gradients.  We used a Monte Carlo approach to
estimate the errors on the slopes. For each star or stellar particle,
we varied their metallicity a thousand times following a Gaussian
distribution centred on their mode metallicity or percentile with a
dispersion corresponding to their uncertainties.  For the observed
galaxies, \feh and their corresponding errors were taken from the SAGA
database \citep{suda2008}.  For the models we adopted a fiducial error
of \feh of $0.2\,\rm{dex}$ which corresponds to a typical
observational uncertainty.  The final slopes $\alpha_{\rm{m}}$ and
$\alpha_{95}$ are the mean and their error bar the standard deviation
of the thousand fits. We note that $\alpha_{95}$ leads to less noisy
behaviour compared to $\alpha_{\rm{m}}$.

For each of the galaxies presented in Fig.~\ref{fig:lgdSphs},
Fig.~\ref{fig:FevsR} provides the time evolution of $\alpha_{\rm{m}}$
and $\alpha_{95}$, as well as the spatial distribution of the stellar
ages and metallicities of the models.  The latter is compared to the
observations. For non-rotating objects, the strength of the gradients
is primarily linked to the length of the star formation history.
Systems with an extended star formation history exhibit stronger
gradients than those which have been early quenched.  Nevertheless, in
case of fusions, the quenched systems can still exhibit relatively
strong gradients, as we will describe further below.  Luminous systems
that formed through sustained star formation do rotate, and exhibit
only very weak gradients.

The first column of Fig.~\ref{fig:FevsR} shows that the radial
distribution of metals traced by $\alpha_{95}$ varies with time, in
relation with the galaxy star formation history.  They stay generally
nearly constant once star formation has
ceased.  In some cases, however, and for isolated galaxies, one sees a
slight decrease of these gradients as the consequence of the mixing of
the different stellar populations induced by their velocity
dispersion.  In the case of a merger event, again once star formation
has stopped, the situation is different.  For example, model
\texttt{h061} merges at about $6\,\rm{Gyr}$ with a lower mass and
lower metallicity system. The  metal-poor stars of this small impacting body is
dispersed in the outer regions of \texttt{h061}.  This increases the
global metallicity gradient for a couple of Gyrs, after which it
decreases again due to the stellar mixing.

\begin{figure*}
        \centering
        \leavevmode  
    \subfigure[\texttt{h050}\,(Andromeda II)]  {\resizebox{0.245\hsize}{!}{\includegraphics[angle=0]{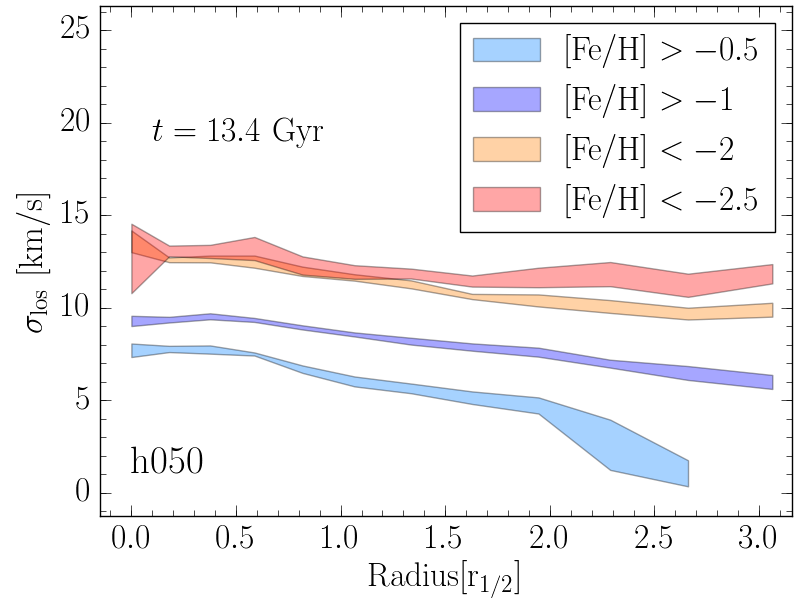}}} 
        \subfigure[\texttt{h070}\,(Sculptor)]  {\resizebox{0.245\hsize}{!}{\includegraphics[angle=0]{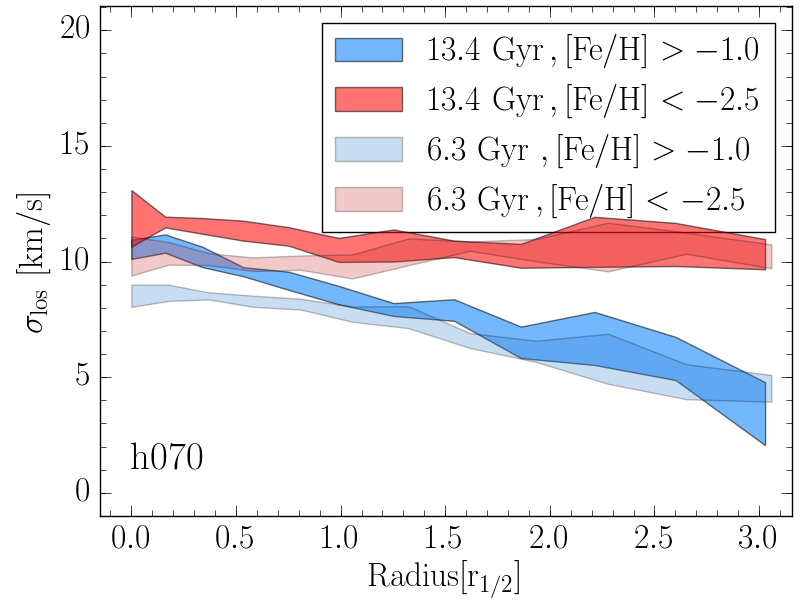}}}   
        \subfigure[\texttt{h177}\,(Sextans)]   {\resizebox{0.24595\hsize}{!}{\includegraphics[angle=0]{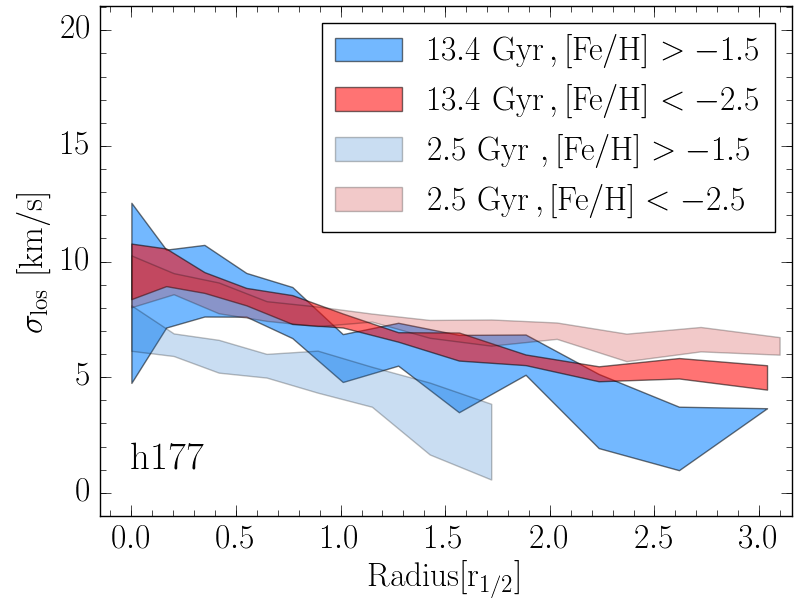}}}   
        \subfigure[\texttt{h141}\,(Ursa Minor)]{\resizebox{0.245\hsize}{!}{\includegraphics[angle=0]{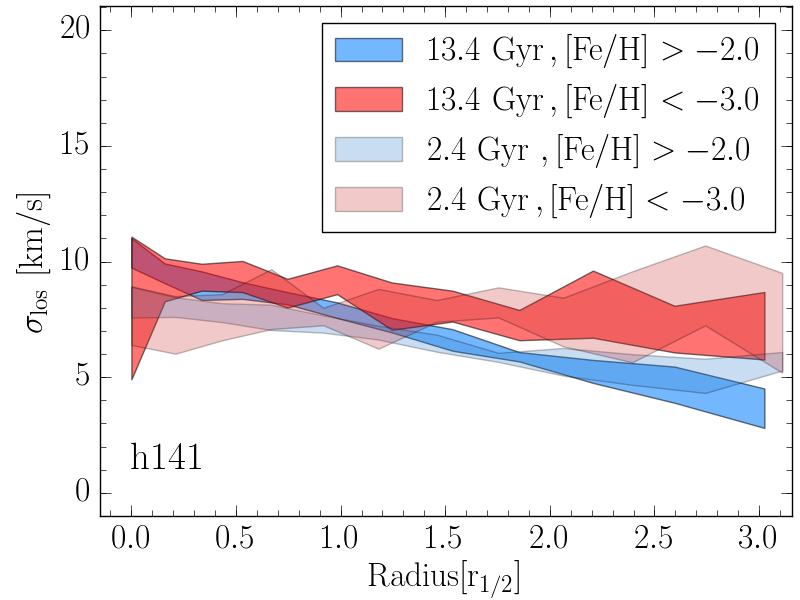}}}   
        \caption{
Line of sight velocity dispersion profile as a function of
metallicity of the four models representing the Sculptor, Sextans,
Ursa Minor, and Draco dSphs. The velocity dispersion of the metal-rich (in
red) and metal-poor stars (in blue) at z=0 is compared to their
values at the end of the star formation activity with transparent colours.}
        \label{fig:kindist2}
\end{figure*}


As long as star formation is active, gradients are strengthened. This
is clearly shown for model \texttt{h070}. It is also seen in cases of
residual formation as in model \texttt{h050} after $t=8\,\rm{Gyr}$ and
even for sustained rotating models where the gradient is weaker as in
model \texttt{h021}. The link between star formation and gradients is
related to the fact that stars form from the gas reservoir in the
centre of the galaxies (see Section~\ref{multiphase}).  This is
illustrated by the stellar age distribution displayed in the second
column of Fig.~\ref{fig:FevsR}.  The young and metal-rich stars are
essentially located in the galaxy centres, while the older and
metal-poor ones are distributed over the entire galaxy.  The scatter
of the oldest stars arises from the gradually shrinking gas reservoir
from which they originate.  Indeed the reservoir is dense and spatially
extended at early times. Star formation, UV-background and supernovae
feedback heating, all result in its depletion and evaporation.

We also note that at given star formation rate and stellar mass,
some galaxies end up with a gradient much more pronounced than others.
The generation of those types of gradients with strength more than
twice as stronger as the others results from their early assembly with
faint metal poor galaxies (at about $3\,\rm{Gyr}$) creating a rather
extended system with low metallicity stars in the outskirts.

The third and fourth columns of Fig.~\ref{fig:FevsR} provide the
radial metallicity distribution for our observed comparison sample and
their corresponding models. Faint galaxies such as Ursa Minor and
Draco, characterized by a star formation that ceased completely
$10\,\rm{Gyr}$ ago \citep{carrera2002,aparicio2001} have marginal
gradients \citep{faria2007,kirby2011,suda2017} in agreement with our
model predictions.  At the other extreme, rotating luminous systems such as
NGC\,6822 have only weak gradients both in the models and the
observations \citep{kirby2013}.  Another clear success is obtained for
Andromeda II \citep{vargas2014,ho2015}.  Two galaxies appear
challenging: both Sculptor and Sextans are lacking metal-rich stars
($\feh \sim-1.5$) outside their half light radius contrary to our
models. The origin of this discrepancy is not fully clear.  In the
case of Sculptor, we are most likely facing the fact that we only have
a limited number of models and even only one for this galaxy for its
corresponding metallicity and luminosity.  As for Sextans, this failure
could be due to the star formation history of the model that is not
extended enough compared to the observations
\citep{lee2009,deboer2012}.


\subsection{Kinematically distinct stellar populations}
\label{stellar_populations}

Kinematically distinct stellar populations as a function of their
metallicity have been mentioned for a number of galaxies, for example
Fornax \citep{battaglia2006}, Sculptor
\citep{tolstoy2004,battaglia2008}, and Sextans \citep{battaglia2011}.
The most spectacular example is Sculptor with the metal-poor stars
($\rm{[Fe/H]}<-1.7$) showing a slowly decreasing velocity dispersion
with distance to the galaxy centre from $12\,\rm{km/s}$ to
$8\,\rm{km/s}$, while its richer counterpart ($\rm{[Fe/H]}>-1.5$)
sharply drops to $2\,\rm{km/s}$.  In Fornax and Sextans, the
difference in velocity dispersion between the most metal-rich
population ($\rm{[Fe/H]}>-1.3$, resp. $\rm{[Fe/H]}>-2.2$) and the
metal-poor ones is of the order of $2$ to $4\,\rm{km/s}$, less
significant than in Sculptor, however still there.

Figure~\ref{fig:kindist2} displays the velocity dispersion profiles
of models \texttt{h050} (Andromeda II), \texttt{h070} (Sculptor),
\texttt{h177} (Sextans) and \texttt{h141} (Uras Minor).  For
\texttt{h070}, \texttt{h177} and \texttt{h141}, we provide the
profiles of the metal-rich and metal-poor populations both at $z=0$
and at the end of the star formation.  This allows us to assess the
impact of the secular dynamical heating.  Given the fact that the star
formation is still on-going for \texttt{h050}, we only show its
profiles at $z=0$. In all cases, younger and more metal-rich stars have
the smallest velocity dispersions and radially decreasing profiles.  The low
velocity dispersion traces that of the cold gas, as discussed in
Section~\ref{multiphase} from which the young stars just emerged.  Older
stellar populations show rather flat profiles, being kinematically
warmer owing to their longer dynamical history.  In the most
distinct cases, \texttt{h050} (Andromeda II) and \texttt{h070}
(Sculptor), a difference of about $5\,\rm{km/s}$ is found  between
the metal-richest and poorest stellar populations, inside
$1.5\,\rm{r_{1/2}}$. This difference increases outwards.  With shorter star
formation histories, the \texttt{h177} (Sextans), \texttt{h141} (Ursa
Minor) models still show a differential kinematics, however to a lower
extent. In all cases, including the cases of sustained star formation such
as model \texttt{h021}, the metal-rich stars are systematically warmer
than their metal poor counterparts, the difference increasing with
radius, in qualitative and quantitative agreement with observations.
The secular heating appears to only affects the central regions
($r<r_{1/2}$), increasing their velocities by about $2\,\rm{km/s}$.

Contrary to \citet{benitez-llambay2016}, in
this work we observed no model with sign of late accretion
events that could restart the star formation activity, generate gradients,
and kinematically distinct stellar populations.  Indeed, once being
washed out by the UV-background heating, most of our dwarf galaxies
evolve as if they were in isolation.



\subsection{Open questions and $\Lambda$CDM paradigm}

\subsubsection{Puzzling cases}

Some galaxies, which benefit from detailed observations, fall outside
the range of our model predictions.  In Section~\ref{detailed_stellar_properties} we have already mentioned the
case of Fornax and Carina. Even more challenging are the properties of
the faint, compact, however gas-rich systems, such as Leo T and Leo P,
which are seen with residual star formation or at least young
stars. 

Despite its rather low luminosity ($1.4\times 10^5\,\rm{L_\odot}$) and
velocity dispersion ($\sigma_{\rm{LOS}}\cong 7.5\,\rm{km/s}$) Leo T is
embedded in an HI cloud of mass of about $2.8\times 10^5\,\rm{M_\odot}$
and displays a rather continuous star formation history extending up
to a recent past \citep{weisz2012}.  Barely brighter, at the edge of
the Local Group, Leo P \citep[$\cong 5\times
  10^5\,\rm{L_\odot}$,][]{giovanelli2013,rhode2013} is also gas-rich
with ongoing star formation \citep{mcquinn2015}.  These two galaxies
seem to have passed re-ionization without a major decrease of
activity.  They call for further investigations both of the strength
and impact of the UV-background heating and the hydrogen
self-shielding.


\subsubsection{The stellar mass - halo mass relation: comparison with previous studies}

%

Our simulations predict that Local Group dwarf galaxies with stellar
masses larger than $10^5\,\rm{M_\odot}$ have dark halo virial masses
between $2\times 10^8$ and $10^{10}\,\rm{M_\odot}$.  These are
at most $3\times 10^9\,\rm{M_\odot}$ for the classical dwarf
spheroidal galaxies with luminosities below $10^7\,\rm{M_\odot}$.

\begin{figure}
        \centering
        \leavevmode   
        \includegraphics[width=9cm]{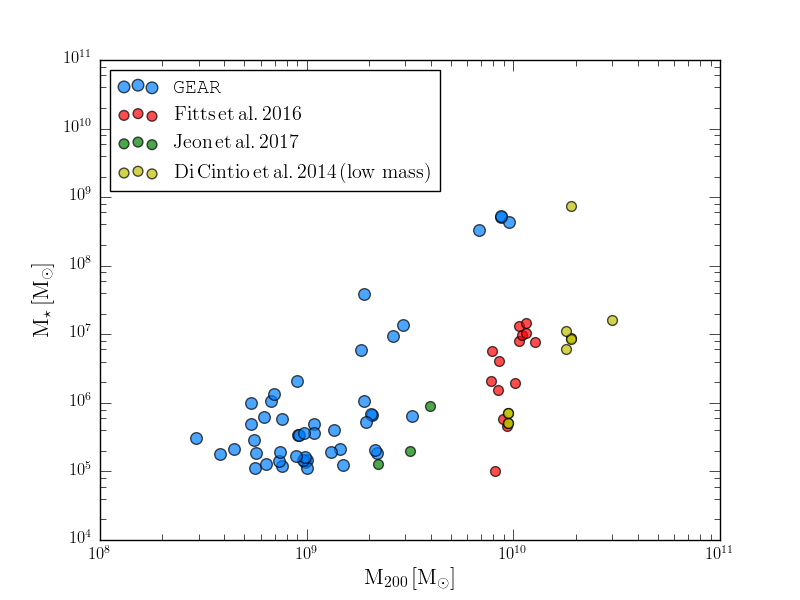}
        \caption{Comparison of the stellar mass $M_\star$ vs. the halo virial mass $M_{200}$ between our models in blue
                and results from other studies based on hydrodynamical codes.}
        \label{fig:MsvsMh}
\end{figure}

These predictions are, at given luminosity or stellar mass,
somewhat on the low tail of the distribution of dark halo masses found
in other studies. 
Fig.~\ref{fig:MsvsMh} compares the stellar mass $M_\star$ vs the halo virial mass $M_{200}$
for different studies.
In our simulations, a $10^5\,\rm{M_\odot}$ stellar mass system has a dark matter halo mass
within $[2\times 10^8 - 3\times 10^9]\,\rm{M_\odot}$.  
For the same stellar mass, \citet{jeon2017} find  dark haloes in the range $[2\times 10^9 - 4\times 10^9]\,\rm{M_\odot}$,
while in the range $[3\times 10^9 - 8\times 10^9]\,\rm{M_\odot}$ for the  FIRE-2 simulations  \citep{fitts2017,hopkins2017}.
The values of the APOSTLE simulations (satellite population), not show in Fig.~\ref{fig:MsvsMh} \citep{sawala2015}
lie in the range $[5\times 10^8 - 6\times 10^9]\,\rm{M_\odot}$.  
At higher stellar masses, the divergence is even more pronounced.  We predict that a $10^7\,\rm{M_\odot}$ stellar mass
system will be hosted by a dark matter halo within $[2\times 10^9 - 5\times 10^9]\,\rm{M_\odot}$, 
while FIRE-2 finds $[8\times 10^9 - 2\times 10^{10}]\,\rm{M_\odot}$, the MaGICC simulations \citep{dicintio2014} find $[2\times 10^{10} - 3\times 10^{10}]\,\rm{M_\odot}$
and the APOSTLE ones $[9\times 10^9 - 10^{10}]\,\rm{M_\odot}$,
Therefore the difference in predicted dark matter halo masses can reach a factor ten.
Before addressing the origin of these differences, it is first important to emphasize that in the context of the the AGORA project \citep{kim2016}, 
for the same cooling and feedback implementation, \texttt{GEAR} predicts very similar results
compared to other particle-based codes like \texttt{Gizmo} and \texttt{Gasoline} used in the FIRE-2 and MaGICC simulations respectively.
The observed differences are thus related to different implementation of physical processes. We discuss in the following what we consider to be the dominant ones:
stellar feedback efficiency, strength of the UV-background, hydrogen self-shielding, and the star formation receipts.

The FIRE-2 simulations implement multiple stellar feedback mechanisms (supernovae, OB/AGB mass loss, photo-heating, radiation pressure) 
with a star formation receipt that let the star form only in very high density regions. These two prescriptions result in a very strong feedback
which easily blow out gas in the smallest haloes which becomes dark. 
Moreover, those simulations use the ionization UV background rates computed by \citet{faucher-giguere2009} which is much stronger
than the \citet{haardt2012} rates we used in our study, for redshifts larger than eight.
Consequently, only massive haloes $[\cong 10^{10}]\,\rm{M_\odot}$ can retain gas and sustain star formation. 
In addition to a supernova feedback comparable to our, the MaGICC simulations use early pre-supernova stellar feedback which boost the
feedback at early time. These simulations do not include hydrogen self-shielding. In such conditions, as predicted by \citet{okamoto2008} 
only massive haloes ($M_{200}\cong 10^{10}\,\rm{M_\odot}$) retain their gas during the re-ionization epoch.
Faint galaxies produced by \citet{jeon2017} are at the edge of our predictions. Their shift towards higher halo mass is certainly due to a stellar
feedback boosted by core-collapse and pair-instability supernovae.


In our framework, lowering the stellar mass for a given halo mass is possible by boosting the feedback or removing the HI self-shielding,  however 
the stellar line of sight velocity dispersion becomes incompatible with observations, with values too high by about $5\,\rm{km/s}$.

\subsubsection{Impact on the missing satellite problem}

The question we could raise is whether our small mass dark matter
haloes hosting stellar systems are in agreement with the observed
number of dwarf satellites around the Milky Way, or in other words,
whether we are suffering from a missing satellite
  problem \citep{klypin1999,moore1999}.  Indeed, compared to the above
mentioned studies, our models populate the upper edge of the abundance
matching predictions \citep{behroozi2013}, meaning that, at  given dark halo
mass, they tend to form more stars. However, as discussed in
Section~\ref{luminosity_velocity}, the large scatter in luminosity at
given observed velocity dispersion hampers any reliable prediction by
this technique.

\begin{figure}
        \centering
        \leavevmode   
        \includegraphics[width=9cm]{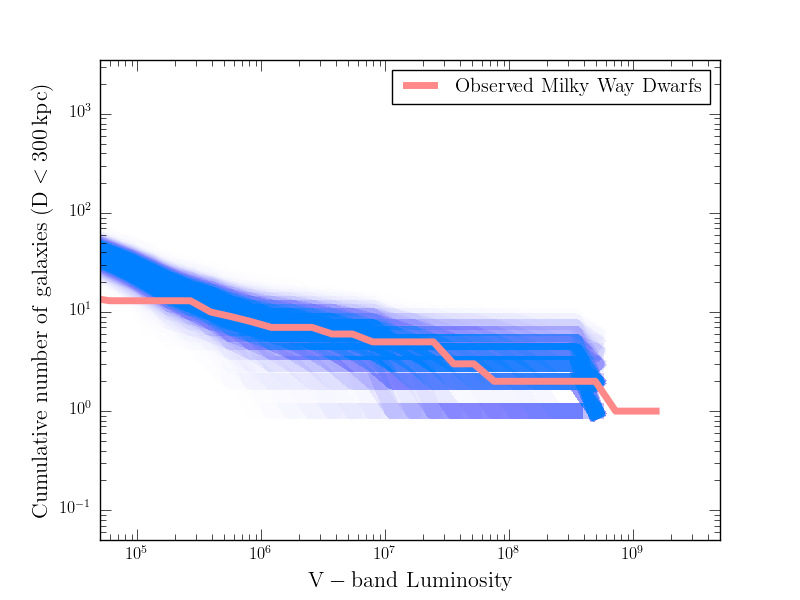}
        \caption{Cumulative number of  Milky Way dwarf satellite galaxies within $300\,\rm{kpc}$ as a function of their V-band luminosity.  
    The blue area shows our model predictions, with the observations in red.}
        \label{fig:NCumvsLv}
\end{figure}

To investigate this issue further, we calculated the expected number of dwarf
satellites around a Milky Way-like galaxy.  Because our cosmological box does
not include such a massive galaxy, we used the cumulative number of dark matter
haloes of \citet{sawala2017}, inside $300\,\rm{kpc}$. From this distribution, we
generated thousand realizations of dark matter haloes.  We then assigned to each
halo mass a stellar luminosity randomly chosen in the corresponding range of
values in our simulations.  Fig.~\ref{fig:NCumvsLv} displays in blue the resulting
cumulative number of dwarfs as a function of their V-band luminosity for each of
these thousand models.  This prediction is compared to the observations inside the same
volume, that is, only considering the Milky Way satellites (red curve).  From a luminosity of
$10^5$ up to $10^9\,\rm{L_{\odot}}$ our predictions are in perfect agreement
with the observed population. Below $10^5\,\rm{L_v}$, the cumulative number of
galaxies rises steeper than the observed curve. We reserve the study of the
regime of the ultra-faint dwarfs for a future study. We note that this agreement
above $10^5\,\rm{L_{\odot}}$ is obtained naturally and the properties of the
dwarfs are also reproduced in detail at $z=0$ with star formation ignited in
\texttt{all} haloes more massive than $2\times 10^8\,\rm{M_\odot}$. This is a
major difference with the studies quoted above in which an important fraction of
the dark haloes have their star formation totally inhibited by reionization,
stellar winds or supernovae feedback.  Would dwarf spheroidal galaxies be hosted
only in dark haloes more massive than a few $10^9\,\rm{M_\odot}$ they would
exhibit too large a stellar velocity dispersion. This is illustrated by model
\texttt{h039} with a luminosity below $3\times 10^5\,\rm{L_\odot}$ and a
massive dark matter halo of $3\times 10^9\,\rm{M_\odot}$.  Its mean
velocity dispersion is $11.7\,\rm{km/s}$ which makes it only marginally
compatible with the observed dwarfs at this luminosity.

\subsubsection{The cusp/core transformation}

Dark matter only simulations predict the density profiles of dark haloes to be
universal, independent of the halo mass, with an inner slope of -1 in logarithmic
scale \citep{navarro1996,navarro1997}.
Such cuspy density profiles have been claimed to be in contradiction with the
observations, leading to the so called cusp/core problem \citep{moore1994}.

Recently, different mechanisms related to the baryonic physics have been
proposed to transform a cuspy profile into a core one \citep{zolotov2012}.  In
agreement with the theoretical predictions of \citet{pontzen2014}, simulations
from the FIRE project have shown that above a stellar mass of about
$10^6\,\rm{M_{\odot}}$, a bursty star formation rate generated by a strong
stellar feedback transforms a central cusp into a core in the inner one kpc
\citep{onorbe2015,chan2015,fitts2017}.  A similar trend has been found by
\citet{read2016a} in slowly rotating dwarf galaxies. These results are however
in contradiction with the APOSTLE simulations of \citet{sawala2016b}.  The real need
of transforming a cusp into a core is indeed still debated.  \citet{pineda2017}
discussed that the rotation curves of dwarf galaxies could mimic a core even if
the galaxy had a cuspy Navarro, Frenk and White (NFW)'s
profile. \citet{strigari2017} found that, contrary to previous studies
\citep[see for example][]{battaglia2008}, the two kinematically distinct stellar
populations in the Sculptor dSph are consistent with populations in equilibrium
within a NFW dark matter potential.  Very recently, \citet{massari2018} showed
that the internal motions of the Sculptor's stars are in agreement with a cuspy
halo.

To check the effect of baryons in our simulations, we have performed, for each
of our zoom-in simulations, a corresponding dark matter only simulation, where
all the baryonic mass was transferred into the non-dissipative dark component.
A one-to-one comparison between haloes of the two simulations confirms that even 
after $14\,\rm{Gyrs}$ of evolution, there is no core beyond the gravitational softening
scale, for haloes that contains baryons.
The halo profiles of the two sets of simulations
remain pretty similar even for the most massive systems that continuously 
formed stars and despite a rather bursty star formation history.


\section{Conclusions}

\label{conclusions_discussions}


We have presented the formation and evolution of dwarf galaxies emerging from a
cosmological $\Lambda$CDM framework.  We first ran a dark matter only simulation
in a $3.4\,\rm{Mpc}$$/h$ box with a resolution of $2\times 512^3$ particles. We
extracted the 198 haloes which at $z=0$ had virial masses between $4 \times
10^8\,\rm{M_\odot}$$/h$ and $10^{10}\,\rm{M_\odot}$$/h$.  We selected the 62
haloes whose initial mass distribution was compact enough to allow a zoom-in
re-simulation up to $z=0$ with a stellar mass resolution of
$1024\,\rm{M_\odot}$, including a full treatment of the baryons.  From those systems, 27 had a final V-band luminosity
larger than $10^5\,\rm{L_\odot}$, corresponding to Local Group classical
spheroidal or irregular dwarfs, and have been studied in detail. These
simulations do not include the presence of a massive central
galaxy. Nevertheless, they allow us to understand the effect of a complex
hierarchical build-up on the final galaxy properties.

We have identified three modes of star formation: quenched, extended and
sustained. These modes result from the rate at which the dark matter haloes assemble in
combination with the UV-background strength and stellar heating.
We have shown that the scaling relations correlating the galaxy integrated
properties such as its total V-band luminosity ($L_{\rm{V}}$), central line
of sight velocity dispersion ($\sigma_{\rm{LOS}}$),  half-light radius
($r_{1/2}$) as well as its gas mass are reproduced over several orders of
magnitude.


We have conducted checks in an unprecedented high level of detail. In
particular, we could reproduce the properties of six Local Group dwarf galaxies
taken as test-beds: NGC 6622, Andromeda II, Sculptor, Sextans, Ursa Minor and
Draco.  This includes their stellar metallicity distributions and abundance ratios
[Mg/Fe], their velocity dispersion profiles, as well as their star formation
histories.  Metallicity gradients form naturally during the star forming period.
Metals accumulate in the centre of the dwarfs where gas and star formation concentrate
with time.  This mechanism is also at the origin of kinematically distinct stellar
populations.

These ensemble of successful results never achieved before by simulations run in a cosmological context,
provides further insights onto galaxy formation:

\begin{itemize}
  \item The observed variety of star formation histories in Local Group dwarfs
    is intrinsic to the hierarchical formation framework.  The interaction with
    a massive central galaxy could be needed  for a handful of cases only, such
    as the Carina (series of star formation bursts) and Fornax (dominant
    intermediate age population) dSphs, whose star formation histories are indeed not
    found in any of our models.

  \item The inclusion of self-shielding appears to be a major ingredient to allow dark matter haloes with masses as low as $3\times 10^8\,\,{M_\odot}$ to host stellar systems 
  with dynamical properties in agreement with the observations.

  \item We confirm the failure of the abundance matching approach at the scale
    of dwarf galaxies. It is caused by the variety of possible merger histories
    leading to a large range of final luminosities at given dark matter halo
    mass, independent of any interaction between the dwarfs and a host massive
    galaxy.

  \item Despite the fact that our simulated dwarf galaxies are embedded in
    haloes with masses standing at the lower boundary of what is found in other
    works, we do not face any missing satellites problem
    down to $10^5\,\rm{L_\odot}$. This is the consequence of an appropriate
    treatment of the star formation and feedback.
  
  \item We do not find evidence for any cusp/core transformation. 

  \item We have uncovered a side effect of the sampling of the dark matter
    haloes by particles. Even at our high resolution, the discrete representation
    of the halo leads to a spurious dynamical heating of the stellar component
    by mutual gravitational interactions. Consequently the galaxy half light
    radii are artificially increased at $z=0$, particularly in systems in which
    the star formation has been quenched. However, this sampling could be
    valid if dark matter were composed of massive black holes.  Otherwise
    the simulations need to reach even higher resolutions to completely avoid this
    bias.

  \item While we learn from the successes of the simulations, the challenges are
    also informative: (i) The existence of faint galaxies ($L_V \sim 5\times
    10^5\,\rm{L_\odot}$) with HI and residual star formation, such as Leo T and
    Leo P, is not predicted in a $\Lambda$CDM framework and a classical
    UV-background prescription.  These cases indicate that further investigation
    of the strength and impact of the UV-background heating and/or the hydrogen
    self-shielding is necessary. (ii) While our models match the observations
    over four dex in luminosity, we have below $10^6\,\rm{L_{\odot}}$ a tail of
    faint dwarfs which stand below the observed range of metallicity. This
    points to the necessity of an improved treatment of the very first
    generations of stars.
  
\end{itemize}

\begin{acknowledgements}

The authors thank the reviewer Stefania Salvadori  for her thorough review and highly appreciate comments and suggestions, 
which significantly contributed to improving the quality of the publication. 
We are indebted to the International Space Science Institute (ISSI), Bern, Switzerland, for supporting and funding the international team `First stars in dwarf galaxies'.
We enjoyed discussions with Pierre North, Daniel Pfenniger, Elena Ricciardelli, Lo\"ic Hausammann, Romain Teyssier and Eline Tolstoy.
We thank Jennifer Schober for her comments and typos corrections.
This work was supported by the Swiss Federal Institute of Technology in Lausanne (EPFL) through the use of the
facilities of its Scientific IT and Application Support Center (SCITAS). The simulations presented here were run on
the Deneb and Bellatrix clusters.
The data reduction and galaxy maps have been
performed using the parallelized Python \texttt{pNbody} package
(\texttt{http://lastro.epfl.ch/projects/pNbody/}).  

\end{acknowledgements}

\bibliographystyle{aa}
\bibliography{bibliography}

\begin{thebibliography}{179}
\expandafter\ifx\csname natexlab\endcsname\relax\def\natexlab#1{#1}\fi

\bibitem[{{Aoki} {et~al.}(2009){Aoki}, {Arimoto}, {Sadakane}, {Tolstoy},
  {Battaglia}, {Jablonka}, {Shetrone}, {Letarte}, {Irwin}, {Hill}, {Francois},
  {Venn}, {Primas}, {Helmi}, {Kaufer}, {Tafelmeyer}, {Szeifert}, \&
  {Babusiaux}}]{aoki2009}
{Aoki}, W., {Arimoto}, N., {Sadakane}, K., {et~al.} 2009, \aap, 502, 569

\bibitem[{{Aparicio} {et~al.}(2001){Aparicio}, {Carrera}, \&
  {Mart{\'{\i}}nez-Delgado}}]{aparicio2001}
{Aparicio}, A., {Carrera}, R., \& {Mart{\'{\i}}nez-Delgado}, D. 2001, \aj, 122,
  2524

\bibitem[{{Atek} {et~al.}(2015){Atek}, {Richard}, {Jauzac}, {Kneib},
  {Natarajan}, {Limousin}, {Schaerer}, {Jullo}, {Ebeling}, {Egami}, \&
  {Clement}}]{atek2015}
{Atek}, H., {Richard}, J., {Jauzac}, M., {et~al.} 2015, \apj, 814, 69

\bibitem[{{Aubert} \& {Teyssier}(2010)}]{aubert2010}
{Aubert}, D. \& {Teyssier}, R. 2010, \apj, 724, 244

\bibitem[{{Bate} \& {Burkert}(1997)}]{bate1997}
{Bate}, M.~R. \& {Burkert}, A. 1997, \mnras, 288, 1060

\bibitem[{{Battaglia} {et~al.}(2008){Battaglia}, {Helmi}, {Tolstoy}, {Irwin},
  {Hill}, \& {Jablonka}}]{battaglia2008}
{Battaglia}, G., {Helmi}, A., {Tolstoy}, E., {et~al.} 2008, \apjl, 681, L13

\bibitem[{{Battaglia} {et~al.}(2011){Battaglia}, {Tolstoy}, {Helmi}, {Irwin},
  {Parisi}, {Hill}, \& {Jablonka}}]{battaglia2011}
{Battaglia}, G., {Tolstoy}, E., {Helmi}, A., {et~al.} 2011, \mnras, 411, 1013

\bibitem[{{Battaglia} {et~al.}(2006){Battaglia}, {Tolstoy}, {Helmi}, {Irwin},
  {Letarte}, {Jablonka}, {Hill}, {Venn}, {Shetrone}, {Arimoto}, {Primas},
  {Kaufer}, {Francois}, {Szeifert}, {Abel}, \& {Sadakane}}]{battaglia2006}
{Battaglia}, G., {Tolstoy}, E., {Helmi}, A., {et~al.} 2006, \aap, 459, 423

\bibitem[{{Behroozi} {et~al.}(2013){Behroozi}, {Wechsler}, \&
  {Conroy}}]{behroozi2013}
{Behroozi}, P.~S., {Wechsler}, R.~H., \& {Conroy}, C. 2013, \apj, 770, 57

\bibitem[{{Ben{\'{\i}}tez-Llambay} {et~al.}(2016){Ben{\'{\i}}tez-Llambay},
  {Navarro}, {Abadi}, {Gottl{\"o}ber}, {Yepes}, {Hoffman}, \&
  {Steinmetz}}]{benitez-llambay2016}
{Ben{\'{\i}}tez-Llambay}, A., {Navarro}, J.~F., {Abadi}, M.~G., {et~al.} 2016,
  \mnras, 456, 1185

\bibitem[{{Bettinelli} {et~al.}(2018){Bettinelli}, {Hidalgo}, {Cassisi},
  {Aparicio}, \& {Piotto}}]{bettinelli2018}
{Bettinelli}, M., {Hidalgo}, S.~L., {Cassisi}, S., {Aparicio}, A., \& {Piotto},
  G. 2018, \mnras, 476, 71

\bibitem[{{Bouwens} {et~al.}(2015){Bouwens}, {Illingworth}, {Oesch}, {Caruana},
  {Holwerda}, {Smit}, \& {Wilkins}}]{bouwens2015}
{Bouwens}, R.~J., {Illingworth}, G.~D., {Oesch}, P.~A., {et~al.} 2015, \apj,
  811, 140

\bibitem[{{Boylan-Kolchin} {et~al.}(2011){Boylan-Kolchin}, {Bullock}, \&
  {Kaplinghat}}]{boylankolchin2011}
{Boylan-Kolchin}, M., {Bullock}, J.~S., \& {Kaplinghat}, M. 2011, \mnras, 415,
  L40

\bibitem[{{Boylan-Kolchin} {et~al.}(2012){Boylan-Kolchin}, {Bullock}, \&
  {Kaplinghat}}]{boylankolchin2012}
{Boylan-Kolchin}, M., {Bullock}, J.~S., \& {Kaplinghat}, M. 2012, \mnras, 422,
  1203

\bibitem[{{Bromm}(2013)}]{bromm2013}
{Bromm}, V. 2013, Reports on Progress in Physics, 76, 112901

\bibitem[{{Brooks} \& {Zolotov}(2014)}]{brooks2014}
{Brooks}, A.~M. \& {Zolotov}, A. 2014, \apj, 786, 87

\bibitem[{{Bullock} \& {Boylan-Kolchin}(2017)}]{bullock2017}
{Bullock}, J.~S. \& {Boylan-Kolchin}, M. 2017, \araa, 55, 343

\bibitem[{{Bullock} {et~al.}(2000){Bullock}, {Kravtsov}, \&
  {Weinberg}}]{bullock2000}
{Bullock}, J.~S., {Kravtsov}, A.~V., \& {Weinberg}, D.~H. 2000, \apj, 539, 517

\bibitem[{{Carrera} {et~al.}(2002){Carrera}, {Aparicio},
  {Mart{\'{\i}}nez-Delgado}, \& {Alonso-Garc{\'{\i}}a}}]{carrera2002}
{Carrera}, R., {Aparicio}, A., {Mart{\'{\i}}nez-Delgado}, D., \&
  {Alonso-Garc{\'{\i}}a}, J. 2002, \aj, 123, 3199

\bibitem[{{Chan} {et~al.}(2015){Chan}, {Kere{\v s}}, {O{\~n}orbe}, {Hopkins},
  {Muratov}, {Faucher-Gigu{\`e}re}, \& {Quataert}}]{chan2015}
{Chan}, T.~K., {Kere{\v s}}, D., {O{\~n}orbe}, J., {et~al.} 2015, \mnras, 454,
  2981

\bibitem[{{Choudhury} {et~al.}(2008){Choudhury}, {Ferrara}, \&
  {Gallerani}}]{choudhury2008}
{Choudhury}, T.~R., {Ferrara}, A., \& {Gallerani}, S. 2008, \mnras, 385, L58

\bibitem[{{Cloet-Osselaer} {et~al.}(2012){Cloet-Osselaer}, {De Rijcke},
  {Schroyen}, \& {Dury}}]{cloet-osselaer2012}
{Cloet-Osselaer}, A., {De Rijcke}, S., {Schroyen}, J., \& {Dury}, V. 2012,
  \mnras, 423, 735

\bibitem[{{Cloet-Osselaer} {et~al.}(2014){Cloet-Osselaer}, {De Rijcke},
  {Vandenbroucke}, {Schroyen}, {Koleva}, \& {Verbeke}}]{cloet-osselaer2014}
{Cloet-Osselaer}, A., {De Rijcke}, S., {Vandenbroucke}, B., {et~al.} 2014,
  \mnras, 442, 2909

\bibitem[{{Cohen} \& {Huang}(2009)}]{cohen2009}
{Cohen}, J.~G. \& {Huang}, W. 2009, \apj, 701, 1053

\bibitem[{{Cohen} \& {Huang}(2010)}]{cohen2010}
{Cohen}, J.~G. \& {Huang}, W. 2010, \apj, 719, 931

\bibitem[{{Coleman} \& {de Jong}(2008)}]{coleman2008}
{Coleman}, M.~G. \& {de Jong}, J.~T.~A. 2008, \apj, 685, 933

\bibitem[{{Cormier} {et~al.}(2014){Cormier}, {Madden}, {Lebouteiller}, {Hony},
  {Aalto}, {Costagliola}, {Hughes}, {R{\'e}my-Ruyer}, {Abel}, {Bayet},
  {Bigiel}, {Cannon}, {Cumming}, {Galametz}, {Galliano}, {Viti}, \&
  {Wu}}]{cormier2014}
{Cormier}, D., {Madden}, S.~C., {Lebouteiller}, V., {et~al.} 2014, \aap, 564,
  A121

\bibitem[{{Crain} {et~al.}(2017){Crain}, {Bah{\'e}}, {Lagos}, {Rahmati},
  {Schaye}, {McCarthy}, {Marasco}, {Bower}, {Schaller}, {Theuns}, \& {van der
  Hulst}}]{crain2017}
{Crain}, R.~A., {Bah{\'e}}, Y.~M., {Lagos}, C.~d.~P., {et~al.} 2017, \mnras,
  464, 4204

\bibitem[{{de Boer} {et~al.}(2012{\natexlab{a}}){de Boer}, {Tolstoy}, {Hill},
  {Saha}, {Olsen}, {Starkenburg}, {Lemasle}, {Irwin}, \&
  {Battaglia}}]{deboer2012}
{de Boer}, T.~J.~L., {Tolstoy}, E., {Hill}, V., {et~al.} 2012{\natexlab{a}},
  \aap, 539, A103

\bibitem[{{de Boer} {et~al.}(2012{\natexlab{b}}){de Boer}, {Tolstoy}, {Hill},
  {Saha}, {Olszewski}, {Mateo}, {Starkenburg}, {Battaglia}, \&
  {Walker}}]{deboer2012b}
{de Boer}, T.~J.~L., {Tolstoy}, E., {Hill}, V., {et~al.} 2012{\natexlab{b}},
  \aap, 544, A73

\bibitem[{{de Boer} {et~al.}(2014){de Boer}, {Tolstoy}, {Lemasle}, {Saha},
  {Olszewski}, {Mateo}, {Irwin}, \& {Battaglia}}]{deboer2014}
{de Boer}, T.~J.~L., {Tolstoy}, E., {Lemasle}, B., {et~al.} 2014, \aap, 572,
  A10

\bibitem[{{De Rijcke} {et~al.}(2013){De Rijcke}, {Schroyen}, {Vandenbroucke},
  {Jachowicz}, {Decroos}, {Cloet-Osselaer}, \& {Koleva}}]{derijcke2013}
{De Rijcke}, S., {Schroyen}, J., {Vandenbroucke}, B., {et~al.} 2013, \mnras,
  433, 3005

\bibitem[{{Di Cintio} {et~al.}(2014){Di Cintio}, {Brook}, {Macci{\`o}},
  {Stinson}, {Knebe}, {Dutton}, \& {Wadsley}}]{dicintio2014}
{Di Cintio}, A., {Brook}, C.~B., {Macci{\`o}}, A.~V., {et~al.} 2014, \mnras,
  437, 415

\bibitem[{{Dolphin}(2002)}]{dolphin2002}
{Dolphin}, A.~E. 2002, \mnras, 332, 91

\bibitem[{{Durier} \& {Dalla Vecchia}(2012)}]{durier2012}
{Durier}, F. \& {Dalla Vecchia}, C. 2012, \mnras, 419, 465

\bibitem[{{Efstathiou}(1992)}]{efstathiou1992}
{Efstathiou}, G. 1992, \mnras, 256, 43P

\bibitem[{{Escala} {et~al.}(2018){Escala}, {Wetzel}, {Kirby}, {Hopkins}, {Ma},
  {Wheeler}, {Kere{\v s}}, {Faucher-Gigu{\`e}re}, \& {Quataert}}]{escala2018}
{Escala}, I., {Wetzel}, A., {Kirby}, E.~N., {et~al.} 2018, \mnras, 474, 2194

\bibitem[{{Fabrizio} {et~al.}(2016){Fabrizio}, {Bono}, {Nonino}, {{\L}okas},
  {Ferraro}, {Iannicola}, {Buonanno}, {Cassisi}, {Coppola}, {Dall'Ora},
  {Gilmozzi}, {Marconi}, {Monelli}, {Romaniello}, {Stetson}, {Th{\'e}venin}, \&
  {Walker}}]{fabrizio2016}
{Fabrizio}, M., {Bono}, G., {Nonino}, M., {et~al.} 2016, \apj, 830, 126

\bibitem[{{Fabrizio} {et~al.}(2011){Fabrizio}, {Nonino}, {Bono}, {Ferraro},
  {Fran{\c c}ois}, {Iannicola}, {Monelli}, {Th{\'e}venin}, {Stetson}, {Walker},
  {Buonanno}, {Caputo}, {Corsi}, {Dall'Ora}, {Gilmozzi}, {James}, {Merle},
  {Pulone}, \& {Romaniello}}]{fabrizio2011}
{Fabrizio}, M., {Nonino}, M., {Bono}, G., {et~al.} 2011, \pasp, 123, 384

\bibitem[{{Faria} {et~al.}(2007){Faria}, {Feltzing}, {Lundstr{\"o}m},
  {Gilmore}, {Wahlgren}, {Ardeberg}, \& {Linde}}]{faria2007}
{Faria}, D., {Feltzing}, S., {Lundstr{\"o}m}, I., {et~al.} 2007, \aap, 465, 357

\bibitem[{{Faucher-Gigu{\`e}re} {et~al.}(2009){Faucher-Gigu{\`e}re}, {Lidz},
  {Zaldarriaga}, \& {Hernquist}}]{faucher-giguere2009}
{Faucher-Gigu{\`e}re}, C.-A., {Lidz}, A., {Zaldarriaga}, M., \& {Hernquist}, L.
  2009, \apj, 703, 1416

\bibitem[{{Ferland} {et~al.}(2013){Ferland}, {Porter}, {van Hoof}, {Williams},
  {Abel}, {Lykins}, {Shaw}, {Henney}, \& {Stancil}}]{ferland2013}
{Ferland}, G.~J., {Porter}, R.~L., {van Hoof}, P.~A.~M., {et~al.} 2013, \rmxaa,
  49, 137

\bibitem[{{Fitts} {et~al.}(2017){Fitts}, {Boylan-Kolchin}, {Elbert}, {Bullock},
  {Hopkins}, {O{\~n}orbe}, {Wetzel}, {Wheeler}, {Faucher-Gigu{\`e}re}, {Kere{\v
  s}}, {Skillman}, \& {Weisz}}]{fitts2017}
{Fitts}, A., {Boylan-Kolchin}, M., {Elbert}, O.~D., {et~al.} 2017, \mnras, 471,
  3547

\bibitem[{{Frebel} {et~al.}(2010){Frebel}, {Simon}, {Geha}, \&
  {Willman}}]{frebel2010b}
{Frebel}, A., {Simon}, J.~D., {Geha}, M., \& {Willman}, B. 2010, \apj, 708, 560

\bibitem[{{Fulbright} {et~al.}(2004){Fulbright}, {Rich}, \&
  {Castro}}]{fulbright2004}
{Fulbright}, J.~P., {Rich}, R.~M., \& {Castro}, S. 2004, \apj, 612, 447

\bibitem[{{Giovanelli} {et~al.}(2013){Giovanelli}, {Haynes}, {Adams}, {Cannon},
  {Rhode}, {Salzer}, {Skillman}, {Bernstein-Cooper}, \&
  {McQuinn}}]{giovanelli2013}
{Giovanelli}, R., {Haynes}, M.~P., {Adams}, E.~A.~K., {et~al.} 2013, \aj, 146,
  15

\bibitem[{{Gullieuszik} {et~al.}(2009){Gullieuszik}, {Held}, {Saviane}, \&
  {Rizzi}}]{gullieuszik2009}
{Gullieuszik}, M., {Held}, E.~V., {Saviane}, I., \& {Rizzi}, L. 2009, \aap,
  500, 735

\bibitem[{{Guo} {et~al.}(2010){Guo}, {White}, {Li}, \&
  {Boylan-Kolchin}}]{guo2010}
{Guo}, Q., {White}, S., {Li}, C., \& {Boylan-Kolchin}, M. 2010, \mnras, 404,
  1111

\bibitem[{{Haardt} \& {Madau}(2012)}]{haardt2012}
{Haardt}, F. \& {Madau}, P. 2012, \apj, 746, 125

\bibitem[{{Hahn} \& {Abel}(2011)}]{hahn2011}
{Hahn}, O. \& {Abel}, T. 2011, \mnras, 415, 2101

\bibitem[{{Harbeck} {et~al.}(2001){Harbeck}, {Grebel}, {Holtzman},
  {Guhathakurta}, {Brandner}, {Geisler}, {Sarajedini}, {Dolphin},
  {Hurley-Keller}, \& {Mateo}}]{harbeck2001}
{Harbeck}, D., {Grebel}, E.~K., {Holtzman}, J., {et~al.} 2001, \aj, 122, 3092

\bibitem[{{Heger} \& {Woosley}(2010)}]{heger2010}
{Heger}, A. \& {Woosley}, S.~E. 2010, \apj, 724, 341

\bibitem[{{Hirai} {et~al.}(2017){Hirai}, {Ishimaru}, {Saitoh}, {Fujii},
  {Hidaka}, \& {Kajino}}]{hirai2017}
{Hirai}, Y., {Ishimaru}, Y., {Saitoh}, T.~R., {et~al.} 2017, \mnras, 466, 2474

\bibitem[{{Ho} {et~al.}(2012){Ho}, {Geha}, {Munoz}, {Guhathakurta}, {Kalirai},
  {Gilbert}, {Tollerud}, {Bullock}, {Beaton}, \& {Majewski}}]{ho2012}
{Ho}, N., {Geha}, M., {Munoz}, R.~R., {et~al.} 2012, \apj, 758, 124

\bibitem[{{Ho} {et~al.}(2015){Ho}, {Geha}, {Tollerud}, {Zinn}, {Guhathakurta},
  \& {Vargas}}]{ho2015}
{Ho}, N., {Geha}, M., {Tollerud}, E.~J., {et~al.} 2015, \apj, 798, 77

\bibitem[{{Hopkins}(2013)}]{hopkins2013}
{Hopkins}, P.~F. 2013, \mnras, 428, 2840

\bibitem[{{Hopkins} {et~al.}(2011){Hopkins}, {Quataert}, \&
  {Murray}}]{hopkins2011}
{Hopkins}, P.~F., {Quataert}, E., \& {Murray}, N. 2011, \mnras, 417, 950

\bibitem[{{Hopkins} {et~al.}(2017){Hopkins}, {Wetzel}, {Keres},
  {Faucher-Giguere}, {Quataert}, {Boylan-Kolchin}, {Murray}, {Hayward},
  {Garrison-Kimmel}, {Hummels}, {Feldmann}, {Torrey}, {Ma}, {Angles-Alcazar},
  {Su}, {Orr}, {Schmitz}, {Escala}, {Sanderson}, {Grudic}, {Hafen}, {Kim},
  {Fitts}, {Bullock}, {Wheeler}, {Chan}, {Elbert}, \&
  {Narananan}}]{hopkins2017}
{Hopkins}, P.~F., {Wetzel}, A., {Keres}, D., {et~al.} 2017, ArXiv e-prints

\bibitem[{{Hurley-Keller} {et~al.}(1998){Hurley-Keller}, {Mateo}, \&
  {Nemec}}]{hurley-keller1998}
{Hurley-Keller}, D., {Mateo}, M., \& {Nemec}, J. 1998, \aj, 115, 1840

\bibitem[{{Ibata} {et~al.}(2013){Ibata}, {Lewis}, {Conn}, {Irwin},
  {McConnachie}, {Chapman}, {Collins}, {Fardal}, {Ferguson}, {Ibata}, {Mackey},
  {Martin}, {Navarro}, {Rich}, {Valls-Gabaud}, \& {Widrow}}]{ibata2013}
{Ibata}, R.~A., {Lewis}, G.~F., {Conn}, A.~R., {et~al.} 2013, \nat, 493, 62

\bibitem[{{Iwamoto} {et~al.}(2005){Iwamoto}, {Umeda}, {Tominaga}, {Nomoto}, \&
  {Maeda}}]{iwamoto2005}
{Iwamoto}, N., {Umeda}, H., {Tominaga}, N., {Nomoto}, K., \& {Maeda}, K. 2005,
  Science, 309, 451

\bibitem[{{Jablonka} {et~al.}(2015){Jablonka}, {North}, {Mashonkina}, {Hill},
  {Revaz}, {Shetrone}, {Starkenburg}, {Irwin}, {Tolstoy}, {Battaglia}, {Venn},
  {Helmi}, {Primas}, \& {Fran{\c c}ois}}]{jablonka2015}
{Jablonka}, P., {North}, P., {Mashonkina}, L., {et~al.} 2015, \aap, 583, A67

\bibitem[{{Jeon} {et~al.}(2017){Jeon}, {Besla}, \& {Bromm}}]{jeon2017}
{Jeon}, M., {Besla}, G., \& {Bromm}, V. 2017, \apj, 848, 85

\bibitem[{{Jin} {et~al.}(2005){Jin}, {Ostriker}, \& {Wilkinson}}]{jin2005}
{Jin}, S., {Ostriker}, J.~P., \& {Wilkinson}, M.~I. 2005, \mnras, 359, 104

\bibitem[{{Kim} {et~al.}(2014){Kim}, {Abel}, {Agertz}, {Bryan}, {Ceverino},
  {Christensen}, {Conroy}, {Dekel}, {Gnedin}, {Goldbaum}, {Guedes}, {Hahn},
  {Hobbs}, {Hopkins}, {Hummels}, {Iannuzzi}, {Keres}, {Klypin}, {Kravtsov},
  {Krumholz}, {Kuhlen}, {Leitner}, {Madau}, {Mayer}, {Moody}, {Nagamine},
  {Norman}, {Onorbe}, {O'Shea}, {Pillepich}, {Primack}, {Quinn}, {Read},
  {Robertson}, {Rocha}, {Rudd}, {Shen}, {Smith}, {Szalay}, {Teyssier},
  {Thompson}, {Todoroki}, {Turk}, {Wadsley}, {Wise}, {Zolotov}, \& {AGORA
  Collaboration29}}]{kim2014}
{Kim}, J.-h., {Abel}, T., {Agertz}, O., {et~al.} 2014, \apjs, 210, 14

\bibitem[{{Kim} {et~al.}(2016){Kim}, {Agertz}, {Teyssier}, {Butler},
  {Ceverino}, {Choi}, {Feldmann}, {Keller}, {Lupi}, {Quinn}, {Revaz},
  {Wallace}, {Gnedin}, {Leitner}, {Shen}, {Smith}, {Thompson}, {Turk}, {Abel},
  {Arraki}, {Benincasa}, {Chakrabarti}, {DeGraf}, {Dekel}, {Goldbaum},
  {Hopkins}, {Hummels}, {Klypin}, {Li}, {Madau}, {Mandelker}, {Mayer},
  {Nagamine}, {Nickerson}, {O'Shea}, {Primack}, {Roca-F{\`a}brega}, {Semenov},
  {Shimizu}, {Simpson}, {Todoroki}, {Wadsley}, {Wise}, \& {AGORA
  Collaboration}}]{kim2016}
{Kim}, J.-h., {Agertz}, O., {Teyssier}, R., {et~al.} 2016, \apj, 833, 202

\bibitem[{{Kirby} {et~al.}(2013){Kirby}, {Cohen}, {Guhathakurta}, {Cheng},
  {Bullock}, \& {Gallazzi}}]{kirby2013}
{Kirby}, E.~N., {Cohen}, J.~G., {Guhathakurta}, P., {et~al.} 2013, \apj, 779,
  102

\bibitem[{{Kirby} {et~al.}(2009){Kirby}, {Guhathakurta}, {Bolte}, {Sneden}, \&
  {Geha}}]{kirby2009}
{Kirby}, E.~N., {Guhathakurta}, P., {Bolte}, M., {Sneden}, C., \& {Geha}, M.~C.
  2009, \apj, 705, 328

\bibitem[{{Kirby} {et~al.}(2010){Kirby}, {Guhathakurta}, {Simon}, {Geha},
  {Rockosi}, {Sneden}, {Cohen}, {Sohn}, {Majewski}, \& {Siegel}}]{kirby2010}
{Kirby}, E.~N., {Guhathakurta}, P., {Simon}, J.~D., {et~al.} 2010, \apjs, 191,
  352

\bibitem[{{Kirby} {et~al.}(2011){Kirby}, {Lanfranchi}, {Simon}, {Cohen}, \&
  {Guhathakurta}}]{kirby2011}
{Kirby}, E.~N., {Lanfranchi}, G.~A., {Simon}, J.~D., {Cohen}, J.~G., \&
  {Guhathakurta}, P. 2011, \apj, 727, 78

\bibitem[{{Klypin} {et~al.}(1999){Klypin}, {Kravtsov}, {Valenzuela}, \&
  {Prada}}]{klypin1999}
{Klypin}, A., {Kravtsov}, A.~V., {Valenzuela}, O., \& {Prada}, F. 1999, \apj,
  522, 82

\bibitem[{{Knebe} {et~al.}(2009){Knebe}, {Wagner}, {Knollmann}, {Diekershoff},
  \& {Krause}}]{knebe2009}
{Knebe}, A., {Wagner}, C., {Knollmann}, S., {Diekershoff}, T., \& {Krause}, F.
  2009, \apj, 698, 266

\bibitem[{{Kobayashi} {et~al.}(2000){Kobayashi}, {Tsujimoto}, \&
  {Nomoto}}]{kobayashi2000}
{Kobayashi}, C., {Tsujimoto}, T., \& {Nomoto}, K. 2000, \apj, 539, 26

\bibitem[{{Koch} {et~al.}(2006){Koch}, {Grebel}, {Wyse}, {Kleyna}, {Wilkinson},
  {Harbeck}, {Gilmore}, \& {Evans}}]{koch2006}
{Koch}, A., {Grebel}, E.~K., {Wyse}, R.~F.~G., {et~al.} 2006, \aj, 131, 895

\bibitem[{{Koch} {et~al.}(2008){Koch}, {McWilliam}, {Grebel}, {Zucker}, \&
  {Belokurov}}]{koch2008}
{Koch}, A., {McWilliam}, A., {Grebel}, E.~K., {Zucker}, D.~B., \& {Belokurov},
  V. 2008, \apjl, 688, L13

\bibitem[{{Kordopatis} {et~al.}(2016){Kordopatis}, {Amorisco}, {Evans},
  {Gilmore}, \& {Koposov}}]{kordopatis2016}
{Kordopatis}, G., {Amorisco}, N.~C., {Evans}, N.~W., {Gilmore}, G., \&
  {Koposov}, S.~E. 2016, \mnras, 457, 1299

\bibitem[{{Kroupa}(2001)}]{kroupa01}
{Kroupa}, P. 2001, \mnras, 322, 231

\bibitem[{{Lardo} {et~al.}(2016){Lardo}, {Battaglia}, {Pancino}, {Romano}, {de
  Boer}, {Starkenburg}, {Tolstoy}, {Irwin}, {Jablonka}, \& {Tosi}}]{lardo2016}
{Lardo}, C., {Battaglia}, G., {Pancino}, E., {et~al.} 2016, \aap, 585, A70

\bibitem[{{Lee} {et~al.}(2009){Lee}, {Yuk}, {Park}, {Harris}, \&
  {Zaritsky}}]{lee2009}
{Lee}, M.~G., {Yuk}, I.-S., {Park}, H.~S., {Harris}, J., \& {Zaritsky}, D.
  2009, \apj, 703, 692

\bibitem[{{Lemasle} {et~al.}(2012){Lemasle}, {Hill}, {Tolstoy}, {Venn},
  {Shetrone}, {Irwin}, {de Boer}, {Starkenburg}, \& {Salvadori}}]{lemasle2012}
{Lemasle}, B., {Hill}, V., {Tolstoy}, E., {et~al.} 2012, \aap, 538, A100

\bibitem[{{Letarte} {et~al.}(2010){Letarte}, {Hill}, {Tolstoy}, {Jablonka},
  {Shetrone}, {Venn}, {Spite}, {Irwin}, {Battaglia}, {Helmi}, {Primas},
  {Fran{\c c}ois}, {Kaufer}, {Szeifert}, {Arimoto}, \&
  {Sadakane}}]{letarte2010}
{Letarte}, B., {Hill}, V., {Tolstoy}, E., {et~al.} 2010, \aap, 523, A17

\bibitem[{{Lovell} {et~al.}(2012){Lovell}, {Eke}, {Frenk}, {Gao}, {Jenkins},
  {Theuns}, {Wang}, {White}, {Boyarsky}, \& {Ruchayskiy}}]{lovell2012}
{Lovell}, M.~R., {Eke}, V., {Frenk}, C.~S., {et~al.} 2012, \mnras, 420, 2318

\bibitem[{{Macci{\`o}} {et~al.}(2017){Macci{\`o}}, {Frings}, {Buck}, {Penzo},
  {Dutton}, {Blank}, \& {Obreja}}]{maccio2017}
{Macci{\`o}}, A.~V., {Frings}, J., {Buck}, T., {et~al.} 2017, \mnras, 472, 2356

\bibitem[{{Massari} {et~al.}(2018){Massari}, {Breddels}, {Helmi}, {Posti},
  {Brown}, \& {Tolstoy}}]{massari2018}
{Massari}, D., {Breddels}, M.~A., {Helmi}, A., {et~al.} 2018, Nature Astronomy,
  2, 156

\bibitem[{{McConnachie}(2012)}]{mcconnachie2012}
{McConnachie}, A.~W. 2012, \aj, 144, 4

\bibitem[{{McQuinn} {et~al.}(2015){McQuinn}, {Skillman}, {Dolphin}, {Cannon},
  {Salzer}, {Rhode}, {Adams}, {Berg}, {Giovanelli}, {Girardi}, \&
  {Haynes}}]{mcquinn2015}
{McQuinn}, K.~B.~W., {Skillman}, E.~D., {Dolphin}, A., {et~al.} 2015, \apj,
  812, 158

\bibitem[{{Moore}(1994)}]{moore1994}
{Moore}, B. 1994, \nat, 370, 629

\bibitem[{{Moore} {et~al.}(1999){Moore}, {Ghigna}, {Governato}, {Lake},
  {Quinn}, {Stadel}, \& {Tozzi}}]{moore1999}
{Moore}, B., {Ghigna}, S., {Governato}, F., {et~al.} 1999, \apjl, 524, L19

\bibitem[{{Moster} {et~al.}(2013){Moster}, {Naab}, \& {White}}]{moster2013}
{Moster}, B.~P., {Naab}, T., \& {White}, S.~D.~M. 2013, \mnras, 428, 3121

\bibitem[{{Moster} {et~al.}(2010){Moster}, {Somerville}, {Maulbetsch}, {van den
  Bosch}, {Macci{\`o}}, {Naab}, \& {Oser}}]{moster2010}
{Moster}, B.~P., {Somerville}, R.~S., {Maulbetsch}, C., {et~al.} 2010, \apj,
  710, 903

\bibitem[{{Navarro} {et~al.}(1996){Navarro}, {Frenk}, \& {White}}]{navarro1996}
{Navarro}, J.~F., {Frenk}, C.~S., \& {White}, S.~D.~M. 1996, \apj, 462, 563

\bibitem[{{Navarro} {et~al.}(1997){Navarro}, {Frenk}, \& {White}}]{navarro1997}
{Navarro}, J.~F., {Frenk}, C.~S., \& {White}, S.~D.~M. 1997, \apj, 490, 493

\bibitem[{{Nichols} {et~al.}(2014){Nichols}, {Revaz}, \&
  {Jablonka}}]{nichols2014}
{Nichols}, M., {Revaz}, Y., \& {Jablonka}, P. 2014, \aap, 564, A112

\bibitem[{{Nichols} {et~al.}(2015){Nichols}, {Revaz}, \&
  {Jablonka}}]{nichols2015}
{Nichols}, M., {Revaz}, Y., \& {Jablonka}, P. 2015, \aap, 582, A23

\bibitem[{{Noh} \& {McQuinn}(2014)}]{noh2014}
{Noh}, Y. \& {McQuinn}, M. 2014, \mnras, 444, 503

\bibitem[{{Norris} {et~al.}(2010){Norris}, {Yong}, {Gilmore}, \&
  {Wyse}}]{norris2010}
{Norris}, J.~E., {Yong}, D., {Gilmore}, G., \& {Wyse}, R.~F.~G. 2010, \apj,
  711, 350

\bibitem[{{O{\~n}orbe} {et~al.}(2015){O{\~n}orbe}, {Boylan-Kolchin}, {Bullock},
  {Hopkins}, {Kere{\v s}}, {Faucher-Gigu{\`e}re}, {Quataert}, \&
  {Murray}}]{onorbe2015}
{O{\~n}orbe}, J., {Boylan-Kolchin}, M., {Bullock}, J.~S., {et~al.} 2015,
  \mnras, 454, 2092

\bibitem[{{O{\~n}orbe} {et~al.}(2014){O{\~n}orbe}, {Garrison-Kimmel}, {Maller},
  {Bullock}, {Rocha}, \& {Hahn}}]{onorbe2014}
{O{\~n}orbe}, J., {Garrison-Kimmel}, S., {Maller}, A.~H., {et~al.} 2014,
  \mnras, 437, 1894

\bibitem[{{Okamoto} {et~al.}(2017){Okamoto}, {Arimoto}, {Tolstoy}, {Jablonka},
  {Irwin}, {Komiyama}, {Yamada}, \& {Onodera}}]{okamoto2017}
{Okamoto}, S., {Arimoto}, N., {Tolstoy}, E., {et~al.} 2017, \mnras, 467, 208

\bibitem[{{Okamoto} {et~al.}(2005){Okamoto}, {Eke}, {Frenk}, \&
  {Jenkins}}]{okamoto2005}
{Okamoto}, T., {Eke}, V.~R., {Frenk}, C.~S., \& {Jenkins}, A. 2005, \mnras,
  363, 1299

\bibitem[{{Okamoto} {et~al.}(2010){Okamoto}, {Frenk}, {Jenkins}, \&
  {Theuns}}]{okamoto2010}
{Okamoto}, T., {Frenk}, C.~S., {Jenkins}, A., \& {Theuns}, T. 2010, \mnras,
  406, 208

\bibitem[{{Okamoto} {et~al.}(2008){Okamoto}, {Gao}, \& {Theuns}}]{okamoto2008}
{Okamoto}, T., {Gao}, L., \& {Theuns}, T. 2008, \mnras, 390, 920

\bibitem[{{Owen} \& {Villumsen}(1997)}]{owen1997}
{Owen}, J.~M. \& {Villumsen}, J.~V. 1997, \apj, 481, 1

\bibitem[{{Pasetto} {et~al.}(2011){Pasetto}, {Grebel}, {Berczik}, {Chiosi}, \&
  {Spurzem}}]{pasetto2011}
{Pasetto}, S., {Grebel}, E.~K., {Berczik}, P., {Chiosi}, C., \& {Spurzem}, R.
  2011, \aap, 525, A99

\bibitem[{{Pawlowski} \& {Kroupa}(2013)}]{pawlowski2013}
{Pawlowski}, M.~S. \& {Kroupa}, P. 2013, \mnras, 435, 2116

\bibitem[{{Pawlowski} {et~al.}(2012){Pawlowski}, {Pflamm-Altenburg}, \&
  {Kroupa}}]{pawlowski2012}
{Pawlowski}, M.~S., {Pflamm-Altenburg}, J., \& {Kroupa}, P. 2012, \mnras, 423,
  1109

\bibitem[{{Pineda} {et~al.}(2017){Pineda}, {Hayward}, {Springel}, \& {Mendes de
  Oliveira}}]{pineda2017}
{Pineda}, J.~C.~B., {Hayward}, C.~C., {Springel}, V., \& {Mendes de Oliveira},
  C. 2017, \mnras, 466, 63

\bibitem[{{Planck Collaboration} {et~al.}(2016){Planck Collaboration}, {Ade},
  {Aghanim}, {Arnaud}, {Ashdown}, {Aumont}, {Baccigalupi}, {Banday},
  {Barreiro}, {Bartlett}, \& et~al.}]{planck2016}
{Planck Collaboration}, {Ade}, P.~A.~R., {Aghanim}, N., {et~al.} 2016, \aap,
  594, A13

\bibitem[{{Pontzen} \& {Governato}(2014)}]{pontzen2014}
{Pontzen}, A. \& {Governato}, F. 2014, \nat, 506, 171

\bibitem[{{Quinn} {et~al.}(1996){Quinn}, {Katz}, \& {Efstathiou}}]{quinn1996}
{Quinn}, T., {Katz}, N., \& {Efstathiou}, G. 1996, \mnras, 278, L49

\bibitem[{{Rahmati} {et~al.}(2013){Rahmati}, {Pawlik}, {Rai{\v c}evic}, \&
  {Schaye}}]{rahmati2013}
{Rahmati}, A., {Pawlik}, A.~H., {Rai{\v c}evic}, M., \& {Schaye}, J. 2013,
  \mnras, 430, 2427

\bibitem[{{Read} {et~al.}(2016){Read}, {Agertz}, \& {Collins}}]{read2016a}
{Read}, J.~I., {Agertz}, O., \& {Collins}, M.~L.~M. 2016, \mnras, 459, 2573

\bibitem[{{Revaz} {et~al.}(2016{\natexlab{a}}){Revaz}, {Arnaudon}, {Nichols},
  {Bonvin}, \& {Jablonka}}]{revaz2016}
{Revaz}, Y., {Arnaudon}, A., {Nichols}, M., {Bonvin}, V., \& {Jablonka}, P.
  2016{\natexlab{a}}, \aap, 588, A21

\bibitem[{{Revaz} \& {Jablonka}(2012)}]{revaz2012}
{Revaz}, Y. \& {Jablonka}, P. 2012, \aap, 538, A82

\bibitem[{{Revaz} {et~al.}(2009){Revaz}, {Jablonka}, {Sawala}, {Hill},
  {Letarte}, {Irwin}, {Battaglia}, {Helmi}, {Shetrone}, {Tolstoy}, \&
  {Venn}}]{revaz2009}
{Revaz}, Y., {Jablonka}, P., {Sawala}, T., {et~al.} 2009, \aap, 501, 189

\bibitem[{{Revaz} {et~al.}(2016{\natexlab{b}}){Revaz}, {Jablonka}, {Teyssier},
  \& {Mayer}}]{revaz2016b}
{Revaz}, Y., {Jablonka}, P., {Teyssier}, R., \& {Mayer}, L. 2016{\natexlab{b}},
  Star Formation in Galaxy Evolution: Connecting Numerical Models to Reality,
  Saas-Fee Advanced Course, Volume 43.~ISBN 978-3-662-47889-9.~Springer-Verlag
  Berlin Heidelberg, 2016., 43

\bibitem[{{Rhode} {et~al.}(2013){Rhode}, {Salzer}, {Haurberg}, {Van Sistine},
  {Young}, {Haynes}, {Giovanelli}, {Cannon}, {Skillman}, {McQuinn}, \&
  {Adams}}]{rhode2013}
{Rhode}, K.~L., {Salzer}, J.~J., {Haurberg}, N.~C., {et~al.} 2013, \aj, 145,
  149

\bibitem[{{Ricotti} \& {Gnedin}(2005)}]{ricotti2005}
{Ricotti}, M. \& {Gnedin}, N.~Y. 2005, \apj, 629, 259

\bibitem[{{Ricotti} {et~al.}(2016){Ricotti}, {Parry}, \&
  {Gnedin}}]{ricotti2016}
{Ricotti}, M., {Parry}, O.~H., \& {Gnedin}, N.~Y. 2016, \apj, 831, 204

\bibitem[{{Robertson} {et~al.}(2015){Robertson}, {Ellis}, {Furlanetto}, \&
  {Dunlop}}]{robertson2015}
{Robertson}, B.~E., {Ellis}, R.~S., {Furlanetto}, S.~R., \& {Dunlop}, J.~S.
  2015, \apjl, 802, L19

\bibitem[{{Robertson} \& {Kravtsov}(2008)}]{robertson2008}
{Robertson}, B.~E. \& {Kravtsov}, A.~V. 2008, \apj, 680, 1083

\bibitem[{{Rocha} {et~al.}(2013){Rocha}, {Peter}, {Bullock}, {Kaplinghat},
  {Garrison-Kimmel}, {O{\~n}orbe}, \& {Moustakas}}]{rocha2013}
{Rocha}, M., {Peter}, A.~H.~G., {Bullock}, J.~S., {et~al.} 2013, \mnras, 430,
  81

\bibitem[{{Sadakane} {et~al.}(2004){Sadakane}, {Arimoto}, {Ikuta}, {Aoki},
  {Jablonka}, \& {Tajitsu}}]{sadakane2004}
{Sadakane}, K., {Arimoto}, N., {Ikuta}, C., {et~al.} 2004, \pasj, 56, 1041

\bibitem[{{Salvadori} {et~al.}(2008){Salvadori}, {Ferrara}, \&
  {Schneider}}]{salvadori2008}
{Salvadori}, S., {Ferrara}, A., \& {Schneider}, R. 2008, \mnras, 386, 348

\bibitem[{{Salvadori} {et~al.}(2015){Salvadori}, {Sk{\'u}lad{\'o}ttir}, \&
  {Tolstoy}}]{salvadori2015}
{Salvadori}, S., {Sk{\'u}lad{\'o}ttir}, {\'A}., \& {Tolstoy}, E. 2015, \mnras,
  454, 1320

\bibitem[{{Salvadori} {et~al.}(2014){Salvadori}, {Tolstoy}, {Ferrara}, \&
  {Zaroubi}}]{salvadori2014}
{Salvadori}, S., {Tolstoy}, E., {Ferrara}, A., \& {Zaroubi}, S. 2014, \mnras,
  437, L26

\bibitem[{{Santana} {et~al.}(2016){Santana}, {Mu{\~n}oz}, {de Boer}, {Simon},
  {Geha}, {C{\^o}t{\'e}}, {Guzm{\'a}n}, {Stetson}, \&
  {Djorgovski}}]{santana2016}
{Santana}, F.~A., {Mu{\~n}oz}, R.~R., {de Boer}, T.~J.~L., {et~al.} 2016, \apj,
  829, 86

\bibitem[{{Saviane} {et~al.}(2000){Saviane}, {Held}, \&
  {Bertelli}}]{saviane2000}
{Saviane}, I., {Held}, E.~V., \& {Bertelli}, G. 2000, \aap, 355, 56

\bibitem[{{Sawala} {et~al.}(2015){Sawala}, {Frenk}, {Fattahi}, {Navarro},
  {Bower}, {Crain}, {Dalla Vecchia}, {Furlong}, {Jenkins}, {McCarthy}, {Qu},
  {Schaller}, {Schaye}, \& {Theuns}}]{sawala2015}
{Sawala}, T., {Frenk}, C.~S., {Fattahi}, A., {et~al.} 2015, \mnras, 448, 2941

\bibitem[{{Sawala} {et~al.}(2016){Sawala}, {Frenk}, {Fattahi}, {Navarro},
  {Bower}, {Crain}, {Vecchia}, {Furlong}, {Helly}, {Jenkins}, {Oman},
  {Schaller}, {Schaye}, {Theuns}, {Trayford}, \& {White}}]{sawala2016b}
{Sawala}, T., {Frenk}, C.~S., {Fattahi}, A., {et~al.} 2016, \mnras, 457, 1931

\bibitem[{{Sawala} {et~al.}(2011){Sawala}, {Guo}, {Scannapieco}, {Jenkins}, \&
  {White}}]{sawala2011}
{Sawala}, T., {Guo}, Q., {Scannapieco}, C., {Jenkins}, A., \& {White}, S. 2011,
  \mnras, 413, 659

\bibitem[{{Sawala} {et~al.}(2017){Sawala}, {Pihajoki}, {Johansson}, {Frenk},
  {Navarro}, {Oman}, \& {White}}]{sawala2017}
{Sawala}, T., {Pihajoki}, P., {Johansson}, P.~H., {et~al.} 2017, \mnras, 467,
  4383

\bibitem[{{Sawala} {et~al.}(2010){Sawala}, {Scannapieco}, {Maio}, \&
  {White}}]{sawala2010}
{Sawala}, T., {Scannapieco}, C., {Maio}, U., \& {White}, S. 2010, \mnras, 402,
  1599

\bibitem[{{Sawala} {et~al.}(2012){Sawala}, {Scannapieco}, \&
  {White}}]{sawala2012}
{Sawala}, T., {Scannapieco}, C., \& {White}, S. 2012, \mnras, 420, 1714

\bibitem[{{Schaye} \& {Dalla Vecchia}(2008)}]{schaye2008}
{Schaye}, J. \& {Dalla Vecchia}, C. 2008, \mnras, 383, 1210

\bibitem[{{Schroyen} {et~al.}(2013){Schroyen}, {De Rijcke}, {Koleva},
  {Cloet-Osselaer}, \& {Vandenbroucke}}]{schroyen2013}
{Schroyen}, J., {De Rijcke}, S., {Koleva}, M., {Cloet-Osselaer}, A., \&
  {Vandenbroucke}, B. 2013, \mnras, 434, 888

\bibitem[{{Schroyen} {et~al.}(2011){Schroyen}, {de Rijcke}, {Valcke},
  {Cloet-Osselaer}, \& {Dejonghe}}]{schroyen2011}
{Schroyen}, J., {de Rijcke}, S., {Valcke}, S., {Cloet-Osselaer}, A., \&
  {Dejonghe}, H. 2011, \mnras, 416, 601

\bibitem[{{Shen} {et~al.}(2014){Shen}, {Madau}, {Conroy}, {Governato}, \&
  {Mayer}}]{shen2014}
{Shen}, S., {Madau}, P., {Conroy}, C., {Governato}, F., \& {Mayer}, L. 2014,
  \apj, 792, 99

\bibitem[{{Shetrone} {et~al.}(2003){Shetrone}, {Venn}, {Tolstoy}, {Primas},
  {Hill}, \& {Kaufer}}]{shetrone2003}
{Shetrone}, M., {Venn}, K.~A., {Tolstoy}, E., {et~al.} 2003, \aj, 125, 684

\bibitem[{{Shetrone} {et~al.}(2001){Shetrone}, {C{\^o}t{\'e}}, \&
  {Sargent}}]{shetrone2001}
{Shetrone}, M.~D., {C{\^o}t{\'e}}, P., \& {Sargent}, W.~L.~W. 2001, \apj, 548,
  592

\bibitem[{{Skillman} {et~al.}(2017){Skillman}, {Monelli}, {Weisz}, {Hidalgo},
  {Aparicio}, {Bernard}, {Boylan-Kolchin}, {Cassisi}, {Cole}, {Dolphin},
  {Ferguson}, {Gallart}, {Irwin}, {Martin}, {Mart{\'{\i}}nez-V{\'a}zquez},
  {Mayer}, {McConnachie}, {McQuinn}, {Navarro}, \& {Stetson}}]{skillman2017}
{Skillman}, E.~D., {Monelli}, M., {Weisz}, D.~R., {et~al.} 2017, \apj, 837, 102

\bibitem[{{Smith} {et~al.}(2017){Smith}, {Bryan}, {Glover}, {Goldbaum}, {Turk},
  {Regan}, {Wise}, {Schive}, {Abel}, {Emerick}, {O'Shea}, {Anninos}, {Hummels},
  \& {Khochfar}}]{smith2017}
{Smith}, B.~D., {Bryan}, G.~L., {Glover}, S.~C.~O., {et~al.} 2017, \mnras, 466,
  2217

\bibitem[{{Spencer} {et~al.}(2017){Spencer}, {Mateo}, {Walker}, {Olszewski},
  {McConnachie}, {Kirby}, \& {Koch}}]{spencer2017}
{Spencer}, M.~E., {Mateo}, M., {Walker}, M.~G., {et~al.} 2017, \aj, 153, 254

\bibitem[{{Spergel} \& {Steinhardt}(2000)}]{spergel2000}
{Spergel}, D.~N. \& {Steinhardt}, P.~J. 2000, Physical Review Letters, 84, 3760

\bibitem[{{Springel}(2005)}]{springel2005}
{Springel}, V. 2005, \mnras, 364, 1105

\bibitem[{{Springel} {et~al.}(2001){Springel}, {White}, {Tormen}, \&
  {Kauffmann}}]{springel2001}
{Springel}, V., {White}, S.~D.~M., {Tormen}, G., \& {Kauffmann}, G. 2001,
  \mnras, 328, 726

\bibitem[{{Starkenburg} {et~al.}(2013){Starkenburg}, {Hill}, {Tolstoy},
  {Fran{\c c}ois}, {Irwin}, {Boschman}, {Venn}, {de Boer}, {Lemasle},
  {Jablonka}, {Battaglia}, {Groot}, \& {Kaper}}]{starkenburg2013}
{Starkenburg}, E., {Hill}, V., {Tolstoy}, E., {et~al.} 2013, \aap, 549, A88

\bibitem[{{Stinson} {et~al.}(2006){Stinson}, {Seth}, {Katz}, {Wadsley},
  {Governato}, \& {Quinn}}]{stinson2006}
{Stinson}, G., {Seth}, A., {Katz}, N., {et~al.} 2006, \mnras, 373, 1074

\bibitem[{{Strigari} {et~al.}(2017){Strigari}, {Frenk}, \&
  {White}}]{strigari2017}
{Strigari}, L.~E., {Frenk}, C.~S., \& {White}, S.~D.~M. 2017, \apj, 838, 123

\bibitem[{{Suda} {et~al.}(2017){Suda}, {Hidaka}, {Aoki}, {Katsuta}, {Yamada},
  {Fujimoto}, {Ohtani}, {Masuyama}, {Noda}, \& {Wada}}]{suda2017}
{Suda}, T., {Hidaka}, J., {Aoki}, W., {et~al.} 2017, \pasj, 69, 76

\bibitem[{{Suda} {et~al.}(2008){Suda}, {Katsuta}, {Yamada}, {Suwa}, {Ishizuka},
  {Komiya}, {Sorai}, {Aikawa}, \& {Fujimoto}}]{suda2008}
{Suda}, T., {Katsuta}, Y., {Yamada}, S., {et~al.} 2008, \pasj, 60, 1159

\bibitem[{{Susa} \& {Umemura}(2004)}]{susa2004}
{Susa}, H. \& {Umemura}, M. 2004, \apj, 600, 1

\bibitem[{{Swan} {et~al.}(2016){Swan}, {Cole}, {Tolstoy}, \&
  {Irwin}}]{swan2016}
{Swan}, J., {Cole}, A.~A., {Tolstoy}, E., \& {Irwin}, M.~J. 2016, \mnras, 456,
  4315

\bibitem[{{Tafelmeyer} {et~al.}(2010){Tafelmeyer}, {Jablonka}, {Hill},
  {Shetrone}, {Tolstoy}, {Irwin}, {Battaglia}, {Helmi}, {Starkenburg}, {Venn},
  {Abel}, {Francois}, {Kaufer}, {North}, {Primas}, \&
  {Szeifert}}]{tafelmeyer2010}
{Tafelmeyer}, M., {Jablonka}, P., {Hill}, V., {et~al.} 2010, \aap, 524, A58

\bibitem[{{Tajiri} \& {Umemura}(1998)}]{tajiri1998}
{Tajiri}, Y. \& {Umemura}, M. 1998, \apj, 502, 59

\bibitem[{{Theler}(2018)}]{theler2018}
{Theler}, R., e. 2018, in prep.

\bibitem[{{Tolstoy} {et~al.}(2009){Tolstoy}, {Hill}, \& {Tosi}}]{tolstoy2009}
{Tolstoy}, E., {Hill}, V., \& {Tosi}, M. 2009, Annual Review of Astron and
  Astrophys, 47, 371

\bibitem[{{Tolstoy} {et~al.}(2001){Tolstoy}, {Irwin}, {Cole}, {Pasquini},
  {Gilmozzi}, \& {Gallagher}}]{tolstoy2001}
{Tolstoy}, E., {Irwin}, M.~J., {Cole}, A.~A., {et~al.} 2001, \mnras, 327, 918

\bibitem[{{Tolstoy} {et~al.}(2004){Tolstoy}, {Irwin}, {Helmi}, {Battaglia},
  {Jablonka}, {Hill}, {Venn}, {Shetrone}, {Letarte}, {Cole}, {Primas},
  {Francois}, {Arimoto}, {Sadakane}, {Kaufer}, {Szeifert}, \&
  {Abel}}]{tolstoy2004}
{Tolstoy}, E., {Irwin}, M.~J., {Helmi}, A., {et~al.} 2004, \apjl, 617, L119

\bibitem[{{Tornatore} {et~al.}(2007){Tornatore}, {Borgani}, {Dolag}, \&
  {Matteucci}}]{tornatore2007}
{Tornatore}, L., {Borgani}, S., {Dolag}, K., \& {Matteucci}, F. 2007, \mnras,
  382, 1050

\bibitem[{{Truelove} {et~al.}(1997){Truelove}, {Klein}, {McKee}, {Holliman},
  {Howell}, \& {Greenough}}]{truelove1997}
{Truelove}, J.~K., {Klein}, R.~I., {McKee}, C.~F., {et~al.} 1997, \apjl, 489,
  L179

\bibitem[{{Tsujimoto} {et~al.}(2015){Tsujimoto}, {Ishigaki}, {Shigeyama}, \&
  {Aoki}}]{tsujimoto2015}
{Tsujimoto}, T., {Ishigaki}, M.~N., {Shigeyama}, T., \& {Aoki}, W. 2015, \pasj,
  67, L3

\bibitem[{{Tsujimoto} {et~al.}(1995){Tsujimoto}, {Nomoto}, {Yoshii},
  {Hashimoto}, {Yanagida}, \& {Thielemann}}]{tsujimoto1995}
{Tsujimoto}, T., {Nomoto}, K., {Yoshii}, Y., {et~al.} 1995, \mnras, 277, 945

\bibitem[{{Valcke} {et~al.}(2008){Valcke}, {de Rijcke}, \&
  {Dejonghe}}]{valcke2008}
{Valcke}, S., {de Rijcke}, S., \& {Dejonghe}, H. 2008, \mnras, 389, 1111

\bibitem[{{Vargas} {et~al.}(2014){Vargas}, {Geha}, \& {Tollerud}}]{vargas2014}
{Vargas}, L.~C., {Geha}, M.~C., \& {Tollerud}, E.~J. 2014, \apj, 790, 73

\bibitem[{{Vazdekis} {et~al.}(1996){Vazdekis}, {Casuso}, {Peletier}, \&
  {Beckman}}]{vazdekis96}
{Vazdekis}, A., {Casuso}, E., {Peletier}, R.~F., \& {Beckman}, J.~E. 1996,
  \apjs, 106, 307

\bibitem[{{Venn} {et~al.}(2012){Venn}, {Shetrone}, {Irwin}, {Hill}, {Jablonka},
  {Tolstoy}, {Lemasle}, {Divell}, {Starkenburg}, {Letarte}, {Baldner},
  {Battaglia}, {Helmi}, {Kaufer}, \& {Primas}}]{venn2012}
{Venn}, K.~A., {Shetrone}, M.~D., {Irwin}, M.~J., {et~al.} 2012, \apj, 751, 102

\bibitem[{{Verbeke} {et~al.}(2015){Verbeke}, {Vandenbroucke}, \& {De
  Rijcke}}]{verbeke2015}
{Verbeke}, R., {Vandenbroucke}, B., \& {De Rijcke}, S. 2015, \apj, 815, 85

\bibitem[{{Vogelsberger} {et~al.}(2014){Vogelsberger}, {Zavala}, {Simpson}, \&
  {Jenkins}}]{vogelsberger2014}
{Vogelsberger}, M., {Zavala}, J., {Simpson}, C., \& {Jenkins}, A. 2014, \mnras,
  444, 3684

\bibitem[{{Walker} {et~al.}(2009{\natexlab{a}}){Walker}, {Mateo}, \&
  {Olszewski}}]{walker2009}
{Walker}, M.~G., {Mateo}, M., \& {Olszewski}, E.~W. 2009{\natexlab{a}}, \aj,
  137, 3100

\bibitem[{{Walker} {et~al.}(2009{\natexlab{b}}){Walker}, {Mateo}, {Olszewski},
  {Pe{\~n}arrubia}, {Wyn Evans}, \& {Gilmore}}]{walker2009b}
{Walker}, M.~G., {Mateo}, M., {Olszewski}, E.~W., {et~al.} 2009{\natexlab{b}},
  \apj, 704, 1274

\bibitem[{{Weisz} {et~al.}(2014{\natexlab{a}}){Weisz}, {Dolphin}, {Skillman},
  {Holtzman}, {Gilbert}, {Dalcanton}, \& {Williams}}]{weisz2014}
{Weisz}, D.~R., {Dolphin}, A.~E., {Skillman}, E.~D., {et~al.}
  2014{\natexlab{a}}, \apj, 789, 147

\bibitem[{{Weisz} {et~al.}(2014{\natexlab{b}}){Weisz}, {Skillman}, {Hidalgo},
  {Monelli}, {Dolphin}, {McConnachie}, {Bernard}, {Gallart}, {Aparicio},
  {Boylan-Kolchin}, {Cassisi}, {Cole}, {Ferguson}, {Irwin}, {Martin}, {Mayer},
  {McQuinn}, {Navarro}, \& {Stetson}}]{weisz2014b}
{Weisz}, D.~R., {Skillman}, E.~D., {Hidalgo}, S.~L., {et~al.}
  2014{\natexlab{b}}, \apj, 789, 24

\bibitem[{{Weisz} {et~al.}(2012){Weisz}, {Zucker}, {Dolphin}, {Martin}, {de
  Jong}, {Holtzman}, {Dalcanton}, {Gilbert}, {Williams}, {Bell}, {Belokurov},
  \& {Wyn Evans}}]{weisz2012}
{Weisz}, D.~R., {Zucker}, D.~B., {Dolphin}, A.~E., {et~al.} 2012, \apj, 748, 88

\bibitem[{{Wetzel} {et~al.}(2016){Wetzel}, {Hopkins}, {Kim},
  {Faucher-Gigu{\`e}re}, {Kere{\v s}}, \& {Quataert}}]{wetzel2016}
{Wetzel}, A.~R., {Hopkins}, P.~F., {Kim}, J.-h., {et~al.} 2016, \apjl, 827, L23

\bibitem[{{Wiersma} {et~al.}(2009){Wiersma}, {Schaye}, {Theuns}, {Dalla
  Vecchia}, \& {Tornatore}}]{wiersma2009}
{Wiersma}, R.~P.~C., {Schaye}, J., {Theuns}, T., {Dalla Vecchia}, C., \&
  {Tornatore}, L. 2009, \mnras, 399, 574

\bibitem[{{Wise} {et~al.}(2014){Wise}, {Demchenko}, {Halicek}, {Norman},
  {Turk}, {Abel}, \& {Smith}}]{wise2014}
{Wise}, J.~H., {Demchenko}, V.~G., {Halicek}, M.~T., {et~al.} 2014, \mnras,
  442, 2560

\bibitem[{{Yajima} {et~al.}(2011){Yajima}, {Choi}, \& {Nagamine}}]{yajima2011}
{Yajima}, H., {Choi}, J.-H., \& {Nagamine}, K. 2011, \mnras, 412, 411

\bibitem[{{Zolotov} {et~al.}(2012){Zolotov}, {Brooks}, {Willman}, {Governato},
  {Pontzen}, {Christensen}, {Dekel}, {Quinn}, {Shen}, \&
  {Wadsley}}]{zolotov2012}
{Zolotov}, A., {Brooks}, A.~M., {Willman}, B., {et~al.} 2012, \apj, 761, 71

\end{thebibliography}
\nocite{*}

\end{document}